\documentclass[a4paper,twoside,10pt,cleardoubleempty,bibtotoc]{scrbook}
\areaset[1cm]{343pt}{640pt}
\usepackage[latin1]{inputenc}
\usepackage[T1]{fontenc}
\usepackage[dvips]{graphicx}
\usepackage{amsfonts,amsmath,booktabs}
\usepackage{ae,aecompl}
\usepackage[tableposition=top,font=small,labelfont=bf,textfont=it,format=hang]{caption}
\usepackage{subfig}
\setkomafont{sectioning}{\usefont{OT1}{phv}{bc}{n}\selectfont}
\usepackage{type1cm}
\usepackage[Lenny]{fncychapcustom}
\usepackage[sort,numbers]{natbib}
\makeatletter
\newcommand*{\c@thmpro}{\c@theorem}
\newcommand*{\p@thmpro}{\p@theorem}

\makeatother 
\makeatletter
\newcommand*{\c@thmlem}{\c@theorem}
\newcommand*{\p@thmlem}{\p@theorem}

\makeatother 
\makeatletter
\newcommand*{\c@thmcor}{\c@theorem}
\newcommand*{\p@thmcor}{\p@theorem}

\makeatother 
\makeatletter
\newcommand*{\c@thmcon}{\c@theorem}
\newcommand*{\p@thmcon}{\p@theorem}

\makeatother 
\makeatletter
\newcommand*{\c@thmexa}{\c@theorem}
\newcommand*{\p@thmexa}{\p@theorem}

\makeatother 
\makeatletter
\newcommand*{\c@thmdef}{\c@theorem}
\newcommand*{\p@thmdef}{\p@theorem}

\makeatother 
\makeatletter
\newcommand*{\c@thmconstr}{\c@theorem}
\newcommand*{\p@thmconstr}{\p@theorem}

\makeatother 
\makeatletter
\newcommand*{\c@thmprob}{\c@theorem}
\newcommand*{\p@thmprob}{\p@theorem}

\makeatother 
\usepackage{amsthm}
\newtheorem{theorem}{Theorem}[chapter]
\newtheorem{lemma}[thmlem]{Lemma}
\newtheorem{proposition}[thmpro]{Proposition}
\newtheorem{corollary}[thmcor]{Corollary}

\theoremstyle{definition}
\newtheorem{example}[thmexa]{Example}
\newtheorem{definition}[thmdef]{Definition}
\newtheorem{construction}[thmconstr]{Construction}
\newtheorem{problem}[thmprob]{Problem}
\theoremstyle{remark}
\newtheorem*{remark}{Remark}
\usepackage[dvips,colorlinks=false,bookmarks=true,bookmarksnumbered,plainpages=false,pdfpagelabels,hypertexnames=false,linktocpage]{hyperref}
\usepackage[chapter]{algorithm}
\makeatletter
\providecommand*{\toclevel@algorithm}{0}
\makeatother 
\usepackage{algpseudocode}
\newcommand\algorithmicto{\textbf{to}}
\algrenewtext{For}[3]{\algorithmicfor\ #1 $\gets$ #2 \algorithmicto\ #3 \algorithmicdo}

\DeclareMathOperator{\rev}{rev}
\DeclareMathOperator{\pre}{pre}
\DeclareMathOperator{\conc}{conc}
\DeclareMathOperator{\GF}{GF}
\DeclareMathOperator{\PC}{PC}
\DeclareMathOperator{\EPC}{EPC}
\DeclareMathOperator{\APC}{APC}

\newcommand{\bra}[1]{\left< #1 \right|}
\newcommand{\ket}[1]{\left| #1 \right>}
\newlength{\skpl}
\hypersetup{
pdftitle = {On Self-Dual Quantum Codes, Graphs, and Boolean Functions},
pdfauthor = {Lars Eirik Danielsen}}

\begin{document}

\titlehead{
The Selmer Center \\
Department of Informatics\\
University of Bergen\\
Norway \\
\vspace{59pt}}
\subject{Master of Science Thesis}
\title{On Self-Dual Quantum Codes, Graphs, and Boolean Functions}
\author{Lars Eirik Danielsen \\ \vspace{59pt} \\ \includegraphics[width=130pt]{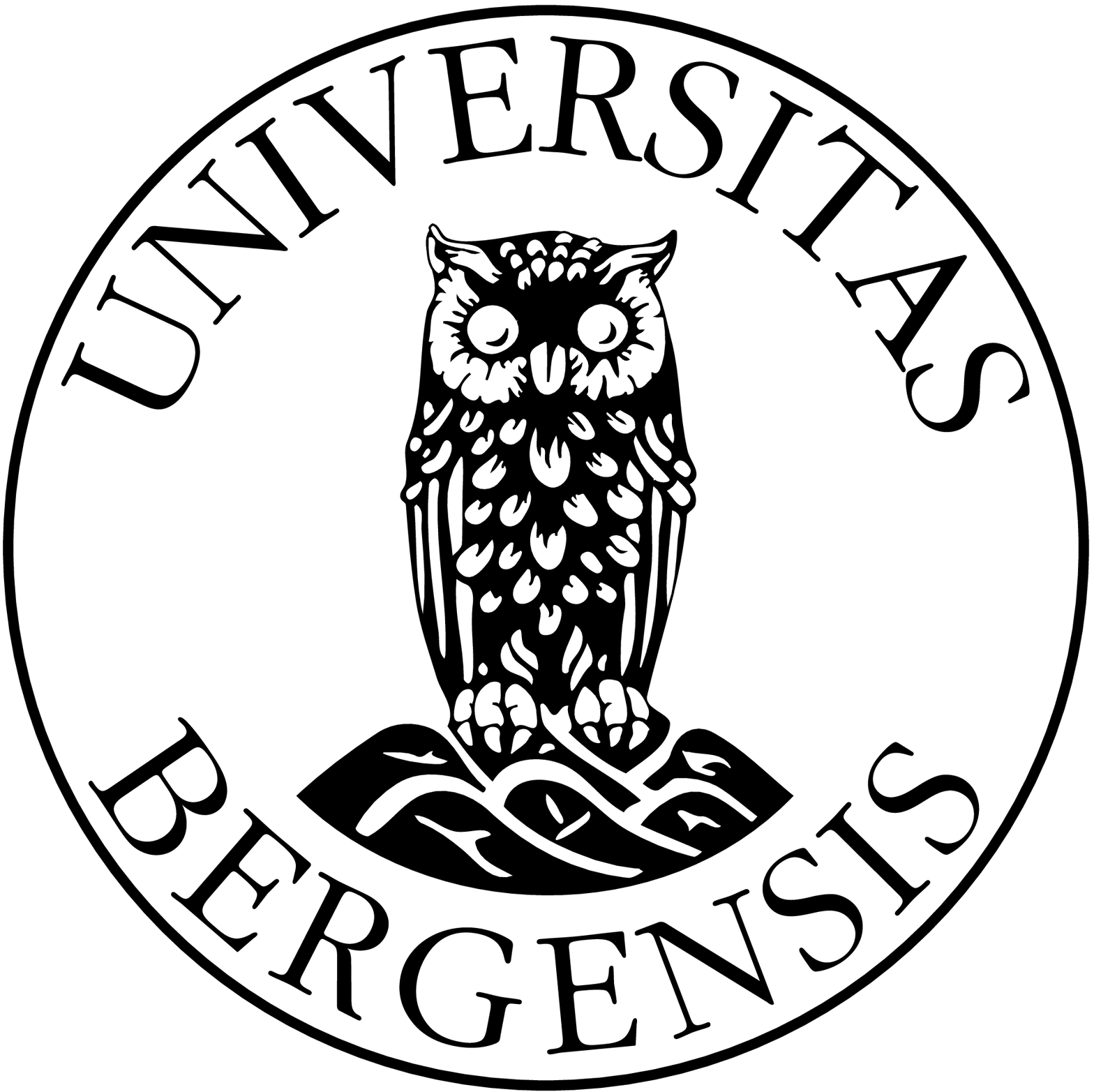} \\ \vspace{59pt}
\\ March 2005}
\date{}

\maketitle

\frontmatter

\chapter{Abstract}

A short introduction to \emph{quantum error correction} is given, and it is shown that
\emph{zero-dimensional quantum codes} can be represented as
\emph{self-dual additive codes over $\GF(4)$} and also as \emph{graphs}.
We show that graphs representing several such codes with high minimum distance can
be described as \emph{nested regular graphs} having \emph{minimum regular vertex degree}
and containing long cycles.
Two graphs correspond to equivalent quantum codes if they are related by
a sequence of \emph{local complementations}.
We use this operation to generate orbits of graphs,
and thus classify all inequivalent self-dual additive
codes over $\GF(4)$ of length up to 12, where previously only all codes
of length up to 9 were known. We show that these codes can be interpreted as 
\emph{quadratic Boolean functions}, and we define \emph{non-quadratic quantum codes}, 
corresponding to Boolean functions of higher degree.
We look at various cryptographic properties of Boolean functions, in particular
the \emph{propagation criteria}. The new \emph{aperiodic propagation criterion} (APC)
and the \emph{APC distance} are then defined.
We show that the distance of a zero-dimensional quantum code is equal to the 
APC distance of the corresponding Boolean function. 
Orbits of Boolean functions with respect to the $\{I,H,N\}^n$ transform set
are generated. We also study the \emph{peak-to-average power ratio} with respect 
to the $\{I,H,N\}^n$ transform set (PAR$_{IHN}$), and prove that
PAR$_{IHN}$ of a quadratic Boolean function is related to the 
size of the maximum independent set over the corresponding orbit of graphs.
A construction technique for non-quadratic Boolean functions with low PAR$_{IHN}$
is proposed. It is finally shown that both PAR$_{IHN}$ and APC distance can
be interpreted as partial entanglement measures.

\chapter{Acknowledgements}
I would like to thank my supervisor, Matthew G. Parker, for all his helpful 
advice and good ideas. I also thank Tor Helleseth and 
the Selmer Center for financial support enabling me to attend
the conference ``Sequences and Their Applications'', SETA'04,
in Seoul, South Korea, where some of the results in this thesis were
presented. Most of the contributions in this thesis are also found in the 
two papers referenced below.

\vskip \baselineskip
\noindent {\sc Danielsen, L.~E. and Parker, M.~G.}: ``Spectral orbits and
  peak-to-average power ratio of Boolean functions with respect to the
  $\{I,H,N\}^n$ transform'', January 2005. To appear in the proceedings of
  Sequences and Their Applications, SETA'04, Lecture Notes in Computer Science,
  Springer-Verlag.
  \newline\url{http://www.ii.uib.no/~larsed/papers/seta04-parihn.pdf}

\vskip \baselineskip
\noindent {\sc Danielsen, L.~E., Gulliver, T.~A., and Parker, M.~G.}: ``Aperiodic
  propagation criteria for Boolean functions'', October 2004. Submitted to
  Information and Computation. 
  \newline\url{http://www.ii.uib.no/~larsed/papers/apc.pdf}

\vskip 3\baselineskip
\noindent \emph{Lars Eirik Danielsen \\ Bergen, March 2005}

\tableofcontents
\listoftables
\listoffigures
\listofalgorithms

\mainmatter

\chapter{Introduction}

\section{Motivation}

In this thesis we will look at a set of objects that can,
under suitable interpretations, be represented as 
\begin{itemize}
\item \emph{zero-dimensional quantum codes},
\item \emph{quantum states},
\item \emph{self-dual additive codes over $\GF(4)$},
\item \emph{isotropic systems},
\item \emph{simple undirected graphs},
\item and \emph{quadratic Boolean functions}.
\end{itemize}
Each interpretation reveals different properties
about the underlying objects and suggests different generalisations.

There has been a lot of interest in \emph{quantum computing} since
the discovery of Shor's algorithm, which can factor an integer in
polynomial time. Practical quantum computers have not yet been built, but it 
is clear that \emph{quantum error correction} must be a crucial part of any
implementation. \emph{Zero-dimensional quantum codes} only represent
single \emph{quantum states}, but are still of interest to physicists since
codes of high distance represent highly \emph{entangled} states which
could be used for testing the decoherence properties of a quantum computer.
It has also been shown that a special type of 
quantum computer can be implemented by performing measurements
on a particular class of entangled states.

Zero-dimensional \emph{quantum stabilizer codes} can be represented as
\emph{self-dual additive codes over $\GF(4)$}. These codes
are of interest to coding theorists, and several construction techniques and
classifications have been published.
A code of this type can be represented by an \emph{isotropic system},
a combinatorical object that has been the subject of much research.
A self-dual additive code over $\GF(4)$ can also be represented by
a \emph{simple undirected graph}. This allows us to use concepts and algorithms
from graph theory to characterise the graphs corresponding to strong codes.
A generalisation to \emph{hypergraphs} is also suggested by this
interpretation.

The same objects are also equivalent to \emph{quadratic Boolean functions},
and the generalisation to Boolean functions of higher degree is natural.
Boolean functions are of great interest to cryptographers, since
they can be used, for instance, to analyse and construct S-boxes in block ciphers
and nonlinear combiners in stream ciphers.
Many criteria for the cryptographic strength of Boolean functions exist,
and it turns out that zero-dimensional quantum codes with high minimum distance 
correspond to Boolean functions that satisfy such a criterion.
This suggests that highly entangled quantum states may correspond to
cryptographically strong Boolean functions. Conversely, various properties 
derived from transformations of Boolean functions can be interpreted
as partial entanglement measures of the corresponding quantum states.

\section{Overview}

\hyperref[l2]{The second chapter} of this thesis gives a very short
introduction to the theory of \emph{quantum computing} and \emph{quantum error correction}.
The properties of \emph{superposition} and \emph{entanglement} are explained, and
we we show how computer algorithms can be implemented by using transformations and measurements of
quantum states. Some of the important discoveries in the theory of quantum computing 
are also mentioned. A few basic concepts from classical coding theory are presented
before we see that error correction is also possible in quantum computers.
Although an infinite number of different errors can affect a quantum state, we
show that quantum codes only need to consider a small set of basis errors.
Quantum codes can be expressed in the \emph{stabilizer formalism}, but
have an equivalent representation as \emph{additive codes over $\GF(4)$}.
We finally introduce the type of codes that will be studied in this thesis,
namely quantum codes of dimension zero, called \emph{self-dual quantum codes},
which represent single quantum states. Some bounds on the distance
of such codes are presented.

\hyperref[l17]{Chapter~3} starts with an introduction to graph theory and defines the
notation that we will use. It is then shown that the computer program \texttt{nauty} can
detect \emph{graph isomorphisms}. By using a simple mapping from \emph{hypergraphs} to
ordinary graphs, \texttt{nauty} can also detect isomorphism of hypergraphs.
A special type of self-dual quantum codes are the \emph{graph codes} which can be
represented by undirected graphs. It can be shown that any self-dual quantum code is equivalent
to a graph code, and therefore that there is a one-to-one correspondence between the set
of simple undirected graphs and the set of self-dual quantum codes.
A method for converting any self-dual quantum code into a graph code is described.
By exploiting the special form of the generator matrix of a graph code, the distance
and \emph{partial weight distribution} of the code can be found by efficient algorithms.
The well-known \emph{Quadratic residue construction} can be used to find 
self-dual quantum codes of high distance. These codes can be represented by a class of
\emph{strongly regular} graphs, called \emph{Paley graphs}. A small modification of such a code
produces a \emph{bordered quadratic residue code}. We construct quadratic residue codes,
and their bordered versions, for all possible lengths up to 30.
For length 18, the quadratic residue construction does not give an optimal code, but
there is a modified technique that does.

In \autoref{l32}, we look at the graphs corresponding to 
two well-known self-dual quantum codes, the Hexacode and the Dodecacode.
Both codes can be represented by graphs with a special nested structure,
which we define as \emph{nested clique graphs}.
We show that there is a lower bound on the vertex degree in graphs representing
self-dual quantum codes, and that graphs with \emph{minimum regular vertex degree}
satisfy this bound with equality.
We perform an exhaustive search of all graph codes with circulant generator matrices,
for lengths up to 30. Many codes with optimal distance and minimum regular vertex degree
are identified, and their nested structures are described. The more general
\emph{nested regular graphs} are also defined. We finally discuss the observation 
that nested regular graphs corresponding to codes of high distance also contain long cycles.

\hyperref[chap:orbits]{Chapter~5} deals with the equivalence of self-dual quantum codes.
We first see that the quantum states represented by equivalent self-dual
quantum codes are related by a simple transformation.
This transformation corresponds to a simple operation, known as \emph{local complementation},
on the graph representations of the codes. In addition to local complementations, 
graph isomorphism must be considered, since isomorphic graphs also 
correspond to equivalent quantum codes. We give three different algorithms for 
generating \emph{LC orbits}, the equivalence classes of self-dual quantum codes
with respect to local complementation and graph isomorphism.
By implementing these algorithms, using various optimisation techniques and
a cluster computer, we are able to generate all LC orbits of codes of length
up to 12. This gives a complete classification of all self-dual additive codes
over $\GF(4)$ of length up to 12, where previously only all codes of length up to 9 had
been classified. A database containing a representative of each LC orbit is also available.
We next look at the LC orbits of some strong codes, and search for 
regular graph structures. The non-existence of any regular graph representation
is established for some codes. Finally, we prove that a single LC operation 
on the graph corresponding to a bordered quadratic residue code produces a regular graph.

\emph{Boolean functions} are introduced in \autoref{chap:boolean}. 
The \emph{algebraic normal form transformation} and the \emph{Walsh-Hadamard transformation}
are defined, and an efficient algorithm for these and other transformations
is described. After defining the \emph{periodic autocorrelation}, we see how Boolean functions
are used in cryptography. The properties of \emph{correlation immunity}, \emph{resilience},
and \emph{perfect nonlinearity} are of particular interest in this context.
We also study the more general \emph{propagation criteria}, and
define the new \emph{aperiodic propagation criterion} (APC), which is related
to the \emph{aperiodic autocorrelation}. We also define the \emph{APC distance}
of a Boolean function. It is explained that Boolean functions can be interpreted
as quantum states, and that quadratic Boolean functions correspond to the self-dual quantum codes
studied in the previous chapters. Boolean functions of higher degree can be represented
by hypergraphs and correspond to a new type of zero-dimensional quantum codes, the \emph{non-quadratic
quantum codes}. We see how errors on a quantum state can be expressed as operations on
the corresponding Boolean function, and show that the distance of
a zero-dimensional quantum code is equal to the APC distance of the corresponding 
Boolean function. The transform set $\{I,H,N\}^n$ is introduced, and it is shown that
the LC orbits of equivalent self-dual quantum codes can be generated by this transform set.
Finally, we define two types of orbits of Boolean functions and enumerate all inequivalent 
functions of up to 5 variables, and all functions of 6 variables with degree up to 3.
We also give examples of non-quadratic quantum codes with high APC distance.

In \autoref{chap:par}, we study another property of Boolean functions and 
their corresponding graphs and quantum states, namely the 
\emph{peak-to-average power ratio} with respect 
to the $\{I,H,N\}^n$ transform (PAR$_{IHN}$). We calculate the PAR$_{IHN}$ of all
quadratic Boolean functions with up to 12 variables, and we prove that the PAR$_{IHN}$
of a quadratic Boolean function equals $2^\lambda$,
where $\lambda$ is the size of the largest independent set in the corresponding 
LC orbit of graphs. We also define $\Lambda_n$, the minimum value of $\lambda$ over
all LC orbits of graphs on $n$ vertices. The values of $\Lambda_n$ for $n$ up to 14
are given, and bounds on $\Lambda_n$ are provided for $n$ up to 21.
A construction technique for non-quadratic Boolean functions with low PAR$_{IHN}$
is proposed, using good quadratic functions as building blocks.
We also look at PAR with respect to other transform sets, in particular
PAR$_{IH}$ and PAR$_{\mathcal{U}}$, and show that PAR$_{\mathcal{U}} =$ PAR$_{IH}$ for 
quadratic Boolean functions corresponding to bipartite graphs.
We show that APC distance and PAR$_{IHN}$ tell something about the degree of entanglement
in a quantum state and briefly mention other measures derived from the $\{I,H,N\}^n$ spectrum.
We also show that PAR$_{IHN}$ is related to an entanglement measure known as the \emph{Schmidt measure}.

We give some final conclusions and present some open problems and ideas for future
research in \autoref{chap:concl}.

While \autoref{l2} and \autoref{l17} of this thesis mostly contain previously 
known results, the later chapters contain many new contributions, and most of these are listed here.
\begin{itemize}
\item An exhaustive search of all circulant graph codes of length up to 30 is performed.
\item It is shown that many self-dual quantum codes of high distance can
      be represented by \emph{nested clique graphs} or \emph{nested regular graphs}, and
      that these graphs also contain long cycles.
\item \emph{Minimum regular vertex degree} is defined, and many graphs with this property 
      are identified, corresponding to self-dual quantum codes of high distance.
\item All self-dual additive codes over $\GF(4)$ of length up to 12 are classified and made available 
      in a \href{http://www.ii.uib.no/~larsed/vncorbits/}{database}~\cite{B14}. 
      Previously only all codes of length up to 9 were known. The new numbers of inequivalent 
      codes have been added to 
      \emph{The On-Line Encyclopedia of Integer Sequences}~\cite{B56}.
\item It is shown that there are no regular graphs corresponding to
      [[11,0,5]] or [[18,0,8]] codes, but that \emph{bordered quadratic residue codes}
      can be transformed into regular graphs by a simple graph operation.
\item The \emph{aperiodic propagation criterion} and the \emph{APC distance} of a Boolean 
      function are defined. It is shown that Boolean functions with APC distance~$d$ 
      can be interpreted as zero-dimensional quantum codes with distance~$d$.
\item We define \emph{non-quadratic quantum codes}, corresponding to hypergraphs and Boolean
      functions of degree higher than two. Several non-quadratic quantum codes with high
      distance are found.
\item We define two types of orbits of Boolean functions and enumerate all inequivalent 
      functions of up to 5 variables, and all functions of 6 variables with degree up to 3.
\item The \emph{peak-to-average power ratio} with respect to the $\{I,H,N\}^n$ transform (PAR$_{IHN}$)
      is studied, and it is shown that the PAR$_{IHN}$ of a quadratic Boolean function
      equals $2^\lambda$, where $\lambda$ is the size of the largest independent set in the corresponding 
      LC orbit of graphs.
\item We define $\Lambda_n$, the minimum value of $\lambda$ over all LC orbits of graphs on $n$ vertices, 
      and give the values of $\Lambda_n$ for $n$ up to 14. Bounds on $\Lambda_n$ are provided for $n$ up to 21.
\item A construction technique for non-quadratic Boolean functions with low PAR$_{IHN}$
      is proposed.
\end{itemize}

\chapter{Quantum Computing and Quantum Codes}\label{l2}

\section{Quantum Computing}\label{l3}

\subsection{Introduction}

We will only give a brief presentation of quantum computing. For more details,
we refer to some of the many good introductions to the topic~\cite{B33, B50, B36}.
Quantum mechanics is a physical theory that describes the behaviour of
elementary particles, such as atoms or photons. The laws of quantum mechanics 
predicts effects which are very different from the physical reality that we
ordinarily observe. Of particular interest are the properties of \emph{superposition}
and \emph{entanglement}.

\subsection{Quantum Superposition}

\begin{figure}[t]
\centering
\subfloat[Setup With Only Filter A]{\includegraphics{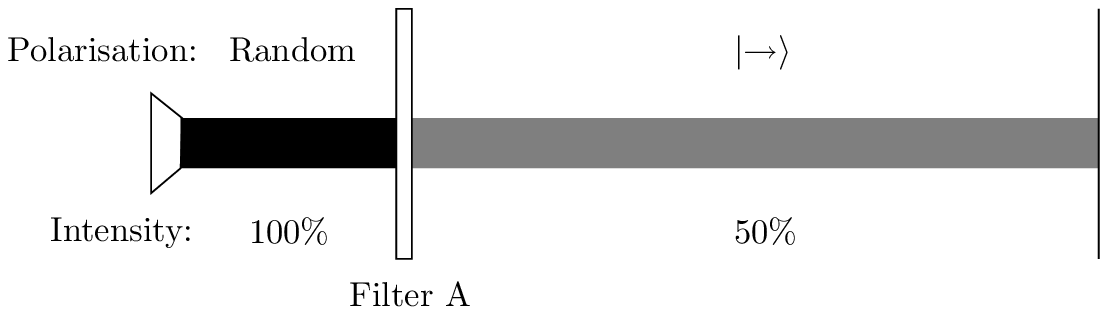}}\label{l1}
\subfloat[Setup With Filters A and B]{\includegraphics{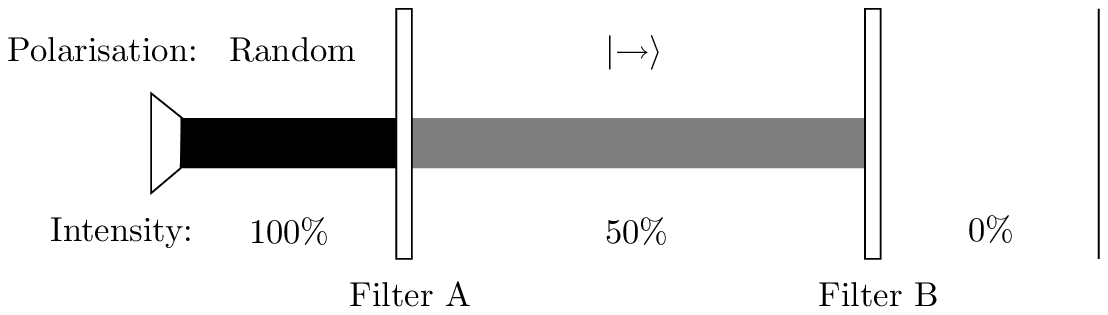}}\label{l4}
\subfloat[Setup With Filters A, C, and B]{\includegraphics{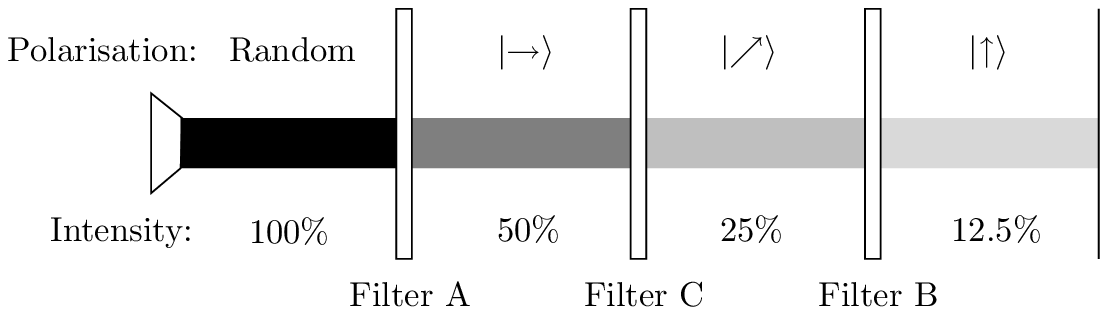}}\label{l5}
\caption{Demonstrating Quantum Effects With Polarisation Filters}\label{l6}
\end{figure}

A simple experiment demonstrating quantum effects uses the polarisation of light.
The light from an ordinary light source consists of photons with a random
polarisation. If we put filter $A$, which has horizontal ($0^\circ$) polarisation, between
the light source and a screen, as shown in \autoref{l1}, 
the intensity of the light reflected from the
screen will be half of the original, and all photons that pass the filter will now have 
horizontal polarisation. This can be verified by adding filter $B$, which has 
vertical ($90^\circ$) polarisation, between filter $A$ and the screen. This time, no light 
reaches the screen at all, as seen in \autoref{l4}.
A most confusing fact is that after adding another filter between $A$ and $B$,
some of the photons do reach the screen. The
filter $C$, with $45^\circ$ polarisation, is added between filters $A$ and $B$, 
and as shown in \autoref{l5}, we will observe light with $\frac{1}{8}$ of the 
original intensity reflected from the screen.

What happens is that the randomly polarised photons are ``measured'' when
they hit filter $A$. Half of them get a horizontal polarisation and pass
through, and the other half get a vertical polarisation and are stopped.
If the horizontally polarised photons then hit filter $B$, they will all
be stopped. But if they hit filter $C$ they will again be ``measured'', but now
with respect to another basis. Half of them will receive a $45^\circ$ polarisation
and pass through. The other half get an orthogonal ($135^\circ$) polarisation and
are stopped. One fourth of the original photons will pass through both filter $A$ and $C$.
If these photons, which now have a $45^\circ$ polarisation, then hit filter $B$, 
they will again be ``measured'', and half of them, $\frac{1}{8}$ of the initial amount,
will pass through.

The results observed in this experiment are due the property of \emph{quantum
superposition}. An object which is in a superposition can be viewed as having two
or more values for an observable quantity at the same time. Once the quantity
is measured, the superposition will randomly collapse into one of the values, according
to probabilities associated with each possible outcome.
A photon could, for instance, have horizontal polarisation with probability $a$
and at the same time vertical polarisation with probability $b$. When this photon
is ``measured'' by a horizontal polarisation filter, it will with probability $a$ 
receive horizontal polarisation and pass through the filter, and with
probability $b$ receive vertical polarisation and be stopped by the filter.

\subsection{Bra/Ket Notation}

The \emph{bra/ket}-notation invented by Dirac is much used in quantum mechanics.
$\bra{\phi}$ is a \emph{bra} (the left side of a bracket), and $\ket{\phi}$ is a \emph{ket}
(the right side of a bracket). Kets are used to describe states. The state of horizontal
polarisation could be described by $\ket{\rightarrow}$, and vertical polarisation
by $\ket{\uparrow}$.
A photon which is in a superposition of these states could be described by
$\alpha\ket{\rightarrow} + \beta\ket{\uparrow}$, where $\alpha$ and $\beta$ are 
complex numbers. $|\alpha|^2$ is then the probability of the state collapsing to 
$\ket{\rightarrow}$ upon measurement, and $|\beta|^2$ is the probability of measuring 
$\ket{\uparrow}$. We must have that $|\alpha|^2 + |\beta|^2 = 1$.

Measurement of a quantum state must be done with respect to a specific basis. In the
experiment with photons we used the bases $\left\{ \ket{\rightarrow},\ket{\uparrow} \right\}$ 
and $\left\{ \ket{\nearrow},\ket{\nwarrow} \right\}$. Since 
$\ket{\uparrow} = \frac{1}{\sqrt{2}} \ket{\nearrow} + \frac{1}{\sqrt{2}} \ket{\nwarrow}$
and $\ket{\rightarrow} = \frac{1}{\sqrt{2}} \ket{\nearrow} - \frac{1}{\sqrt{2}} \ket{\nwarrow}$,
vertically and horizontally polarised photons will have probability $\frac{1}{2}$ of getting through a
filter with $45^\circ$ polarisation, which is what we observed in the experiment.
The negative sign in the expression 
$\ket{\rightarrow} = \frac{1}{\sqrt{2}} \ket{\nearrow} - \frac{1}{\sqrt{2}} \ket{\nwarrow}$
denotes \emph{phase}. Information about the phase is lost once the superposition collapses,
and in this example it can be ignored.
With the basis states of a measurement basis we associate orthonormal basis vectors.
For example, $\ket{\rightarrow} = \binom{1}{0}$ and $\ket{\uparrow} = \binom{0}{1}$.
The state $\alpha \ket{\rightarrow} + \beta \ket{\uparrow}$ can then be described by the vector 
$\binom{\alpha}{\beta}$. Each ket has a corresponding bra, $\bra{\phi} = \ket{\phi}^\dag$, 
where the operator $\dag$ first conjugates and then transposes a vector,
e.g., $\binom{\alpha}{\beta}^\dag = (\overline{\alpha}, \overline{\beta})$.
The \emph{inner product}, $\bra{\phi}\ket{\psi}$ (also written $\left< \phi | \psi \right>$),
is a scalar, and is equal to zero if the vectors associated with $\ket{\phi}$ and $\ket{\psi}$ 
are orthogonal. The \emph{outer product}, $\ket{\phi}\bra{\psi}$, is a matrix which can be used
to express transformations on quantum states.

\subsection{Quantum Bits}
A quantum bit, or \emph{qubit}, has two possible states, labelled $\ket{0}$ and $\ket{1}$.
All measurements will be done with respect to the basis $\left\{ \ket{0},\ket{1} \right\}$.
A qubit can be represented by any two-level quantum system. Using vertical and horizontal polarisation 
of a photon, we could assign $\ket{0}=\ket{\rightarrow}$ and $\ket{1}=\ket{\uparrow}$, or
$\ket{0}=\ket{\nearrow}$ and $\ket{1}=\ket{\nwarrow}$. Other possible implementations are the
up/down spin of an electron or two energy levels of an atom.

Unlike a classical bit, a qubit can be in a superposition of $\ket{0}$ and $\ket{1}$.
The state of a general qubit can be denoted $\alpha\ket{0} + \beta\ket{1}$, with
$|\alpha|^2$ being the probability of getting the result $\ket{0}$ when measuring
the qubit, and $|\beta|^2$ the probability of getting a $\ket{1}$.
Several qubits can be combined to form a \emph{quantum register}. The state of a two-qubit
register can be denoted $\alpha\ket{00} + \beta\ket{01} + \gamma\ket{10} + \delta\ket{11}$,
or equivalently by the vector $(\alpha, \beta, \gamma, \delta)^T$,
where $|\alpha|^2$ is the probability of measuring both qubits as zero, and so on.
It is also possible to measure only one of the two qubits. If we measure the first qubit, 
the probability of getting $\ket{0}$ is $|\alpha|^2 + |\beta|^2$, and the probability of getting
$\ket{1}$ is $|\gamma|^2 + |\delta|^2$. Upon measurement, the state will collapse, so
later measurements of the same qubit will always yield the same value as the first time.
If the first qubit is measured as $\ket{0}$, the remaining state is
\[
\frac{\alpha}{\sqrt{|\alpha|^2 + |\beta|^2}}\ket{00} + 
\frac{\beta}{\sqrt{|\alpha|^2 + |\beta|^2}}\ket{01}.
\]
If the first qubit is measured as $\ket{1}$, the remaining state is
\[
\frac{\gamma}{\sqrt{|\gamma|^2 + |\delta|^2}}\ket{10} + 
\frac{\delta}{\sqrt{|\gamma|^2 + |\delta|^2}}\ket{11}.
\]

If $\alpha$, $\beta$, $\gamma$ and $\delta$ are all equal to $\frac{1}{2}$, the
first qubit we measure will be $\ket{0}$ or $\ket{1}$ with probability $\frac{1}{2}$,
and the second qubit we measure will also be $\ket{0}$ or $\ket{1}$ with probability 
$\frac{1}{2}$. But this is not the general case.
Consider the state $\frac{1}{\sqrt{2}}\ket{00} + \frac{1}{\sqrt{2}}\ket{11}$.
If we measure the first qubit to be $\ket{0}$, the state will collapse to
$\ket{00}$, and if we measure the first qubit to be $\ket{1}$, the state
will collapse to $\ket{11}$. We see that the value of the second qubit is determined
when we measure the first, and that the two qubits will always have the same value.
We have observed another fundamental property of quantum mechanics, namely
\emph{quantum entanglement}.

\subsection{The Tensor Product}

\begin{definition}
The \emph{tensor product} (also known as the Kronecker product)
of the $n \times m$ matrix $A$, and the $k \times l$ matrix $B$, gives the $nk \times ml$ matrix
\begin{equation}
A \otimes B 
= \begin{pmatrix}AB_{0,0} & AB_{0,1} & \cdots & AB_{0,l-1}\\ AB_{1,0} & AB_{1,1} & \cdots & AB_{0,l-1}\\
\vdots & \vdots & \ddots & \vdots \\ AB_{k-1,0} & AB_{k-1,1} & \cdots & AB_{k-1,l-1}\end{pmatrix},
\end{equation}
where $B_{i,j}$ is the value in row $i$ and column $j$ of $B$.
\end{definition}

The tensor product of $\mathbf{u}$, a column vector of length $n$, and $\mathbf{v}$, a column vector
of length $k$, is a column vector of length $nk$,
\begin{equation}
\mathbf{u} \otimes \mathbf{v} = \begin{pmatrix}\mathbf{u} v_1\\\mathbf{u} v_2\\\vdots\\\mathbf{u} v_k\end{pmatrix}.
\end{equation}
When we write $\ket{01}$, it is in fact shorthand notation for $\ket{0} \otimes \ket{1}$, where
we take the tensor product of the basis vectors associated with the quantum states,
\[
\ket{01} = \ket{0} \otimes \ket{1} = \begin{pmatrix}0\\1\end{pmatrix} \otimes 
\begin{pmatrix}1\\0\end{pmatrix}
= \begin{pmatrix}\begin{pmatrix}0\\1\end{pmatrix} 1 \vspace{1pt}\\
\begin{pmatrix}0\\1\end{pmatrix} 0\end{pmatrix}
=
\begin{pmatrix}0\\1\\0\\0\end{pmatrix}.
\]

\subsection{Quantum Entanglement}

An entangled quantum state is a multi-qubit state where the values of the qubits are
not independent. There is no classical counterpart to this situation.
Qubits that are not entangled can be separated and described independently, 
using the tensor product.
For example, $\frac{1}{2}(\ket{00}+\ket{01}+\ket{10}+\ket{11}) = 
\frac{1}{\sqrt{2}}(\ket{0}+\ket{1)} \otimes \frac{1}{\sqrt{2}}(\ket{0}+\ket{1)}$. 
Measuring one of these qubits will not affect the outcome of the other.
The state $\frac{1}{\sqrt{2}}\ket{00} + \frac{1}{\sqrt{2}}\ket{11}$, however,
can not be factorised in this way, and we have already seen
that this is an entangled state.

Classical computers only use the tensor-factorisable space, and a
register of $n$ classical bits will at any time be in one of $2^n$ possible states.
A register of $n$ qubits in a quantum computer, however, has a state
space defined by $2^n$ basis vectors, which is an exponentially
larger space than in the classical case. The state space also
grows exponentially with the number of qubits. It is these properties
that give quantum computers their advantage.

A fascinating fact is that one can generate two maximally entangled
qubits, $\frac{1}{\sqrt{2}}(\ket{00} + \ket{11})$, (called
an EPR pair), and then separate the two particles by an arbitrary distance.
When we then measure one of the qubits, the combined state changes
instantaneously, and a later measurement of the second qubit will always
give the same value as the that of the first qubit. This effect of quantum
mechanics was thought to be a paradox, but it has been proved that it is not possible to
use entangled particles to communicate faster than the speed
of light, so there is no violation of the fundamental laws of physics.

In addition to being entangled with each other, it is possible that qubits
in a quantum register could be entangled with the environment, i.e., any
particles outside the register. Quantum states
that are entangled with the environment are called \emph{mixed states}, and
can be described by \emph{density operators}.
We will, however, only consider \emph{pure states}, i.e., quantum states that are
not entangled with the environment. 

\subsection{Quantum Transformations}

\begin{definition}
A matrix $U$ is a \emph{unitary matrix} if $UU^\dag = I$, where $\dag$ means conjugate
transpose and $I$ is the identity matrix.
\end{definition}

\begin{definition}
A matrix that can be written as a tensor product of $2 \times 2$ unitary matrices
is a \emph{local unitary matrix}.
\end{definition}

In addition to measurements, we can perform transformations on quantum states.
A quantum transformation must be reversible, and it can be shown that it must 
therefore be defined by a unitary transformation matrix.
Transformations given by local unitary matrices operate independently on each qubit, 
and therefore do not change the overall entanglement properties of the quantum state.
We can think of local unitary transformations as ``rotations'' which enables us to 
look at the same quantum state ``from another angle'', without changing its properties.
In the bra/ket notation, transformations can be described by
outer products. For instance,
\begin{equation}
\ket{0}\bra{1} + \ket{1}\bra{0}
= \binom{0}{1} (1,0) + \binom{1}{0} (0,1) 
= \begin{pmatrix}0 & 0\\1 & 0\end{pmatrix}
+ \begin{pmatrix}0 & 1\\0 & 0\end{pmatrix}
= \begin{pmatrix}0 & 1\\1 & 0\end{pmatrix}
\end{equation}
is the transformation that maps $\ket{0}$ to $\ket{1}$
and $\ket{1}$ to $\ket{0}$,
\begin{equation}
(\ket{0}\bra{1} + \ket{1}\bra{0}) (\alpha\ket{0} + \beta\ket{1})
= \begin{pmatrix}0 & 1\\1 & 0\end{pmatrix} \binom{\alpha}{\beta}
= \binom{\beta}{\alpha}
= \alpha\ket{1} + \beta\ket{0}.
\end{equation}
This is called the \emph{bit-flip} or ``not'' transformation.
It is easy to verify that it is unitary and self-inverse.

A local unitary transformation on an $n$-qubit quantum state is given by
a tensor product of $n$ $2 \times 2$ unitary matrices,
\begin{equation}
U = U_0 \otimes U_1 \otimes \cdots \otimes U_{n-1}.
\end{equation}
To express that the transform $U_0$ should be applied to qubit
number $i$, we can write $U_0^{(i)}$.
The transformation that applies $U_0$ to the $i$th qubit and
$U_1$ to all other qubits is then
\begin{equation}
U = U_0^{(i)} \bigotimes_{j \ne i} U_1^{(j)}.
\end{equation}
Note the factors must be placed in the correct order
before this tensor multiplication can be carried out.

\begin{definition}\label{l7}
We define the \emph{Pauli matrices},
\[
\sigma_x = \begin{pmatrix} 0 & 1 \\ 1 & 0 \end{pmatrix},\quad
\sigma_z = \begin{pmatrix} 1 & 0 \\ 0 & -1 \end{pmatrix},\quad
\sigma_y = \begin{pmatrix} 0 & -i \\ i & 0 \end{pmatrix},\quad
I = \begin{pmatrix} 1 & 0 \\ 0 & 1 \end{pmatrix}.
\]
\end{definition}

The Pauli matrices is a useful set of quantum transformations.
$\sigma_x$ represents a \emph{bit-flip}, $\sigma_z$ is a \emph{phase-flip},
and $\sigma_y$ is a combination of both, since  $\sigma_y = i\sigma_x\sigma_z$.
The factor $i$ in the definition of $\sigma_y$ makes some manipulations easier,
but in most cases the overall phase factor of
a quantum state can be ignored. We also include the identity 
matrix, $I$, which makes no change to the qubit it is applied to.
We will later see that Pauli matrices can be used to represent errors
on quantum states.

\begin{definition}
The \emph{Hadamard} transformation is defined by the matrix
\[
H = \frac{1}{\sqrt{2}} \begin{pmatrix} 1 & 1 \\ 1 & -1 \end{pmatrix}.
\]
\end{definition}

Observe that by applying $H$, we transform the state $\ket{0}$ into the superposition
$\frac{1}{\sqrt{2}}(\ket{0}+\ket{1})$, while the state $\ket{1}$ is transformed into
$\frac{1}{\sqrt{2}}(\ket{0}-\ket{1})$.
The \emph{Walsh-Hadamard} transformation applies $H$ to every qubit of a state. If
we apply the Walsh-Hadamard transformation to an $n$-qubit all-zero state, $\ket{00\cdots 0}$,
we get a superposition of all the $2^n$ basis states, each with the same probability.

We have already seen that measurements of quantum states destroy much of the 
information the states contain. It is also easy to show that it is impossible
to make perfect copies of a quantum state, since there is no unitary, and thus non-destructive,
transformation which performs this copying.

\begin{theorem}[Dieks, and Wootters and Zurek]
A quantum state can not be cloned, i.e., there is no operation
that takes $\ket{\phi}$ to $\ket{\phi\phi}$, where
$\ket{\phi}$ is any quantum state.
\end{theorem}
\begin{proof}
Assume that such an operation exists, and let $\ket{\phi}$ and $\ket{\psi}$ be
two distinct quantum states. Then the cloning operation gives
\begin{align}
\ket{\phi} &\to \ket{\phi\phi},\label{l8}\\
\ket{\psi} &\to \ket{\psi\psi},\label{l9}\\
\ket{\phi} + \ket{\psi} &\to (\ket{\phi} + \ket{\psi}) \otimes (\ket{\phi} + \ket{\psi})
= \ket{\phi\phi} + \ket{\psi\psi} + \ket{\phi\psi} + \ket{\psi\phi}.\label{l10}
\end{align}
But, since all quantum operations must be linear, it follows from
\eqref{l8} and \eqref{l9} that
\begin{equation}
\ket{\phi} + \ket{\psi} \to \ket{\phi\phi} + \ket{\psi\psi},
\end{equation}
which is a contradiction of \eqref{l10}.
\end{proof}

\subsection{Quantum Computers}

The idea of using quantum mechanical effects to perform computations
was first introduced by Feynman in the 1980s, when he discovered
that classical computers could not simulate all aspects of
quantum physics efficiently.
In 1985, Deutsch showed that it is possible to implement
any function which is computable by a classical computer
using registers of entangled qubits and
arrays of \emph{quantum gates}, each performing a unitary quantum transformation.

The advantage of quantum computers, compared to classical computers,
is the property of \emph{quantum parallelism}. We have seen
that an $n$-qubit quantum register can be in a superposition of 
all its $2^n$ basis states. A function of $n$ variables, implemented by
an array of quantum gates, can therefore be applied to all the basis states
simultaneously, and the result will be a superposition of the function's 
$2^n$ possible outputs. If we try to measure the result directly, the superposition will collapse, and
we will only observe one random value of the function, which is not very useful.
The advantage of quantum computers comes from the discovery that
appropriate transformations on a superposition of states
enables us to observe a common property of all the states.
This makes it possible, for instance, to find the period
of a function by applying the function once to a superposition of all possible input values.
Another way to make use of quantum parallelism is
to use transformations that amplify the probability of desired results.

\emph{Shor's algorithm}, discovered in 1994, can factor an integer in polynomial time.
For classical computers, all known algorithms require a running
time that grows exponentially with the number of bits in 
the integer to be factored. Interest in quantum computing increased with 
the discovery of Shor's algorithm, since the security of many popular public-key cryptography schemes
is based on the assumed infeasibility of factoring large integers.
The factoring problem can be reduced to the problem of finding the period of a function.
In Shor's algorithm, this is accomplished by applying the \emph{quantum Fourier transform} to a
superposition of all values of the function.

\emph{Grover's algorithm}, discovered in 1996, can be used to
search for an element in an unsorted list in
running time of order $O(\sqrt{n})$. Classical computers
can not do better than $O(n)$. This is another example of
the advantages of quantum computing, although not as impressive
as the exponential gain of Shor's algorithm.
Grover's algorithm finds a value for which a given
statement is true. This is done by evaluating the statement
for a superposition of all possible values, and
then repeatedly using a transformation that increases the
probability of the state that satisfies the statement.
When we finally read the value of the quantum register, we
will with very high probability observe the desired state.

Many different techniques for the construction of quantum computers are being 
researched, but the best implementations so far only operate on 2 or 3
qubits. Although there are many interesting theoretical results about quantum
computers, a practical and scalable implementation is not possible 
with the technology available today.

Quantum computing should not be confused with the concept
of \emph{quantum key distribution}, although both exploit
the property of quantum superposition.
In quantum key distribution, a sequence of qubits, typically 
represented by the 
polarisation of photons, is sent over an insecure quantum channel. 
Only the sender knows which basis each qubit is encoded with.
It is impossible for an eavesdropper to clone the qubits, and if
he tries to measure one using the wrong basis, its state will change.
Eavesdropping can later be detected, when the choices of encoding bases
are made public. Quantum key distribution requires much simpler technology than
quantum computing. Working
systems for quantum key distribution, using up to 150 kilometres 
of optical cables, have been successfully implemented.

\section{Classical Error Correction}

We here give a short introduction to some basic
concepts of error correcting codes that will be useful when
we later discuss quantum error correction.

\begin{definition}
Let $\mathcal{A}$ be an alphabet, and let $\mathcal{A}^n$ be the
set of all $n$-tuples of elements from $\mathcal{A}$. A
\emph{code}, $\mathcal{C}$, over $\mathcal{A}$ of length $n$ is a subset of 
$\mathcal{A}^n$, $\mathcal{C} \subset \mathcal{A}^n$.
\end{definition}

A code maps a vector $\boldsymbol{v}$ of $k$ symbols to a 
vector $\boldsymbol{u}$ of $n$ symbols, called a codeword, where
$n > k$. Any alphabet of symbols may be chosen, but the binary alphabet
$\{0,1\}$, where the symbols are called bits, is often used.
The $n-k$ extra symbols added by the encoding
process provides redundancy. If a codeword is changed by a transmission error, 
this redundancy may enable us to determine the original codeword, or at least to
detect that an error occurred.
An error in this context is an operation that change one or more symbols of a codeword into other symbols.
If the binary alphabet is used, an error flips the value of one or more bits, $0 \mapsto 1$ or $1 \mapsto 0$.

\begin{definition}
The code $\mathcal{C}$ can be defined by a $k \times n$ matrix $G$, called a
\emph{generator matrix}, where $\mathcal{C} = \{\boldsymbol{v}G \mid \boldsymbol{v}\in \mathcal{A}^k\}$.
\end{definition}

Encoding is a simple process once we know $G$, the generator matrix of a code $\mathcal{C}$, 
since the codeword corresponding to $\boldsymbol{v}$ is $\boldsymbol{u} = \boldsymbol{v}G$.
The rows of $G$ are called the basis codewords of $\mathcal{C}$, since any codeword
is a linear combination of these rows.

\begin{definition}
Let $\mathcal{C}$ be a code over the alphabet $\mathcal{A}$, where $\mathcal{A}=\GF(q)$ is a finite field,
and let $G$ be the generator matrix of $\mathcal{C}$.
If all linear combinations of the rows of $G$ are codewords in $\mathcal{C}$, then
$\mathcal{C}$ is a vector space and a $k$-dimensional subspace of $\GF(q)^n$.
A code that fulfils these criteria is called a \emph{linear code}.
\end{definition}

\begin{definition}
Let $\mathcal{C}$ be a code over a finite field with generator matrix $G$.
If any sum of the rows of $G$, i.e., any $\GF(2)$-linear combination, is a codeword in $\mathcal{C}$,
and all codewords in $\mathcal{C}$ are $\GF(2)$-linear combinations of the rows of $G$,
then $\mathcal{C}$ is an \emph{additive code}.
If the binary alphabet is used, all additive codes are linear codes,
but this is not true for the general case.
\end{definition}

\begin{definition}
The code $\mathcal{C}$ can also be defined by an $(n-k) \times n$ matrix $H$,
called the \emph{parity check matrix} of $\mathcal{C}$.
$\mathcal{C} = \{\boldsymbol{u} \in \mathcal{A}^n \mid \boldsymbol{u}H^T = \boldsymbol{0}\}$,
where $\boldsymbol{0}$ is the all-zero vector.
\end{definition}

Given the parity check matrix $H$ of a code $\mathcal{C}$, it is easy to check whether
a vector $\boldsymbol{u}$ is codeword by checking if $\boldsymbol{u}H^T = \boldsymbol{0}$.
If we receive a vector that does not satisfy this criteria, we know that an error has occurred.
If the codeword $\boldsymbol{u}$ is transmitted and $\boldsymbol{u}'$ is the received vector, then
we can write $\boldsymbol{u}' = \boldsymbol{u} + \boldsymbol{e}$, where $\boldsymbol{e}$ is the
transmission error. It is easy to verify that
$\boldsymbol{u}'H^T = \boldsymbol{u}H^T + \boldsymbol{e}H^T = \boldsymbol{0} + \boldsymbol{e}H^T = \boldsymbol{e}H^T$.
We see that the value of $\boldsymbol{u}'H^T$ only depends on the error $\boldsymbol{e}$,
and we therefore call this value the \emph{syndrome} of $\boldsymbol{e}$.
Given a set of errors with distinct syndromes, we can determine which of the errors has occurred
by using the syndrome calculated from the received vector.

\begin{definition}
The \emph{Hamming weight} of a vector $\boldsymbol{a}$ of length $n$, 
denoted $w_H(\boldsymbol{a})$, is the number of non-zero coordinates of $\boldsymbol{a}$, i.e.,
$w_H(\boldsymbol{a}) = |\{a_i \ne 0 \mid i \in \mathbb{Z}_n\}|$, where
$a_i$ is the $i$th coordinate of $\boldsymbol{a}$.
\end{definition}

\begin{definition}
The \emph{Hamming distance} between two vectors $\boldsymbol{a}$ and $\boldsymbol{b}$,
both of length $n$, denoted $d(\boldsymbol{a}, \boldsymbol{b})$, is the number of coordinates 
where the two vectors have different values, i.e., $d(\boldsymbol{a}, \boldsymbol{b}) =
|\{a_i \ne b_i \mid i \in \mathbb{Z}_n\}|$.
\end{definition}

\begin{definition}
The \emph{minimum distance} of a code $\mathcal{C}$, denoted $d(\mathcal{C})$, is 
the smallest number of symbol errors needed to change one codeword into another,
i.e., $d(\mathcal{C}) = \min\{d(\boldsymbol{a}, \boldsymbol{b}) \mid  
\boldsymbol{a}, \boldsymbol{b} \in \mathcal{C}, \boldsymbol{a} \ne \boldsymbol{b}\}$.
\end{definition}

\begin{proposition}\label{l11}
A code can detect $s$ errors if $d(\mathcal{C}) \ge s+1$. A code can correct $t$ errors if
$d(\mathcal{C}) \ge 2t+1$.
\end{proposition}

\begin{definition}
A code of length $n$ containing $M$ codewords and having minimum distance $d$ is
called an $(n,M,d)$ code.
For linear codes, the notation $[n,k,d]$ may also be used, where
$k$ is the dimension of the code. The number of codewords in a linear code
over $\GF(q)$ is then $q^k$.
\end{definition}

\begin{proposition}
The distance of a linear code $\mathcal{C}$ can easily be found as
$d(\mathcal{C}) = \min \{ w_H(\boldsymbol{u}) \mid \boldsymbol{u} \in \mathcal{C} \backslash \{\boldsymbol{0}\}\}$,
i.e., the weight of the minimum weight non-zero codeword in $\mathcal{C}$.
\end{proposition}

\begin{definition}
Every code $\mathcal{C}$ over $\GF(q)$ has a \emph{dual code},
$\mathcal{C}^\perp = \{\boldsymbol{u} \in \GF(q) \mid 
 \boldsymbol{u} \cdot \boldsymbol{c} = 0, \forall \boldsymbol{c} \in \mathcal{C}\}$.
If  $\mathcal{C} \subseteq \mathcal{C}^\perp$, then $\mathcal{C}$ is a
\emph{self-orthogonal} code. If $\mathcal{C} = \mathcal{C}^\perp$, then $\mathcal{C}$ is a
\emph{self-dual} code.
\end{definition}

If $\mathcal{C}$ has generator matrix $G$ and parity check matrix $H$,
then the dual code, $\mathcal{C}^\perp$, has generator matrix $H$ and
parity check matrix $G$.

\section{Quantum Error Correction}\label{l12}

\subsection{Introduction}

For more detailed information about quantum error correction,
we refer to some of the many introductions to the subject~\cite{B32, B25, B36}.

A major problem for the implementation of quantum computers is
that it is impossible to totally isolate a few qubits 
from the rest of the world. The qubits will rapidly interact
with the environment, and entanglement will be destroyed in 
a process known as \emph{decoherence}. Because of decoherence,
the state of a quantum register will not remain stable for
long enough time to do any useful computations.
We have seen that we can not observe a quantum state without 
destroying entanglement, and that we can not make copies of it. 
Surprisingly, it was shown by Steane and Shor that
quantum error correction is still possible.
It can even be shown that it is possible to process quantum information
arbitrarily accurately, given that the effects of decoherence
can be kept under a certain threshold for each step of
the computation.

A classical bit can only have one of two values, 0 or 1, and the
only possible error is a bit-flip. A qubit
has a continuous state space, since $\alpha$ and $\beta$ in the
expression $\alpha\ket{0} + \beta\ket{1}$ can take any complex values.
Since any $2 \times 2$ unitary matrix describes a possible transformation, 
an infinite number of different errors may affect a single qubit.

\begin{proposition}
The set of  Pauli matrices, introduced in \autoref{l7},
span the space of $2 \times 2$ unitary matrices. Any error on a single qubit,
$\ket{\phi} \to E\ket{\phi}$,
may therefore be expressed as a linear combination of the Pauli matrices,
\begin{equation}
\ket{\phi} \rightarrow (aI + b\sigma_x + c\sigma_y + d\sigma_z)\ket{\phi}
= a\ket{\phi} + b\sigma_x\ket{\phi} + c\sigma_y\ket{\phi} + d\sigma_z\ket{\phi}.\label{l13}
\end{equation}
\end{proposition}

We will see that
the error correction process causes the superposition in \eqref{l13} to collapse into one of four
states, so that we observe no error with probability $|a|^2$, a bit-flip error with
probability $|b|^2$, a phase-flip error with probability $|c|^2$, and a
combined bit-flip and phase-flip error with probability $|d|^2$. The process will also
determine which Pauli error has occurred. We can then recover the state $\ket{\phi}$
by applying the same Pauli transformation, since all the Pauli matrices are self-inverse.
This procedure is performed without observing the state $\ket{\phi}$ directly, 
but by comparing the values of several qubits.
A comparison of two qubits can be done without learning the value of either
qubit, and therefore without collapsing their superpositions.

As in the classical case, quantum error correction is done by 
adding redundant qubits which are used to detect or correct errors.
A quantum code encodes $k$ qubits using $n$ qubits. It has $2^k$ basis
codewords, but any linear combination of those is also a valid codeword,
since the code must be able to encode all superpositions of the basis states.
We assume that errors affect each qubit independently, which
may in reality not be the case.
We describe the errors by error operators, which are tensor products of Pauli matrices.
The \emph{weight} of an error operator is the number of positions in which it
is different from identity. For instance, the error operator
$I \otimes \sigma_x \otimes \sigma_z \otimes I \otimes I$ has weight 2. Note in
particular that a combined bit-flip and phase-flip error only count as one error.
If the errors described by all Pauli error operators of weight up to $t$ 
can be corrected by a code, then the code can
correct an arbitrary error affecting up to $t$ qubits. 
If a code should be able to correct the two errors $E_a$ and $E_b$, then the code must be able to 
tell the difference between $E_a\ket{\phi_i}$ and $E_b\ket{\phi_j}$, the
two errors operating on two different basis codewords.
To guarantee that this is possible, the vectors corresponding to the states 
$E_a\ket{\phi_i}$ and $E_b\ket{\phi_j}$ must be orthogonal.

\begin{example}\label{l14}
The repetition code is a simple classical code that
encodes a bit by making a number of copies of it. Decoding
is achieved by a majority rule.
Quantum coding is not that easy, since qubits can not be copied,
but it is possible to encode, for instance, one qubit using three qubits by mapping 
the basis states $\ket{0}$ to $\ket{000}$ and $\ket{1}$ to $\ket{111}$.
The state $\ket{\phi} = \alpha\ket{0}+\beta\ket{1}$ would in
that case be encoded into $\ket{\psi} = \alpha\ket{000}+\beta\ket{111}$. 
Note that these three qubits are highly entangled, and not three independent copies 
of $\ket{\phi}$.
This code can correct any single bit-flip, but does not correct phase-flips.
Consider the error
$(\sigma_x \otimes I \otimes I)\ket{\psi} = \alpha\ket{100}+\beta\ket{011}$.
We can not observe the value of any qubit, but it is possible to compare
two qubits and learn if they have the same value. Comparing the first
and second qubit tells us that they are different, so one of them must
be wrong. When we find that the second and
third qubits are equal, we know that the error is in the first qubit, assuming only one error has occurred.
We bit-flip the first qubit to correct the error.
This code is also able to correct any linear combination of single bit-flip
error operators. Consider, for instance, the error
\[
(\frac{3}{5} \sigma_x \otimes I \otimes I + \frac{4}{5} I \otimes \sigma_x \otimes I)\ket{\psi}
= \frac{3}{5}(\alpha\ket{100}+\beta\ket{011}) + \frac{4}{5}(\alpha\ket{010}+\beta\ket{101}).
\]
When we compare the values of the qubits, this superposition will collapse into
$\alpha\ket{100}+\beta\ket{011}$ with probability $\frac{9}{25}$ and 
$\alpha\ket{010}+\beta\ket{101}$ with probability $\frac{16}{25}$.
The results of the comparisons will be according to the chosen state,
and the error correction proceeds as in the previous case.
\end{example}

In order to learn what error we must correct, a number of extra qubits,
known as an \emph{ancilla}, are added temporarily. After appropriate transformations,
we may read a \emph{syndrome}, which tells us what error has occurred, from these qubits.
It is in fact the act of measuring the syndrome that collapses the superposition
of errors into a single error.

\begin{example}\label{l15}
A code that can correct both bit-flips and phase-flips is 
Shor's ``nine-qubit repetition code''. This code maps
$\ket{0}$ to $(\ket{000}+\ket{111})(\ket{000}+\ket{111})(\ket{000}+\ket{111})$ 
and $\ket{1}$ to $(\ket{000}-\ket{111})(\ket{000}-\ket{111})(\ket{000}-\ket{111})$.
Bit-flips are corrected by the inner layer of this code, by exactly the
same procedure as in \autoref{l14}. By comparing the signs of the three 
outer blocks, we may correct any single phase-flip. The two steps are actually
independent, so both one phase-flip and one bit-flip can always be corrected.
Note that a phase-flip on the first qubit followed by a phase-flip on the
second qubit, i.e., the error operator $\sigma_z \otimes \sigma_z \otimes I
\otimes \cdots \otimes I$, leaves a codeword
unchanged. The code need not be able to correct this error, nor to tell
which qubit in each block of three has been affected in case of a phase-flip.
Codes where any error operator has this property are called \emph{degenerate codes}.
\end{example}

\begin{definition}
The minimum distance, $d$, of a quantum code, is the minimum weight error operator
that gives an errored state not orthogonal to the original state, and
therefore not guaranteed to be detectable.
\end{definition}

It follows from \autoref{l11} that a quantum code with 
$d \ge s+1$ can detect $s$ errors, and that a quantum code with $d \ge 2t+1$
can correct $t$ errors.

\begin{definition}
A quantum code that encodes $k$ qubits using $n$ qubits and have distance $d$ is
called an $[[n,k,d]]$ code. 
\end{definition}

The double brackets helps us distinguish a quantum code from a classical code.
The nine-qubit code described in \autoref{l15} is a $[[9,1,3]]$ code.

\subsection{Stabilizer Codes}

An $[[n,k,d]]$ quantum code can be described by a \emph{stabilizer} given
by a set of $n-k$ error operators.
Such codes are called \emph{stabilizer codes}~\cite{B24, B25}.

Consider the nine-qubit code from \autoref{l15}. When we compare
the two first qubits to detect a possible bit-flip in one of them, 
what we really do is to measure the eigenvalue of the 
operator $M_1 = \sigma_z \otimes \sigma_z \otimes I \otimes \cdots \otimes I$,
i.e., we find the value $m$ in $M_1\ket{\phi} = m\ket{\phi}$.
If the qubits have the same value, then the result is $+1$, and otherwise it is
$-1$.
To compare the first two three-qubit blocks to detect a phase-flip
in one of them, we measure the eigenvalue of 
$\sigma_x \otimes \sigma_x \otimes \sigma_x \otimes \sigma_x \otimes \sigma_x \otimes \sigma_x
\otimes I \otimes I \otimes I$.
The complete stabilizer for Shor's nine-qubit code is given by the 8 operators,
\begin{alignat*}{9}
M_1 &= \sigma_z &&\otimes \sigma_z &&\otimes I &&\otimes I &&\otimes I &&\otimes 
   I &&\otimes I &&\otimes I &&\otimes I, \\
M_2 &= I &&\otimes \sigma_z &&\otimes \sigma_z &&\otimes I &&\otimes I &&\otimes 
   I &&\otimes I &&\otimes I &&\otimes I, \\
M_3 &= I &&\otimes I &&\otimes I &&\otimes \sigma_z &&\otimes \sigma_z &&\otimes  
   I &&\otimes I &&\otimes I &&\otimes I, \\
M_4 &= I &&\otimes I &&\otimes I &&\otimes I &&\otimes \sigma_z &&\otimes \sigma_z 
   &&\otimes I &&\otimes I &&\otimes I, \\
M_5 &= I &&\otimes I &&\otimes I &&\otimes I &&\otimes I &&\otimes I &&\otimes 
   \sigma_z &&\otimes \sigma_z &&\otimes I, \\
M_6 &= I &&\otimes I &&\otimes I &&\otimes I &&\otimes I &&\otimes I &&\otimes I 
   &&\otimes \sigma_z &&\otimes \sigma_z, \\
M_7 &= \sigma_x &&\otimes \sigma_x &&\otimes \sigma_x &&\otimes \sigma_x &&\otimes 
   \sigma_x &&\otimes \sigma_x &&\otimes I &&\otimes I &&\otimes I, \\
M_8 &= I &&\otimes I &&\otimes I &&\otimes \sigma_x &&\otimes \sigma_x &&\otimes \sigma_x 
   &&\otimes \sigma_x &&\otimes \sigma_x &&\otimes \sigma_x.
\end{alignat*}
The error we want to detect \emph{anticommutes} with the operator we
actually measure, since the Pauli operators anticommute, i.e.,
$AB = -BA$, where $A,B \in \{\sigma_x,\sigma_y,\sigma_z\}$ and $A \ne B$.
For a valid codeword, it must be true for all $i$ that $M_i\ket{\phi} = \ket{\phi}$, i.e., 
that the eigenvalue of all operators is $+1$.
If a correctable error has occurred, the set of operators that give eigenvalues $-1$ will identify
the error.
Consider, for instance, the error $E = \sigma_x \otimes I \otimes \cdots \otimes I$, a bit-flip error on 
the first qubit, which takes $\ket{\phi}$ to $E\ket{\phi}$.
$E$ will anticommute with $M_1$, so $M_1E\ket{\phi} = -EM_1\ket{\phi} = -E\ket{\phi}$, and the resulting
eigenvalue is $-1$. Likewise, $E$ will anticommute with $M_2$, but will commute with the other six operators.
This gives us a set of eigenvalues uniquely identifying the error $E$.

The \emph{stabilizer}, $S$, is an Abelian group generated by the set of $n-k$ operators.
(An Abelian group is a group where all elements commute.)
$S$ consists of all operators $M$ for which $M\ket{\phi} = \ket{\phi}$ for 
all codewords $\ket{\phi}$.
Two given errors can be corrected if there exists an operator in $S$ that can
distinguish them, i.e., measuring the eigenvalue of the operator gives different 
values for the two errors.
Let the \emph{centraliser} of $S$, $C(S)$, be the set of errors that commute
with all the $n-k$ generators of $S$.
$C(S) \backslash S$ is then the set of errors that are not detectable.
Hence, the distance, $d$, of a stabilizer code is the minimum weight of any
operator in $C(S) \backslash S$.

In \autoref{l15}, we studied a $[[9,1,3]]$ stabilizer code. This code does not represent
an optimal way of encoding one qubit with the possibility of correcting any
single error. In fact, there exists a $[[5,1,3]]$ code. The 4 operators 
generating its stabilizer are given by the following matrix, each row
corresponding to one operator.
\[
S =
\begin{pmatrix}
\sigma_x & \sigma_z & \sigma_z & \sigma_x & I \\
I & \sigma_x & \sigma_z & \sigma_z & \sigma_x \\
\sigma_x & I & \sigma_x & \sigma_z & \sigma_z \\
\sigma_z & \sigma_x & I & \sigma_x & \sigma_z
\end{pmatrix}
\]
An alternate representation of the stabilizer $S$ uses two binary matrices, 
the bit-flip matrix $X$ and the phase-flip matrix $Z$.
Let $X_{i,j} = 1$ when $S_{i,j} = \sigma_x$ or $S_{i,j} = \sigma_y$, and $X_{i,j} = 0$ otherwise.
Let $Z_{i,j} = 1$ when $S_{i,j} = \sigma_z$ or $S_{i,j} = \sigma_y$, and $Z_{i,j} = 0$ otherwise.
We combine the two matrices to make the $n \times 2n$ binary stabilizer matrix
$S_b = ( Z \mid X )$.
For the $[[5,1,3]]$ code, we get
\[
S_b =
\left(
\begin{array}{ccccc|ccccc}
0 & 1 & 1 & 0 & 0 & 1 & 0 & 0 & 1 & 0\\
0 & 0 & 1 & 1 & 0 & 0 & 1 & 0 & 0 & 1\\
0 & 0 & 0 & 1 & 1 & 1 & 0 & 1 & 0 & 0\\
1 & 0 & 0 & 0 & 1 & 0 & 1 & 0 & 1 & 0
\end{array}
\right).
\]

\subsection{Quantum Codes over GF(4)}

\begin{proposition}[Calderbank et~al.~\cite{B10}]
We can consider a quantum error correcting code as an additive
code over the finite field $\GF(4)$, by identifying 
the four Pauli matrices with the elements of $\GF(4)$.
We denote $\GF(4) = \{0,1,\omega,\omega^2\}$, where $\omega^2 = \omega + 1$.
The mappings used are $I \mapsto 0$, $\sigma_z \mapsto 1$, $\sigma_x  \mapsto 
\omega$, and $\sigma_y \mapsto \omega^2$.
\end{proposition}

As an example, the $[[5,1,3]]$ code previously described can be represented by
the additive code over $\GF(4)$ generated by the matrix
\[
C=
\begin{pmatrix}
\omega & 1 & 1 & \omega & 0 \\
0 & \omega & 1 & 1 & \omega \\
\omega & 0 & \omega & 1 & 1 \\
1 & \omega & 0 & \omega & 1
\end{pmatrix}.
\]

\begin{definition}
\emph{Conjugation} in $\GF(4)$ is defined by $\overline{x} = x^2$.
The \emph{trace map}, $\text{tr} : \GF(4) \mapsto \GF(2)$, is defined by
$\text{tr}(x) = x + \overline{x}$.
The \emph{trace inner product} of two vectors of length $n$ over $\GF(4)$, 
$\boldsymbol{u}$ and $\boldsymbol{v}$, is given by
$\boldsymbol{u} * \boldsymbol{v} = \sum_{i=1}^n tr(u_i \overline{v_i})$.
\end{definition}

In addition to replacing the symbols we use, we must make sure that
the properties of a stabilizer code are preserved in a code over $\GF(4)$.
A stabilizer is a group generated by $n-k$ operators.
This corresponds to an additive subset of $GF(4)^n$, generated by $n-k$ vectors.
The stabilizer is an Abelian group, which means that any two operators in the 
stabilizer commute. The corresponding property of an additive code over 
$\GF(4)$, $\mathcal{C}$, is that any two codewords, 
$\boldsymbol{u}, \boldsymbol{v} \in \mathcal{C}$, must have trace inner product 
$\boldsymbol{u} * \boldsymbol{v} = 0$.
This is equivalent to saying that the code must be \emph{self-orthogonal}
with respect to the trace inner product, or that $\mathcal{C} \subseteq \mathcal{C}^\perp$,
where $\mathcal{C}^\perp = \{ \boldsymbol{u} \in \GF(4)^n \mid
\boldsymbol{u}*\boldsymbol{c}=0, \forall \boldsymbol{c} \in \mathcal{C} \}$.
If the stabilizer $S$ corresponds to $\mathcal{C}$, 
then the centraliser $C(S)$, the set of errors that commute
with all generators of $S$, corresponds to $\mathcal{C}^\perp$.
The set of undetectable errors, $C(S) \backslash S$, corresponds to
$\mathcal{C}^\perp \backslash \mathcal{C}$. Hence, the weight of the minimum 
weight non-zero vector
in $\mathcal{C}^\perp \backslash \mathcal{C}$ is the distance of a quantum code
over $\GF(4)$.

\subsection{Self-Dual Quantum Codes}

The codes studied in this thesis will be of the special case
where the dimension $k=0$. A zero-dimensional stabilizer code with high distance
represents a single quantum state which is robust to error,
sometimes called a \emph{stabilizer state}. Codes of higher dimension can be 
constructed from zero-dimensional quantum codes, but identifying
stabilizer states is also an interesting application in itself,
since the states corresponding to codes of high distance
will be highly entangled. Highly entangled quantum states could be used 
for testing the decoherence properties of a quantum computer,
and it has also been shown that a \emph{one-way quantum computer}
can be implemented by performing measurements on a particular class 
of entangled states, known as \emph{cluster states}~\cite{B48,B8,B49}.
An $[[n,0,d]]$ code is nondegenerate by definition, and
is generated by an $n \times n$ generator matrix, corresponding to
an $(n,2^n,d)$ classical code. The $\GF(4)$-representation
of such codes will be \emph{self-dual}, i.e., $\mathcal{C} = \mathcal{C}^\perp$,
and we therefore call zero-dimensional quantum codes of this type \emph{self-dual quantum codes}.
The distance of a self-dual quantum code is simply the minimum distance 
of $\mathcal{C}$, i.e., the weight of the minimum weight codeword in $\mathcal{C}$.

\begin{example}
As an example, consider the self-dual quantum code with
generator matrix
\[
C=
\begin{pmatrix}
  \omega & 0 & 0 & 1 & 1 & 1\\
  0 & \omega & 0 & \omega^2 & 1 & \omega\\
  0 & 0 & \omega & \omega^2 & \omega & 1\\
  0 & 1 & 0 & \omega & \omega^2 & 1\\
  0 & 0 & 1 & \omega & 1 & \omega^2\\
  1 & \omega^2 & 0 & \omega & 0 & 0
\end{pmatrix}.
\]
There are 64 $\GF(2)$-linear combinations of the 6 rows of $C$.
In addition to the all-zero codeword, we have 45 codewords of weight
4 and 18 of weight 6. This is therefore a $[[6,0,4]]$ code.
\end{example}

\begin{definition}
We distinguish between two types of self-dual quantum codes.
A code is of \emph{type II} if all codewords have even weight,
otherwise it is of \emph{type I}.
It can be shown that a type II code must have even length.
\end{definition}

\begin{theorem}[Rains and Sloane~\cite{B47}]
Let $d_I$ be the minimum distance of a type I code of length n. Then
$d_I$ is upper-bounded by
\begin{equation}
d_I \le \begin{cases}
2 \left\lfloor \frac{n}{6} \right\rfloor + 1, \quad \text{if } n \equiv 0 \text{ } (\text{mod } 6) \\ 
2 \left\lfloor \frac{n}{6} \right\rfloor + 3, \quad \text{if } n \equiv 5 \text{ } (\text{mod } 6) \\ 
2 \left\lfloor \frac{n}{6} \right\rfloor + 2, \quad \text{otherwise.}
\end{cases}
\end{equation}
There is a similar bound on $d_{II}$, the distance
of a type II code of length n,
\begin{equation}
d_{II} \le 2 \left\lfloor \frac{n}{6} \right\rfloor + 2.
\end{equation}
\end{theorem}

A code that meets the appropriate bound is called \emph{extremal}.
Calderbank et~al. also use a linear programming bound~\cite{B10} on
the distance of self-dual quantum codes and give a table of the 
best bounds. This table has later been extended by Grassl~\cite{B27}.
For some lengths, no code meeting the best upper bound on distance has been discovered,
so it remains uncertain whether such a code exists. In particular, for $n=24$, the
best known self-dual quantum code has distance 8, while the upper bound is 10.
Let $d_m$ be the highest attainable distance for self-dual additive codes 
over $\GF(4)$. (Non-additive codes~\cite{B46} with higher distance may exist.)
\autoref{l16} shows, for lengths up to 30, the values of $d_I$, $d_{II}$ 
and $d_m$.
Note that type II codes where the length is a multiple of 6, i.e., 6, 12, 18, 24
and 30, are particularly strong codes.

\begin{table}
\centering
\caption{Bounds on the Distance of Self-Dual Quantum Codes\label{l16}}
\begin{tabular}{cccc}
\toprule
$n$ & $d_I$ & $d_{II}$ & $d_m$ \\
\midrule
 2 &  2 &  2 & 2 \\
 3 &  2 &    & 2 \\
 4 &  2 &  2 & 2 \\
 5 &  3 &    & 3 \\
 6 &  3 &  4 & 4 \\
 7 &  4 &    & 3 \\
 8 &  4 &  4 & 4 \\
 9 &  4 &    & 4 \\
10 &  4 &  4 & 4 \\
11 &  5 &    & 5 \\
12 &  5 &  6 & 6 \\
13 &  6 &    & 5 \\
14 &  6 &  6 & 6 \\
15 &  6 &    & 6 \\
16 &  6 &  6 & 6 \\
17 &  7 &    & 7 \\
18 &  7 &  8 & 8 \\
19 &  8 &    & 7 \\
20 &  8 &  8 & 8 \\
21 &  8 &    & 8 \\
22 &  8 &  8 & 8 \\
23 &  9 &    & 8--9 \\
24 &  9 & 10 & 8--10 \\
25 & 10 &    & 8--9 \\
26 & 10 & 10 & 8--10 \\
27 & 10 &    & 9--10 \\
28 & 10 & 10 & 10 \\
29 & 11 &    & 11 \\
30 & 11 & 12 & 12 \\
\bottomrule
\end{tabular}
\end{table}

\chapter{Quantum Codes and Graphs}\label{l17}

\section{Introduction to Graph Theory}

A \textbf{graph} is a pair $G=(V,E)$, where $V = \{v_0, v_1, \ldots, v_{n-1}\}$ is a set of 
$n$ \textbf{vertices} (or \textbf{nodes}), and $E$ is a set of distinct pairs 
of elements from $V$, i.e., $E \subseteq V \times V$. A pair $\{v_i,v_j\} \in E$ is
called an \textbf{edge}. We will only consider 
\textbf{undirected graphs}, which are graphs where $E$ is a set 
of distinct \emph{unordered} pairs of elements from $V$.
Furthermore, the graphs we will look at will all be \textbf{simple graphs},
which are graphs with no \emph{self-loops}, $\{v_i,v_i\} \not\in E$.
A graph $G'=(V',E')$ that satisfies $V' \subseteq V$ and 
$E' \subseteq E$ is a \textbf{subgraph} of $G$, denoted $G' \subseteq G$.
Given a subset of vertices $A \subseteq V$, the \textbf{induced subgraph} 
$G(A) \subseteq G$ has vertices $A$ and edges $\{\{v_i,v_j\} \in E \mid v_i,v_j \in A\}$,
i.e., all edges from $E$ whose endpoints are both in $A$.
The \textbf{complement graph} $\overline{G}$ has
vertices $\overline{V} = V$ and edges $\overline{E} = V \times V - E$,
i.e., the edges in $E$ are changed to non-edges, and the non-edges to edges.

Two \textbf{isomorphic} graphs are structurally equal, but the labelling of the vertices may differ.
More formally, two graphs $G=(V,E)$ and $G'=(V,E')$ are isomorphic iff
there exists a permutation $\pi$ of $V$ such that $\{v_i,v_j\} \in E \iff \{\pi(v_i), \pi(v_j)\}
\in E'$.

A graph may be represented by an \textbf{adjacency matrix} $\Gamma$. This is a
$|V| \times |V|$ matrix where $\Gamma_{i,j} = 1$ if $\{v_i,v_j\} \in E$, and
$\Gamma_{i,j} = 0$ otherwise.
For simple graphs, the adjacency matrix must have 0s on the diagonal, i.e., $\Gamma_{i,i} = 0$.
The adjacency matrix of an undirected graph will be \emph{symmetric},
i.e., $\Gamma_{i,j} = \Gamma_{j,i}$.

\begin{figure}
\centering
\reflectbox{\includegraphics[width=.5\linewidth]{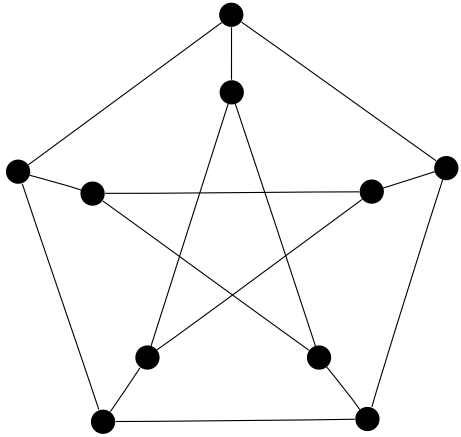}}
\caption{The Strongly Regular Petersen Graph}\label{l18}
\end{figure}

Two vertices are called \textbf{adjacent} (or \textbf{neighbours}) if they
are joined by an edge. The \textbf{neighbourhood} of a vertex $v$, denoted
$N_v$, is the set of vertices that are adjacent to $v$.
The \textbf{vertex degree} (or \textbf{valency}) of a vertex is the 
number of neighbours it has. A \textbf{regular graph} is a graph where 
all vertices have the same degree.
A regular graph where all vertices have degree $k$ is called
a \textbf{\emph{k}-regular graph}.
We will also denote any \emph{k}-regular graph on $n$ vertices $R^k_n$.
Note that $R^k_n$ does not uniquely define a graph; there may be several
non-isomorphic graphs $R^k_n$ for the same value of $k$ and $n$.
A \textbf{strongly regular graph}~\cite{B11} 
with parameters $(n,k,\lambda,\mu)$ is a $k$-regular graph on $n$ vertices,
with the additional property that any two adjacent vertices 
have $\lambda$ common neighbours, and any 
two non-adjacent vertices have $\mu$ common neighbours.
An example of a strongly regular graph with parameters $(10,3,0,1)$ is the well-known
\emph{Petersen graph}, shown in \autoref{l18}.

A \textbf{complete graph} is a graph where all pairs of vertices are
connected by an edge. The complete graph on $n$ vertices has
$\binom{n}{2}$ undirected edges. A \textbf{clique} is a complete subgraph.
A $k$-clique is a clique consisting of $k$ vertices.
We will use the notation $K_n$ for both complete graphs on $n$ vertices and $n$-cliques.
Note that $K_n = R^{n-1}_n$ and that $K_n$ does define a unique graph, up to isomorphism.
An \textbf{independent set} is the complement of a clique, i.e., a subgraph with no edges.
The \textbf{independence number}, $\alpha(G)$, is the size of the largest independent set in $G$.
A \textbf{bipartite graph} is a graph where the vertices can be partitioned into two independent sets,
i.e., $V = A \cup B$ where the induced subgraphs $G(A)$ and $G(B)$ both contain no edges.

A \textbf{path} is a sequence of vertices $(u_0,u_1,\ldots,u_{k-1})$ where
$\{u_0,u_1\}, \{u_1,u_2\},$ $\ldots, \{u_{k-2},u_{k-1}\} \in E$.
A \textbf{connected graph} is a graph where there is path from every vertex to all other vertices.
A \textbf{simple path} never visits the same vertex more than once.
A \textbf{cycle} (or \textbf{circuit}) is a path $(u_0,u_1,\ldots,u_{k-1},u_0)$, 
i.e., a path that starts and ends at the same vertex.
A \textbf{simple cycle} never visits the same vertex more than once, except the first vertex, 
which is also visited last.
Let $C_n$ denote the graph consisting of only a simple cycle on $n$ vertices.
Note that $C_n = R^2_n$ and that $C_n$ does define a unique graph, up to isomorphism.
A \textbf{Hamiltonian path} is a simple path that visits every vertex of the graph once.
If there is also an edge between the first and the last vertex of a
Hamiltonian path, we have a \textbf{Hamiltonian cycle}.

A \textbf{hypergraph}, $G=(V,E)$, is a generalised graph where an edge may connect more than two vertices.
An edge, $e \in E$, of a hypergraph is given by a set of at least two vertices, 
$e = \{u_0, u_1, \ldots, u_{k-1}\}$. Edges on more than two vertices are called \textbf{hyperedges}.

\section[Graph Isomorphism with \texttt{nauty}]{Graph Isomorphism with nauty}\label{l19}

Determining whether two graphs are isomorphic is considered to be a hard problem, but
an efficient algorithm has been developed by McKay and implemented in the program 
\texttt{nauty}~\cite{nauty}. \texttt{nauty} can also produce a \emph{canonical representative} 
of a graph. The canonical representative is isomorphic to the original graph, but
may have a different vertex labelling. This labelling is arbitrary with no special 
properties, but it is chosen in a consistent way such that all isomorphic graphs will have
the same canonical representative.
\texttt{nauty} also includes a utility called \texttt{geng} which can generate all
non-isomorphic graphs on a given number of vertices.

Checking for hypergraph isomorphism is not directly supported by \texttt{nauty}, but 
\texttt{nauty} can detect isomorphism of graphs where the vertices have been
divided into a set of disjoint partitions, $V = P_1 \cup P_2 \cup \cdots \cup P_{k}$.
Two such partitioned graphs are isomorphic if their partitions are of the same sizes, and if
a relabelling of the vertices of one of the graphs produces the other, with the restriction that
labels can only be exchanged within partitions.
There is one-to-one mapping between a hypergraph, $G=(V,E)$, on $n$ vertices with $m$ hyperedges
and an ordinary graph, $G'=(V',E')$, on $n+m$ vertices with partitions of size $n$ and $m$.
We first add all vertices in $V$ to $V'$ and all simple graph edges in $E$ to $E'$. For
each hyperedge, $e_i \in E$, $0 \le i < m$, we add an extra vertex, $v_{n+i}$, to $V'$.
If $e_i = \{u_0, u_1, \ldots, u_{k-1}\}$, we add the $k$ edges $\{u_0, v_{n+i}\}, \ldots, 
\{u_{k-1}, v_{n+i}\}$ to $E'$. The canonical representative of $G'$ is found by \texttt{nauty},
and the result is mapped back to a hypergraph, which is the canonical representative of $G$.

\begin{example}
We have the hypergraph $G=(V,E)$ with $V=\{v_0, v_1, v_2, v_3\}$ and
$E=\{\{v_0,v_1,v_2\}$, $\{v_1,v_2,v_3\}$, $\{v_1,v_2\}$, $\{v_1,v_3\}\}$.
We map this hypergraph to the graph $G'$ on 6 vertices with edges
$E'=\{\{v_1,v_2\}$, $\{v_1,v_3\}$, $\{v_0,v_4\}$, $\{v_1,v_4\}$, $\{v_2,v_4\}$, $\{v_1,v_5\}$, 
 $\{v_2,v_5\}$, $\{v_3,v_5\}\}$,
where the vertices are partitioned into the sets $\{v_0, v_1, v_2, v_3\}$ and $\{v_4, v_5\}$.
We use \texttt{nauty} to find the canonical labelling of this partitioned graph, and we then map
the resulting graph to the hypergraph $G''=(V,E'')$, where 
$E''=\{\{v_0,v_2,v_3\}$, $\{v_1,v_2,v_3\}$, $\{v_1,v_3\}$, $\{v_2,v_3\}\}$.
Any hypergraph isomorphic to $G$ will also have canonical representative $G''$.
\end{example}

\section{Graph Codes}

\begin{definition}
A \emph{graph code} is a self-dual additive code over $\GF(4)$ with generator matrix
$C=\Gamma + \omega I$, where $I$ is the identity matrix and $\Gamma$ is the adjacency matrix of 
a simple undirected graph, which must be symmetric with 0s along the diagonal.
\end{definition}

\begin{figure}
 \centering
 \subfloat[The ``2-clique of 3-cliques'']
 {\includegraphics[width=.40\linewidth]{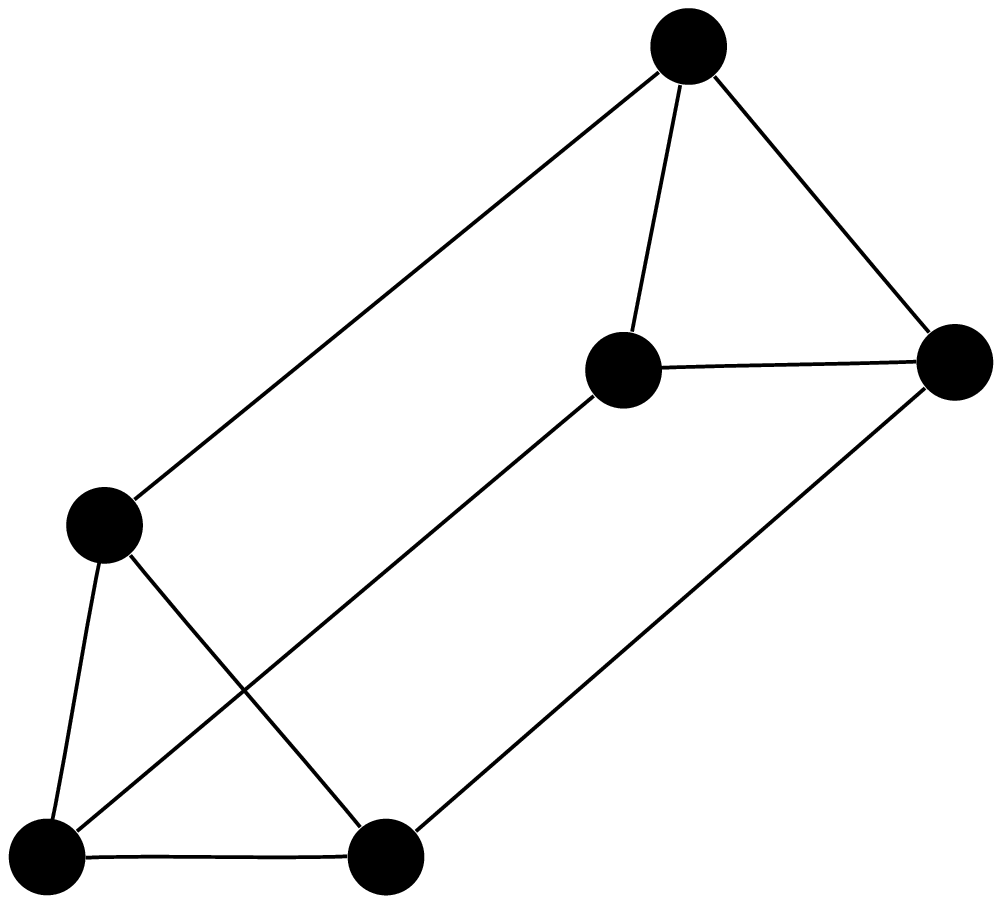}\label{l20}}
 \quad
 \subfloat[The ``Wheel Graph'']
 {\includegraphics[width=.40\linewidth]{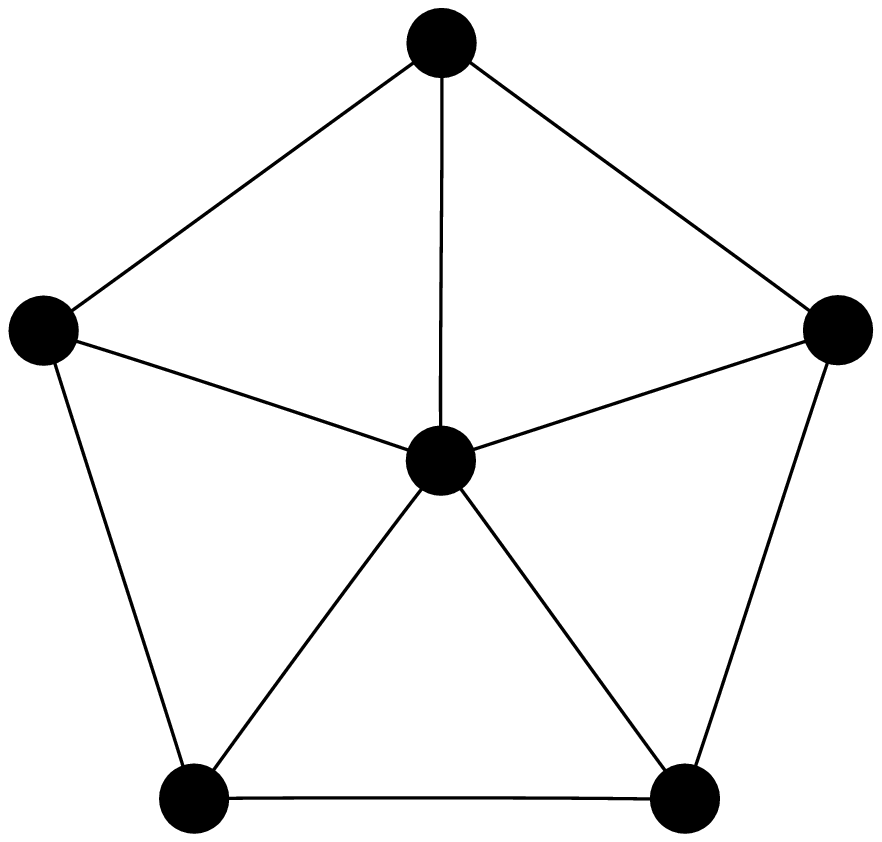}\label{l21}}
 \caption{Two Graph Representations of the [[6,0,4]] Hexacode}\label{l22}
\end{figure}

\begin{example}
Consider the graph shown in \autoref{l20}. 
This graph has adjacency matrix
\[
\Gamma =
\begin{pmatrix}
0 & 1 & 1 & 1 & 0 & 0 \\
1 & 0 & 1 & 0 & 1 & 0 \\
1 & 1 & 0 & 0 & 0 & 1 \\
1 & 0 & 0 & 0 & 1 & 1 \\
0 & 1 & 0 & 1 & 0 & 1 \\
0 & 0 & 1 & 1 & 1 & 0
\end{pmatrix}.
\]
The corresponding self-dual additive code over $\GF(4)$ is
generated by the matrix
\[
C = \Gamma + \omega I =
\begin{pmatrix}
\omega & 1 & 1 & 1 & 0 & 0 \\
1 & \omega & 1 & 0 & 1 & 0 \\
1 & 1 & \omega & 0 & 0 & 1 \\
1 & 0 & 0 & \omega & 1 & 1 \\
0 & 1 & 0 & 1 & \omega & 1 \\
0 & 0 & 1 & 1 & 1 & \omega
\end{pmatrix}.
\]
The same code can also be described using stabilizer formalism.
The stabilizer code is generated by operators given by 
the rows of the matrix
\[
S=
\begin{pmatrix}
\sigma_x & \sigma_z & \sigma_z & \sigma_z & I & I \\
\sigma_z & \sigma_x & \sigma_z & I & \sigma_z & I \\
\sigma_z & \sigma_z & \sigma_x & I & I & \sigma_z \\ 
\sigma_z & I & I & \sigma_x & \sigma_z & \sigma_z \\
I & \sigma_z & I & \sigma_z & \sigma_x & \sigma_z \\ 
I & I & \sigma_z & \sigma_z & \sigma_z & \sigma_x
\end{pmatrix}.
\]
Stabilizer codes of this type are known as graph codes, and
the single quantum states they encode are called \emph{graph states}.
\end{example}

Schlingemann and Werner~\cite{B54} studied quantum codes associated with graphs,
and first proved the following theorem.
Briegel and Raussendorf~\cite{B8} had previously studied arrays of entangled particles,
which can be modelled by graphs.

\begin{theorem}[Schlingemann and Werner~\cite{B54}, Grassl et~al.~\cite{B28},
Glynn~\cite{B22}, and Van~den~Nest et~al.~\cite{nest}]\label{l23}
For any self-dual quantum code, there is an equivalent graph code.
This means that there is a one-to-one correspondence between the set
of simple undirected graphs and the set of self-dual additive codes over $\GF(4)$.
\end{theorem}

It follows from \autoref{l23} that, without loss of generality, we
can restrict our study of self-dual additive codes over $\GF(4)$ to
those with generator matrices of the form $\Gamma + \omega I$.

Van~den~Nest et~al.~\cite{nest} describe the following algorithm for transforming any
stabilizer code into a graph code.
We will operate on the transpose of the binary stabilizer matrix,
$T = S_b^T = \binom{Z^T}{X^T} = \binom{A}{B}$. 
It is easy to see that a graph code given by the adjacency matrix $\Gamma$
corresponds to the binary stabilizer $S_b = ( \Gamma \mid I )$,
and to the transpose binary stabilizer $T = \binom{\Gamma}{I}$.
Our goal is to convert $T = \binom{A}{B}$, the transpose binary stabilizer of a given code,
into $T' = \binom{A'}{I}$, the transpose binary stabilizer of an equivalent graph code.
$A'$ will then be the adjacency matrix of the corresponding graph.
Right-multiplying $T$ with an invertible $n \times n$ matrix
will perform a basis change, an operation that gives us an equivalent stabilizer code.
If $B$ is an invertible matrix, we can simply multiply $T$ by the inverse of $B$ and get
$TB^{-1} = \binom{AB^{-1}}{I}$. $AB^{-1}$ will then be the resulting adjacency matrix.
If this matrix has elements on the diagonal that are not 0, those elements may simply be
changed to 0.
In some cases $B$ may not be invertible. It has been proved by Van~den~Nest et~al.~\cite{nest}
that $T$ can then always be transformed into an equivalent code 
$T' = \binom{A'}{B'}$, where $B'$ is invertible. They also show how the appropriate 
transformation is found.

\begin{example}
We are given the following generator matrix of a $[[6,0,4]]$ stabilizer code.
\[
S=
\begin{pmatrix}
\sigma_x & I & I & \sigma_z & \sigma_z & \sigma_z \\
I & \sigma_x & I & \sigma_y & \sigma_z & \sigma_x \\
I & I & \sigma_x & \sigma_y & \sigma_x & \sigma_z \\
I & \sigma_z & I & \sigma_x & \sigma_y & \sigma_z \\
I & \sigma_z & \sigma_z & I & \sigma_x & \sigma_x \\
\sigma_z & \sigma_z & I & \sigma_z & I & \sigma_x
\end{pmatrix}
\]
The corresponding binary stabilizer is
\[
S_b = (Z \mid X) =
\left(
\begin{array}{cccccc|cccccc}
0 & 0 & 0 & 1 & 1 & 1 & 1 & 0 & 0 & 0 & 0 & 0\\
0 & 0 & 0 & 1 & 1 & 0 & 0 & 1 & 0 & 1 & 0 & 1 \\
0 & 0 & 0 & 1 & 0 & 1 & 0 & 0 & 1 & 1 & 1 & 0\\
0 & 1 & 0 & 0 & 1 & 1 & 0 & 0 & 0 & 1 & 1 & 0\\
0 & 1 & 1 & 0 & 0 & 0 & 0 & 0 & 0 & 0 & 1 & 1 \\
1 & 1 & 0 & 1 & 0 & 0 & 0 & 0 & 0 & 0 & 0 & 1\\
\end{array}
\right).
\]
Let $A=Z^T$ and $B=X^T$. The transpose binary stabilizer is then $T = \binom{A}{B}$.
Since $B$ is invertible, 
\[
TB^{-1} = \binom{AB^{-1}}{I} =
\begin{pmatrix}
0 & 0 & 0 & 1 & 1 & 1 \\
0 & 0 & 1 & 1 & 0 & 1 \\
0 & 1 & 0 & 1 & 1 & 0 \\
1 & 1 & 1 & 1 & 1 & 1 \\
1 & 0 & 1 & 1 & 0 & 0 \\
1 & 1 & 0 & 1 & 0 & 0 \\
\hline
1 & 0 & 0 & 0 & 0 & 0 \\
0 & 1 & 0 & 0 & 0 & 0 \\
0 & 0 & 1 & 0 & 0 & 0 \\
0 & 0 & 0 & 1 & 0 & 0 \\
0 & 0 & 0 & 0 & 1 & 0 \\
0 & 0 & 0 & 0 & 0 & 1
\end{pmatrix}.
\]
We set the nonzero diagonal element in $AB^{-1}$ to 0 and get the 
adjacency matrix of the simple undirected graph shown in \autoref{l21}.
\end{example}

\section{Efficient Algorithms for Graph Codes}

We have seen that a graph code, $\mathcal{C}$, is a self-dual additive code
over $\GF(4)$ whose generator matrix is of the form $\Gamma + \omega I$.
It can be shown that the additive code over $\mathbb{Z}_4$ given by $2\Gamma + I$ has
the same weight distribution as $\mathcal{C}$. For graph codes, but not in the general case, we
may therefore replace the elements from $\GF(4)$ with elements from 
$\mathbb{Z}_4$ by the mappings $0 \mapsto 0$, $1 \mapsto 2$, $\omega \mapsto 1$, $\omega^2 \mapsto 3$.

\begin{example}
A self-dual additive code over $\GF(4)$ generated by
\[
\begin{pmatrix}
\omega & 0 & 1 & 0 & 1\\
0 & \omega & 0 & 1 & 0\\
1 & 0 & \omega & 0 & 1\\
0 & 1 & 0 & \omega & 0\\
1 & 0 & 1 & 0 & \omega
\end{pmatrix},
\]
has the same weight distribution as the additive code over $\mathbb{Z}_4$ generated by
\[
\begin{pmatrix}
1 & 0 & 2 & 0 & 2\\
0 & 1 & 0 & 2 & 0\\
2 & 0 & 1 & 0 & 2\\
0 & 2 & 0 & 1 & 0\\
2 & 0 & 2 & 0 & 1
\end{pmatrix}.
\]
\end{example}

The interpretation as a code over $\mathbb{Z}_4$ is an advantage when we write computer programs to operate
on such codes, since $\mathbb{Z}_4$ arithmetic may be faster and simpler to
implement. 

\begin{proposition}\label{l24}
Let $\mathcal{C}$ be a self-dual additive code over $\GF(4)$ with generator 
matrix $C = \Gamma + \omega I$.
Let $\boldsymbol{s} \in \mathcal{C}$ be a codeword formed by adding $k$ different rows
of $C$. Then it must be true that $w_H(\boldsymbol{s}) \ge k$.
\end{proposition}
\begin{proof}
Each row of $C$ has an element $\omega$, and this element is in a different
position in each row. All other elements in $C$ are 0 or 1.
Let row number $i$ of $C$ be one of the rows we added to get $\boldsymbol{s}$.
Element $s_i$ will then be $a_0 + a_1 + \cdots + a_k$, where
$a_i = \omega$ and $a_j \in \{0,1\}$, $\forall j \ne i$.
It follows that $s_i \in \{\omega, \omega+1 \}$.
$\boldsymbol{s}$ will have $k$ elements of the same form
and therefore $w_H(\boldsymbol{s}) \ge k$.
\end{proof}

The special form of the generator matrix 
of a graph code makes it easier to find the distance of the code.
An $[[n,0,d]]$ code has $2^n$ codewords, but if the generator matrix is given in
graph form, it is not necessary to check all the codewords to find the distance of the code.
If we have found a codeword $\boldsymbol{s}$, where $w_H(\boldsymbol{s}) \le e$, we
know that no codeword formed by adding $e$ or more rows of the generator matrix 
can have lower weight. This fact is used in \autoref{l25}.
A similar technique can also be used to find the weight distribution of a code.
To find $w_p$, the number of codewords of weight $p$, only codewords formed by adding
$p$ or fewer rows of the generator matrix needs to be considered.
This approach is used by \autoref{l26}.
To find the complete weight distribution, $\boldsymbol{w} = \{w_0, w_1, \ldots, w_n\}$, we
must generate all codewords, but a \emph{partial weight distribution},
$\boldsymbol{w}_p = \{w_0, w_1, \ldots, w_p\}$, where $p < n$, can be found more efficiently.
We will later use the partial weight distribution to distinguish inequivalent codes.

\begin{algorithm}
\caption{Finding the Distance of a Graph Code}\label{l25} 
\begin{algorithmic}
\State \hskip -6pt
\begin{tabular}{ll}
\textbf{Input} 
& $C$: a generator matrix in graph form\\
\textbf{Output} 
& $d$: the distance of the code generated by $C$
\end{tabular}
\Statex
\Procedure{FindDistance}{$C$}
  \State $d \gets \infty$
  \State $i \gets 1$
  \While{i < $d$}
    \ForAll{codewords $\boldsymbol{s}$, such that $\boldsymbol{s}$ is a sum of $i$ rows}
      \If{$w_H(\boldsymbol{s}) < d$}
        \State $d \gets w_H(\boldsymbol{s})$
        \If{$d = i$}
          \State \textbf{return} $d$
        \EndIf
      \EndIf
    \EndFor
    \State $i \gets i + 1$
  \EndWhile
  \State \textbf{return} $d$
\EndProcedure
\end{algorithmic} 
\end{algorithm}

\begin{algorithm}
\caption{Finding the Number of Codewords of a Given Weight}\label{l26} 
\begin{algorithmic}
\State \hskip -6pt
\begin{tabular}{ll}
\textbf{Input} 
& $C$: a generator matrix in graph form\\
& $p$: weight of the codewords we want to count\\
\textbf{Output} 
& $w_p$: the number of codewords of weight $p$
\end{tabular}
\Statex
\Procedure{CountWeight}{$C$, $p$}
  \State $w_p \gets 0$
  \For{i}{0}{p}
    \ForAll{codewords $\boldsymbol{s}$, such that $\boldsymbol{s}$ is a sum of $i$ rows}
      \If{$w_H(\boldsymbol{s}) = p$}
        \State $w_p \gets w_p + 1$
      \EndIf
    \EndFor
  \EndFor
  \State \textbf{return} $w_p$
\EndProcedure
\end{algorithmic} 
\end{algorithm}

\section{Quadratic Residue Codes}\label{l27}

\begin{definition}
A \emph{Paley graph}, $G=(V,E)$, is constructed as follows.
Given a prime power $m$, such that $m \equiv 1 \pmod{4}$,
let the elements of the finite field $\GF(m)$ be the set of vertices, $V$.
Let two vertices, $i$ and $j$, be joined by an edge, $\{i,j\} \in E$,
iff their difference is a quadratic residue (square) in $\GF(m) \backslash \{0\}$,
i.e., there exists an $x \in \GF(m) \backslash \{0\}$ such that $x^2 \equiv i-j$.
\end{definition}

\begin{proposition}
A Paley graph is a strongly regular graph~\cite{B11}
with parameters $(4t+1, 2t, t-1, t)$, i.e., 
it has $4t + 1$ vertices, each with degree $2t$, and
the properties that any two adjacent vertices have $t-1$ common neighbours and 
any two non-adjacent vertices have $t$ common neighbours.
\end{proposition}

We will study graph codes based on Paley graphs. Some bounds on the distance 
of self-dual quantum codes constructed from strongly 
regular graphs in general have been given by Tonchev~\cite{B57}. 

\begin{definition}
A self-dual additive code over $\GF(4)$ with generator matrix $\Gamma + \omega I$,
where $\Gamma$ is the adjacency matrix of a Paley graph,
is a type of \emph{quadratic residue code}~\cite{B43,B23}.
\end{definition}

When $p$ is a prime, the adjacency matrix of the Paley graph
on $\GF(p)$ will be circulant, with each row being the cyclic shift
of a \emph{Legendre sequence}.
Let $QR$ be the set of all quadratic residues modulo $p$.
$a$ is a quadratic residue modulo $p$ iff $a \not\equiv 0 \pmod{p}$
and the congruence $y^2 \equiv a \pmod{p}$ has a solution
$y \in \mathbb{Z}_p$.
The Legendre sequence of length $p$, $\boldsymbol{l}_p = (l_0, l_1, \ldots, l_{p-1})$, 
is a binary sequence with $l_i = 1$ if $i \in QR$, and $l_i = 0$ otherwise.
Let $\boldsymbol{l}_p \gg s$ be a Legendre sequence cyclically shifted $s$ times to the right.
Form the $p \times p$ matrix $\Gamma$ by letting row $i$ be $\boldsymbol{l}_p \gg i$,
for $0 \le i < p$.
It can be shown that $p$ must be a prime of the form $4k+1$ for $\Gamma$ to be
symmetric, which is a requirement for the adjacency matrix of an undirected graph.

\begin{definition}
To get a \emph{bordered quadratic residue code}~\cite{B43,B23} of length $m+1$, first
construct the quadratic residue code of length $m$. 
Then add a top row of $m$ 1s, $(1,1,\ldots,1)$, to the generator matrix.
Finally, add a leftmost column with an $\omega$ followed by $m$ 1s,
$(\omega,1,1,1,\ldots,1)^T$, to the generator matrix.
\end{definition}

\begin{figure}
\centering
\subfloat[The $C_5$ Graph]
  {\reflectbox{\includegraphics[width=.37\linewidth]{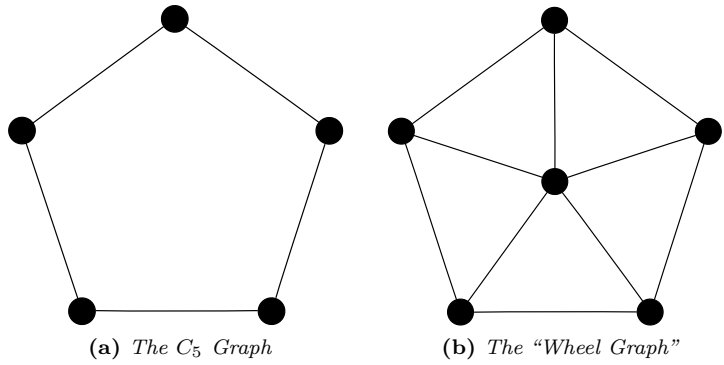}}
  \label{l28}}
\quad
\subfloat[The ``Wheel Graph'']
  {\reflectbox{\includegraphics[width=.37\linewidth]{f7.eps}}
  \label{l29}}
\caption{Graphs of the QR and BQR Codes for $m=5$}
\label{l30}
\end{figure}

\begin{example}
We will construct the quadratic residue code of length 5 and
bordered quadratic residue code of length 6.
The quadratic residues modulo 5 are $QR=\{1,4\}$, and
the Legendre sequence of length 5 is $\boldsymbol{l} = (0,1,0,0,1)$.
From this sequence we construct the matrix
\[
\Gamma =
\begin{pmatrix}
0&1&0&0&1\\
1&0&1&0&0\\
0&1&0&1&0\\
0&0&1&0&1\\
1&0&0&1&0
\end{pmatrix}.
\]
This is the adjacency matrix of the Paley graph on 5 vertices,
which is the graph $C_5$, shown in \autoref{l28}.
$\Gamma + \omega I$ is the generator matrix of 
a $[[5,0,3]]$ quantum code.
We border the matrix and get
\[
\Gamma' =
\begin{pmatrix}
0&1&1&1&1&1\\
1&0&1&0&0&1\\
1&1&0&1&0&0\\
1&0&1&0&1&0\\
1&0&0&1&0&1\\
1&1&0&0&1&0
\end{pmatrix},
\]
the adjacency matrix of the ``wheel graph'', shown in \autoref{l29},
which represents the extremal $[[6,0,4]]$ code, also known as the
\emph{Hexacode}.
\end{example}

The integers $m \le 30$ where $m$ is a prime of the form
$4k+1$ are 5, 13, 17, and 29. The quadratic residue codes
for $m=5$, 13 and 29, and their bordered extensions, achieve
the highest possible distance, as given by \autoref{l16} on
page~\pageref*{l16}.
For $m=17$ the construction gives a $[[17,0,5]]$ code,
but there exists a code with distance 7. The bordered extension
is a $[[18,0,6]]$ code, when distance 8 is achievable.
It can be shown (under some constraints) 
that there exists a unique $[[18,0,8]]$ code~\cite{B4}.
This code was constructed by MacWilliams et al.~\cite{B34} using
a construction similar to bordered quadratic residue codes.
Glynn et al.~\cite{B23} present the following technique for constructing
the $[[18,0,8]]$ quantum code.
For a prime $p=4k+1$, generate a set, $K$, of all
powers of 4 modulo $p$. For instance, for $p=5$, $K=\{4^0=1, 4^1=4\}$,
which is also the set of quadratic residues modulo 5. In general, if
2 is a primitive root modulo $p$, we will construct a quadratic residue code.
This is the case for $p=5$, 13 and 29, but for $p=17$ we get
$K=\{4^0=1,4^1=4,4^2=16,4^3=13\}$. We then find a partition of 
$\mathbb{Z}_{17} \backslash \{0\}$
into sets that are multiples of $K$, $\mathbb{Z}_{17} \backslash \{0\} =
K \cup 2K \cup 3K \cup 6K$, where $2K = \{2,8,15,9\}$, 
$3K = \{3,12,14,5\}$, and $6K = \{6,7,11,10\}$. We must combine two of these
sets to make the set $H$, which must satisfy $2H + H = \mathbb{Z}_{17} \backslash \{0\}$.
The valid combinations are $H = K \cup 3K$ and $H = K \cup 6K$.
Note that the quadratic residue set, $K \cup 2K$ is not valid.
The two valid combinations generate the sequences
$(0,1,0,1,1,1,0,0,0,0,0,0,1,1,1,0,1)$ and $(0,1,0,0,1,0,1,1,0,0,1,1,0,1,0,0,1)$.
We make two circulant adjacency matrices by cyclically shifting the
sequences, in exactly the same way as we did with Legendre sequences.
The two matrices correspond to two $[[17,0,7]]$ codes, which can be
shown to be equivalent. Bordering either of the matrices will
generate the $[[18,0,8]]$ code.

Paley graphs and quadratic residue codes can be constructed 
for any prime power $m = p^n$ where $m \equiv 1 \pmod{4}$.
The only possible non-prime lengths below 30 are 9 and 25.
The generator matrices of these codes are not circulant, but
they are composed of circulant submatrices.
When $n=2$, there will be $p^2$ $p \times p$ circulant matrices.
For $m=9$, we generate the adjacency matrix
\[
\left(
\begin{array}{ccc|ccc|ccc}
0&1&1&0&0&1&0&1&0 \\
1&0&1&1&0&0&0&0&1 \\
1&1&0&0&1&0&1&0&0 \\
\hline
0&1&0&0&1&1&0&0&1 \\
0&0&1&1&0&1&1&0&0 \\
1&0&0&1&1&0&0&1&0 \\
\hline
0&0&1&0&1&0&0&1&1 \\
1&0&0&0&0&1&1&0&1 \\
0&1&0&1&0&0&1&1&0
\end{array}
\right).
\]
This corresponds to a $[[9,0,3]]$ code. The optimal distance for length 9
is 4. The bordered $[[10,0,4]]$ code is, however, extremal.
We have also generated the quadratic residue code for $m=25$. It is
a $[[25,0,5]]$ code, and the bordered extension is a $[[26,0,6]]$ code.
In both cases the best achievable distance is at least 8.
This suggests that quadratic residue codes of non-prime lengths
are not as strong as for prime lengths.
\autoref{l31} summarises the distance of quadratic
residue and bordered quadratic residue codes for lengths up to 30.

\begin{table}
\centering
\caption{Distance ($d$) of Quadratic Residue Codes of Length $m$ and Bordered Quadratic Residue Codes
of Length $m+1$\label{l31}}
\begin{tabular}{cccc}
\toprule
\multicolumn{2}{c}{QR codes} & 
\multicolumn{2}{c}{BQR codes} \\
\midrule
$m$ & $d$ & $m+1$ & $d$ \\
\cmidrule(r){1-2}\cmidrule(l){3-4}
5 & 3 & 6 & 4 \\
9 & 3 & 10 & 4 \\
13 & 5 & 14 & 6 \\
17 & 5 & 18 & 6 \\
25 & 5 & 26 & 6 \\
29 & 11 & 30 & 12 \\
\bottomrule
\end{tabular}
\end{table}

\chapter{Nested Regular Graph Codes}\label{l32}

\section{The Hexacode and the Dodecacode}

\begin{figure}
\centering
\subfloat[{$K_2[K_3]$} Graph of the {$[[6,0,4]]$} Code]
  {\includegraphics[width=.47\linewidth]{f6.eps}
  \label{l33}}
\quad
\subfloat[{$K_3[K_3]$} Graph of a {$[[9,0,4]]$} Code]
  {\includegraphics[width=.47\linewidth]{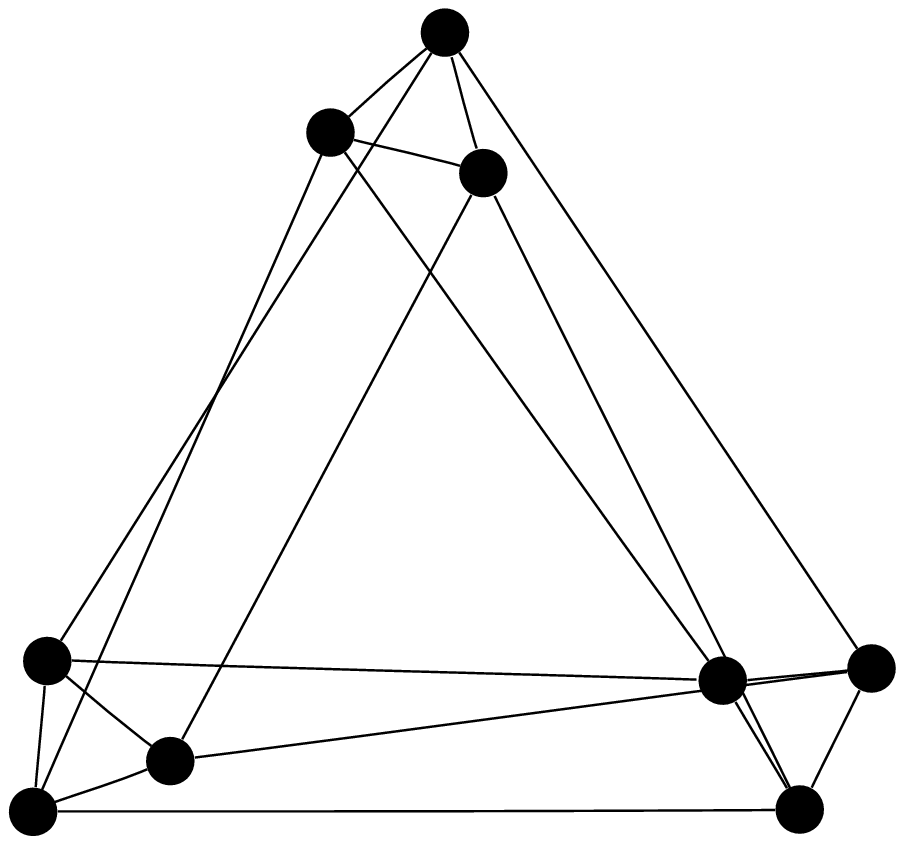}
  \label{l34}}
\\
\subfloat[{$K_3[K_4]$} Graph of the {$[[12,0,6]]$} Code]
  {\includegraphics[width=.47\linewidth]{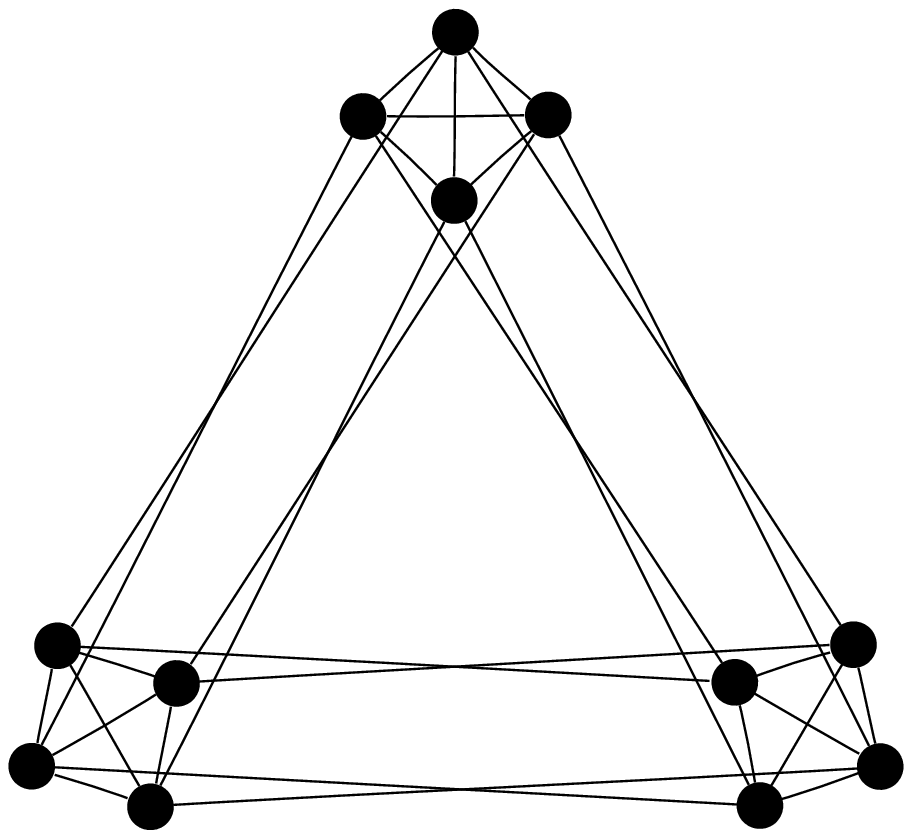}
  \label{l35}}
\quad
\subfloat[{$K_2[K_3[K_3]]$} Graph of a {$[[18,0,6]]$} Code]
  {\includegraphics[width=.47\linewidth]{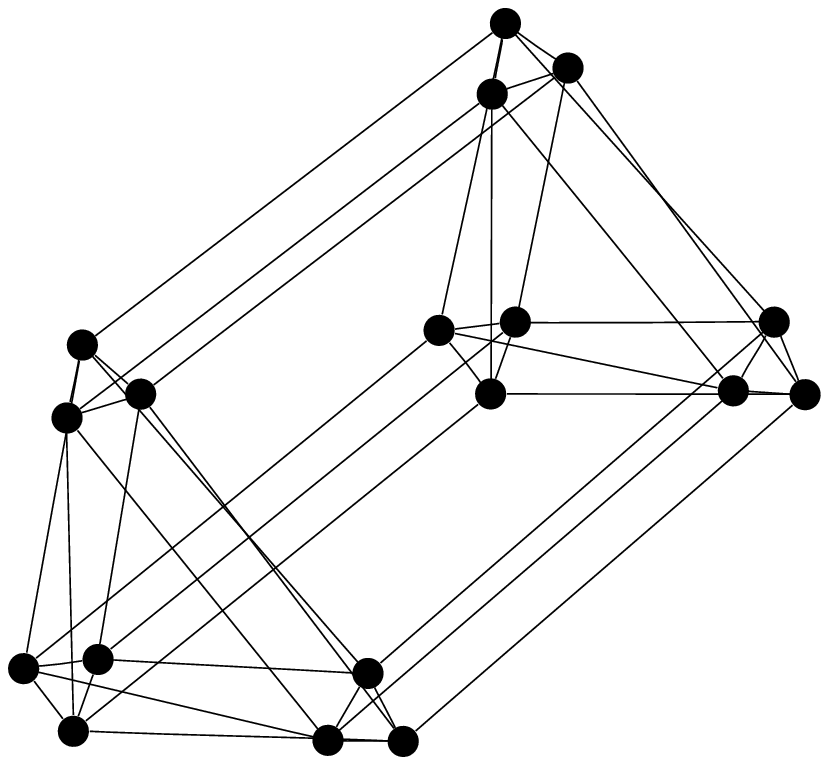}}
\\
\subfloat[{$K_5[K_4]$} Graph of a {$[[20,0,8]]$} Code]
  {\includegraphics[width=.47\linewidth]{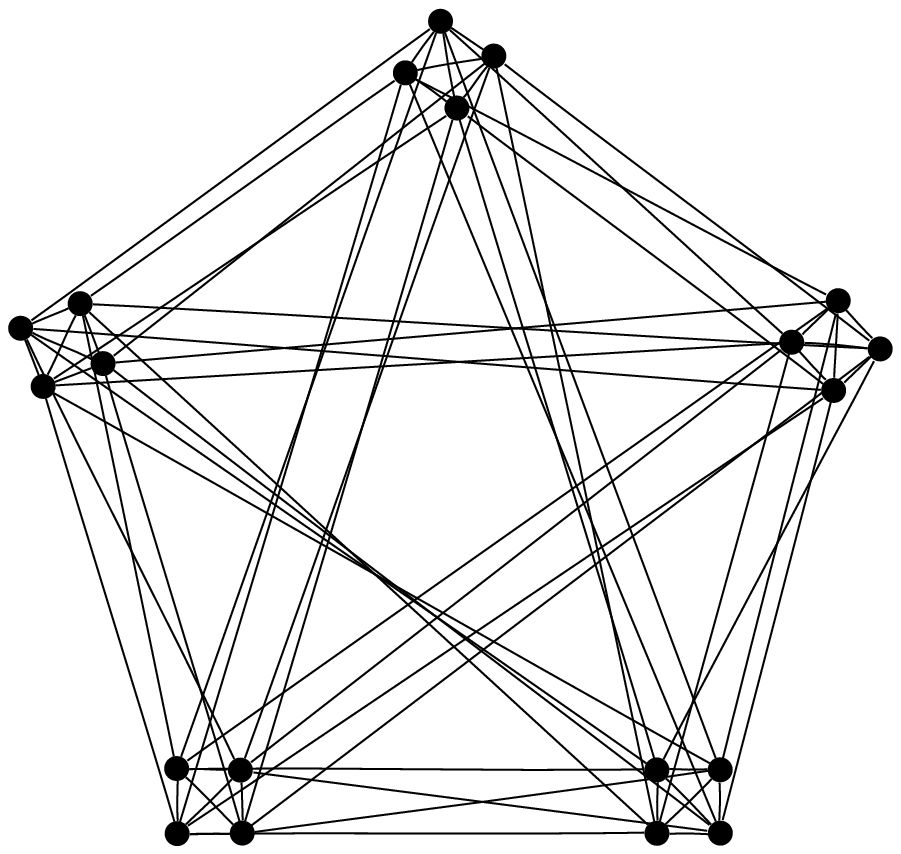}
  \label{l36}}
\quad
\subfloat[{$K_5[K_5]]$} Graph of a {$[[25,0,8]]$} Code]
  {\includegraphics[width=.47\linewidth]{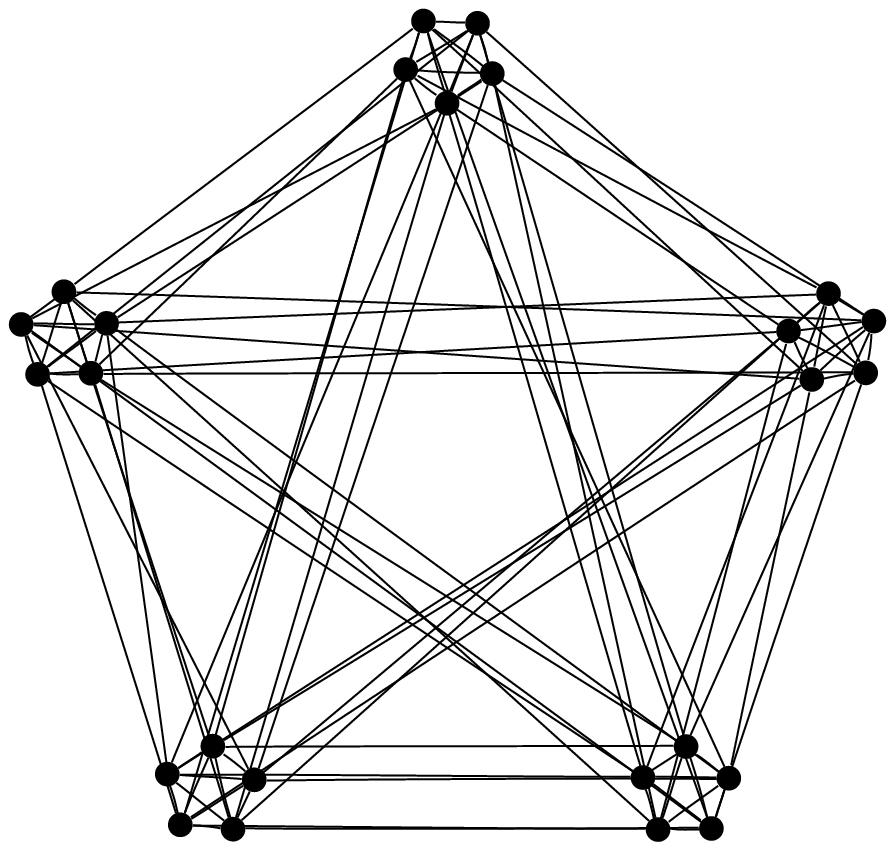}
  \label{l37}}
\caption{Nested Clique Graphs}
\label{l38}
\end{figure}

The unique extremal $[[6,0,4]]$ quantum code is also known as the \emph{Hexacode}.
As all self-dual quantum codes, it can be represented by a graph with adjacency matrix 
$\Gamma$, such that $\Gamma + \omega I$ is the generator matrix of a self-dual 
additive $[[6,0,4]]$ code over $\GF(4)$. One such representation of the Hexacode 
is given by the generator matrix
\[
C=
\begin{pmatrix}
\omega&0&1&1&1&0\\
0&\omega&0&1&1&1\\
1&0&\omega&0&1&1\\
1&1&0&\omega&0&1\\
1&1&1&0&\omega&0\\
0&1&1&1&0&\omega
\end{pmatrix}.
\]
The corresponding graph is shown in \autoref{l33}.
The figure emphasises the fact that the graph consists of
two instances of the complete graph $K_3$, or equivalently,
that there are two 3-cliques in the graph that partition all six vertices
into two disjoint sets.
We also see that each vertex in each 3-clique is connected to
exactly one vertex in the other 3-clique. We will view these connections as an
``outer'' 2-clique; the rationale behind this will become clear.
We call the whole graph a ``2-clique of 3-cliques'', or $K_2[K_3]$ for short.
We also note that $C$ is a circulant matrix, and that all vertices
have vertex degree 3.

Another well-known unique extremal quantum code is the $[[12,0,6]]$ \emph{Dodecacode}.
It can be represented as a graph code with a circulant generator matrix
whose first row is $(\omega00101110100)$.
The corresponding graph is shown in \autoref{l35}. 
We see how this graph can be called a ``3-clique of 4-cliques'', or $K_3[K_4]$.

\begin{definition}
A general \emph{nested clique graph}, $G=(V,E)$, can be described as an
``$n_1$-clique of $n_2$-cliques of $\cdots$ of $n_l$-cliques'', denoted
$K_{n_1}[K_{n_2}[\cdots[K_{n_l}]]]$.
The number of vertices in the graph is $n = n_1 n_2 \cdots n_l$.
It must be possible to partition the vertices into $\frac{n}{n_l}$ disjoint
subsets of size $n_l$, $V=V_1 \cup V_2 \cup \cdots \cup V_{\frac{n}{n_l}}$, 
such that the induced subgraph on each subset is the complete graph $K_{n_l}$.
Let $E_i$ be the edges of the induced subgraph on $V_i$.
Let $E_{ij}$ be the set of edges $\{u,v\}$ where $u \in V_i$ and $v \in V_j$.
$|E_{ij}|$ must be either 0 or $n_l$. If $|E_{ij}|$ is $n_l$, every vertex
in $V_i$ must be connected to one vertex in $V_j$, and every vertex in $V_j$ 
must be connected to one vertex in $V_i$. If this is the case, we 
say that the sets $V_i$ and $V_j$ are connected.
All edges in $E$ must be part of a clique or a connection between cliques, i.e,
$\bigcup_{i} E_i \bigcup_{i,j} E_{ij} = E$.
We then form the graph $G'=(V',E')$ on $\frac{n}{n_l}$ vertices, each vertex 
$v'_i \in V'$ corresponding to a subset $V_i \subset V$.
Let there be an edge between $v'_i$ and $v'_j$ iff $V_i$ and $V_j$
are connected.
$G'$ must be the nested clique graph $K_{n_1}[K_{n_2}[\cdots[K_{n_{l-1}}]]]$,
or, if $l=2$, $G'$ must be the complete graph $K_{n_1}$.
\end{definition}

There may be many non-isomorphic $K_{n_1}[K_{n_2}[\cdots[K_{n_l}]]]$ graphs,
so the nested clique characterisation does not uniquely identify a graph,
but only partially describes its structure. In particular, 
the connections between the inner cliques of a nested clique graph are not defined.
For instance, in the $K_3[K_4]$ graph shown in \autoref{l35},
the 12 edges that are not part of an inner 4-clique form a Hamiltonian cycle.
Another $K_3[K_4]$ graph, where these 12 edges form four disjoint cycles of length three
(or four 3-cliques), corresponds to a $[[12,0,4]]$ code.
This means that a highly regular nested structure is not enough to 
guarantee optimal distance.

\section{Graph Codes with Minimum Regular Vertex Degree}

\begin{proposition}
All vertices of a graph corresponding to an $[[n,0,d]]$ quantum code have
a vertex degree of at least $d-1$.
\end{proposition}
\begin{proof}
If a graph has a vertex $v_i$ of degree $\delta$, then the $i$th row of the adjacency
matrix, $\Gamma$, will have weight $\delta$. The $i$th row of the generator matrix,
$\Gamma + \omega I$, of the corresponding self-dual additive code
over $\GF(4)$, will then define a codeword of weight $\delta + 1$.
A vertex with degree less than $d-1$ would therefore correspond to a codeword
of weight less than $d$, which is impossible.
\end{proof}

\begin{proposition}\label{l40}
In a $k$-regular graph with an odd number of vertices, $k$ must be even.
\end{proposition}
\begin{proof}
Let $\Delta$ be the sum of the vertex degrees of all vertices in a graph.
Since each edge of a graph is incident on two vertices, $\Delta = 2 |E|$, i.e.,
twice the number of edges.
It follows that $\Delta$ must always be an even number.
In a $k$-regular graph on $n$ vertices, $\Delta = kn$, and
thus either $k$ or $n$ must be an even number.
\end{proof}

\begin{definition}
If a graph corresponding to an $[[n,0,d]]$ quantum code has a regular vertex
degree of $d-1$, then it has \emph{minimum regular vertex degree}.
By \autoref{l40}, if $n$ is odd and $d$ is even, a regular vertex degree
of $d-1$ is impossible. In this case, a regular vertex 
degree of $d$ is the minimum regular vertex degree.
\end{definition}

The graph representation of the Hexacode shown in \autoref{l33}
is 3-regular, and the graph representation of the Dodecacode 
shown in \autoref{l35} is 5-regular. 
With the distances of the two codes being 4 and 6, respectively, we conclude
that the graph representations of both codes have minimum regular vertex degree.
This also implies that no other graph representation of these two codes can
have a smaller number of edges. When $n$ is odd and $d$ is even, a graph representation 
with minimum regular vertex degree is not necessarily the graph representation with 
the fewest edges.

\begin{proposition}
The nested clique graph $K_{n_1}[K_{n_2}[\cdots[K_{n_l}]]]$ is a $k$-regular graph,
where $k = {(n_1 - 1)}+{(n_2 - 1)}+\cdots+{(n_l - 1)}$.
\end{proposition}
\begin{proof}
Every vertex of the graph has $n_l - 1$ neighbours as part of an $n_l$-clique.
Each vertex must also be connected to one vertex in $n_{l-1}-1$ other $n_l$-cliques,
which contributes $n_{l-1} - 1$ to its degree.
The same must be true for the other layers of nesting.
\end{proof}

\section{Other Nested Regular Graph Codes}

We have observed that the extremal $[[6,0,4]]$ code corresponds to
a $K_2[K_3]$ graph, and that the extremal $[[12,0,6]]$ code corresponds
to a $K_3[K_4]$ graph. The intuitive next step is to search for an
extremal $[[20,0,8]]$ code among $K_4[K_5]$ graphs.
An exhaustive computer search of all $K_4[K_5]$ graphs did, however, 
not find such a code. The best result was a $[[20,0,6]]$ code, but with
only 3 codewords of weight 6 and none of weight 7.

\begin{figure}
\centering
\subfloat[{$C_4[K_4]$} Graph of a {$[[16,0,6]]$} Code]
  {\includegraphics[width=.47\linewidth]{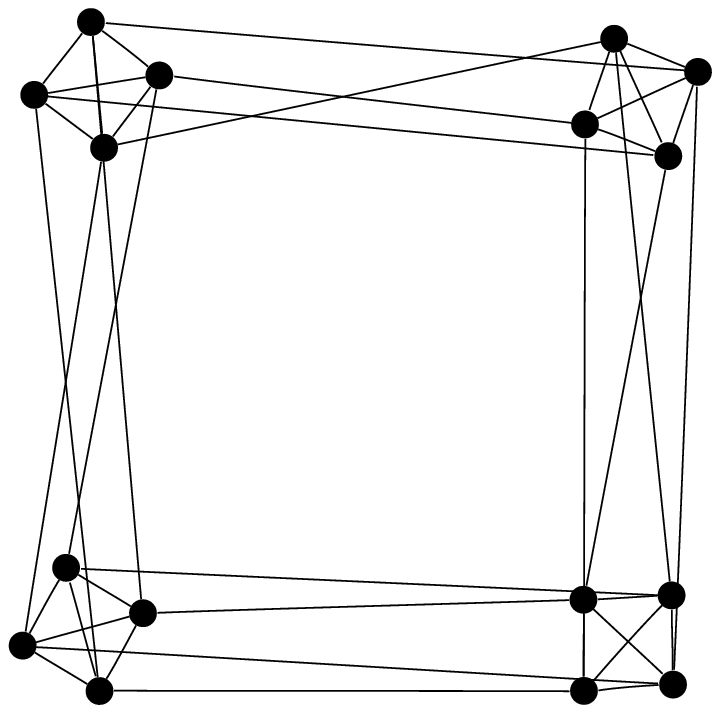}
  \label{fig:4cl4lc}}
\quad
\subfloat[{$R^4_6[K_4]$} Graph of a {$[[24,0,8]]$} Code]
  {\includegraphics[width=.47\linewidth]{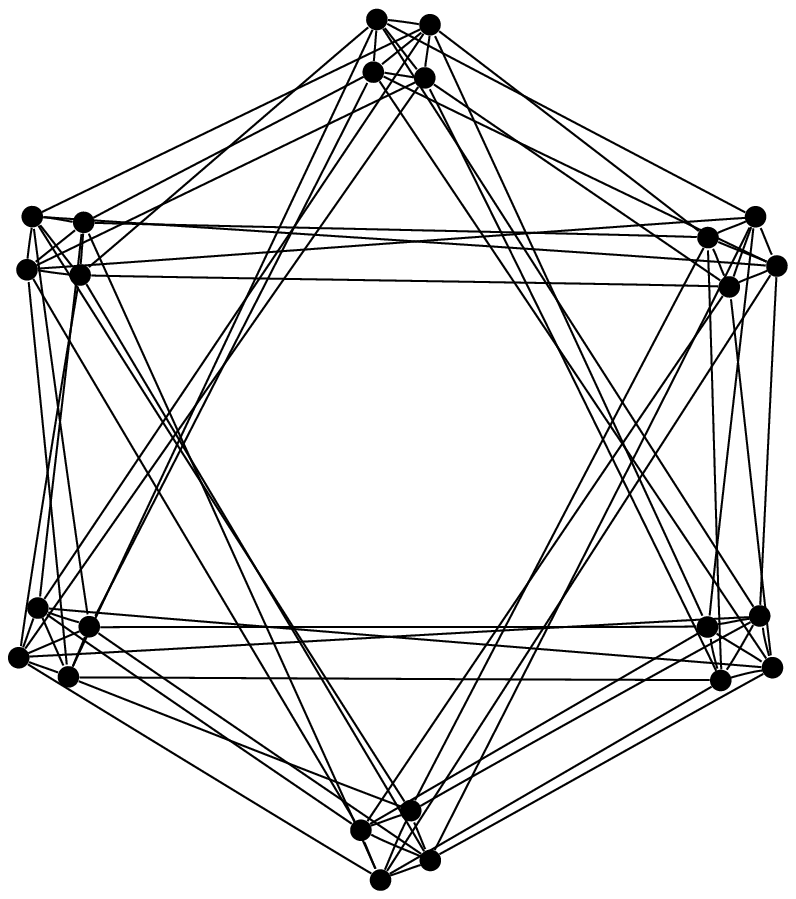}
  \label{fig:6x4cl}}
\\
\subfloat[{$R^6_9[K_3]$} Graph of a {$[[27,0,8]]$} Code]
  {\includegraphics[width=.47\linewidth]{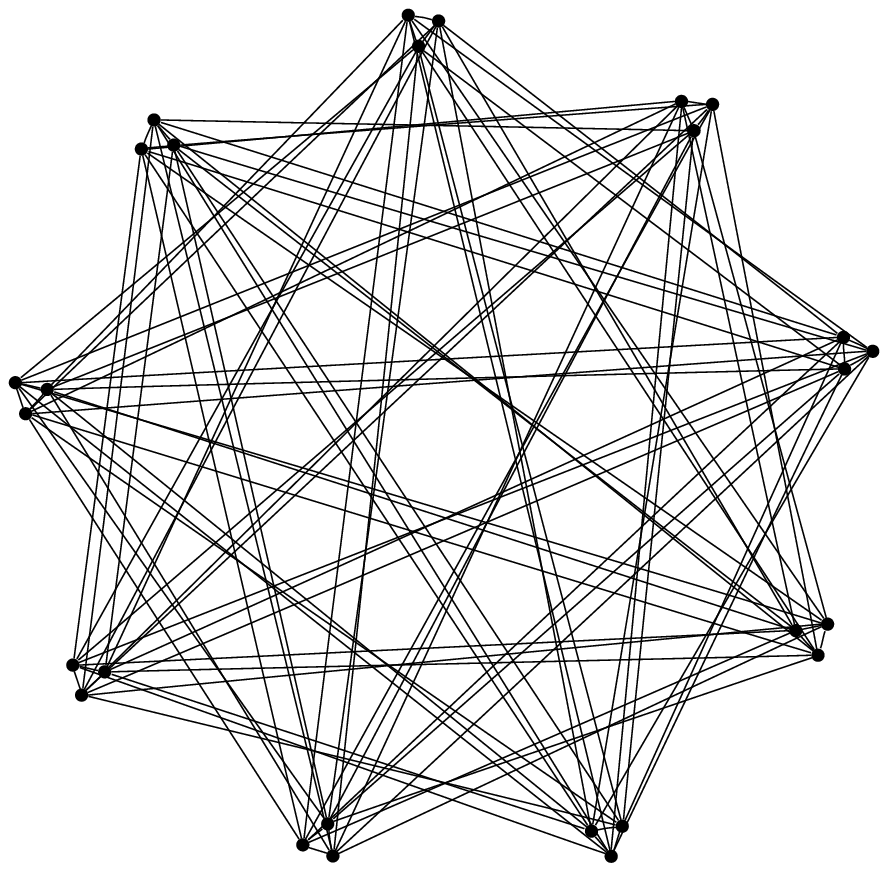}
  \label{fig:9x3cl}}
\caption{Nested Regular Graphs}
\label{fig:cliquegraphs2}
\end{figure}

Searching through all $2^{\binom{n}{2}}$ undirected graphs
on $n$ vertices is infeasible for graphs with more than a few vertices.
We have seen that the Hexacode and the Dodecacode have 
graph representations with circulant adjacency matrices, and
this is also true for all quadratic residue codes of prime length.
It therefore seems reasonable to restrict our search to the
$2^{\left\lceil\frac{n-1}{2}\right\rceil}$ circulant
symmetric adjacency matrices of graphs on $n$ vertices.
We have performed an exhaustive search of these graphs for
$n \le 30$. The most important parameter to optimise is the distance
of the quantum code. For each length $n$, we
identify the circulant adjacency matrices corresponding 
to $[[n,0,d]]$ codes with optimal distance, as listed in 
\autoref{l16} on page~\pageref*{l16},
or highest possible distance if no codes with optimal distance are found.
We next want to minimise the regular vertex degree.
Among the graphs corresponding to codes with highest possible distance
and with lowest possible regular vertex degree, we try to
identify structures similar to the nested clique representations
of the Hexacode and the Dodecacode.
Algorithms for finding all cliques in a graph have a running time
that increases exponentially with the number of vertices.
But we will only consider graphs of up to 30 vertices, 
and finding all cliques in such graphs can be done quickly
with a suitable algorithm~\cite{B9}.
\autoref{tab:cycliccodes} summarises the results of the search by showing
data about one code of each length.
Distances and degrees marked $*$ in the table are not optimal.
Degrees marked $\dag$ are equal to $d$, but still optimal
according to \autoref{l40}.
Some of the nested regular structures identified can be seen in
\autoref{l38} and \autoref{fig:cliquegraphs2}.

\begin{table}
\centering
\caption{Nested Regular Graphs with Degree $\delta$ Corresponding to 
Circulant Graph Codes of Length $n$ and Distance $d$\label{tab:cycliccodes}}
\begin{tabular}{cccll}
\toprule
$n$ & $d$ & $\delta$ & Graph & First row of generator matrix \\
\midrule
 2 & 2         & 1            & $K_2$           & $\omega1$ \\ 
 3 & 2         & \hspace{\skpl}$2^\dag$ & $K_3$           & $\omega11$ \\
 4 & 2         & 1            & $K_2+K_2$       & $\omega010$ \\
 5 & 3         & 2            & $C_5$           & $\omega0110$ \\
 6 & 4         & 3            & $K_2[K_3]$      & $\omega01110$ \\
 7 & 3         & 2            & $C_7$           & $\omega001100$ \\
 8 & 4         & 3            & $K_2[C_4]$      & $\omega0011100$ \\
 9 & 4         & \hspace{\skpl}$4^\dag$ & $K_3[K_3]$      & $\omega00111100$ \\
10 & 4         & 3            & $K_2[C_5]$      & $\omega000111000$ \\
11 & \hspace{\skpl}$4^*$ & \hspace{\skpl}$4^\dag$ &                 & $\omega0001111000$ \\
12 & 6         & 5            & $K_3[K_4]$      & $\omega00101110100$ \\
13 & 5         & 4            &                 & $\omega000101101000$ \\
14 & 6         & 5            &                 & $\omega0001011101000$ \\
15 & 6         & \hspace{\skpl}$8^*$    &                 & $\omega01110011001110$ \\
16 & 6         & 5            & $C_4[K_4]$      & $\omega000100111001000$ \\
17 & 7         & \hspace{\skpl}$8^*$    &                 & $\omega0100011111100010$ \\
18 & \hspace{\skpl}$6^*$ & 5            & $K_2[K_3[K_3]]$ & $\omega00000101110100000$ \\
19 & 7         & 6            &                 & $\omega000101001100101000$ \\
20 & 8         & 7            & $K_5[K_4]$      & $\omega0000100111110010000$ \\
21 & \hspace{\skpl}$7^*$ & 6            &                 & $\omega00001000111100010000$ \\
22 & 8         & 7            &                 & $\omega000001001111100100000$ \\
23 & \hspace{\skpl}$8^*$ & \hspace{\skpl}$10^*$   &                 & $\omega0000011101111011100000$ \\
24 & \hspace{\skpl}$8^*$ & 7            & $R^4_6[K_4]$    & $\omega00000100011111000100000$ \\
25 & \hspace{\skpl}$8^*$ & \hspace{\skpl}$8^\dag$ & $K_5[K_5]$      & $\omega000010001101101100010000$ \\
26 & \hspace{\skpl}$8^*$ & 7            &                 & $\omega0000000100111110010000000$ \\
27 & \hspace{\skpl}$8^*$ & \hspace{\skpl}$8^\dag$ & $R^6_9[K_3]$    & $\omega00000001100111100110000000$ \\
28 & 10        & \hspace{\skpl}$11^*$   &                 & $\omega000001110100111001011100000$ \\
29 & 11        & \hspace{\skpl}$14^*$   &                 & $\omega0110000101110110111010000110$ \\
30 & 12        & \hspace{\skpl}$17^*$   &                 & $\omega01100001101111111110110000110$ \\
\bottomrule
\end{tabular}
\end{table}

\begin{proposition}\label{prop:always2}
A graph, $G$, with no isolated vertex, i.e., no vertex with degree~0, corresponds
to a self-dual $[[n,0,d]]$ quantum code, $\mathcal{C}$, with minimum distance $d \ge 2$.
\end{proposition}
\begin{proof}
It follows from \autoref{l24} that any codeword with non-zero weight
formed by adding 2 or more rows of the generator matrix of $\mathcal{C}$ will
have weight higher than 2.
A codeword of weight less than 2 must therefore be a row of the generator matrix.
A vertex in $G$ with degree $\delta \ge 1$ corresponds to a row in the generator
matrix of $\mathcal{C}$, and hence a codeword, of weight $\delta+1 \ge 2$.
\end{proof}

\begin{figure}
\centering
\subfloat[{$K_2[C_4]$} Graph]
  {\reflectbox{\includegraphics[width=.40\linewidth]{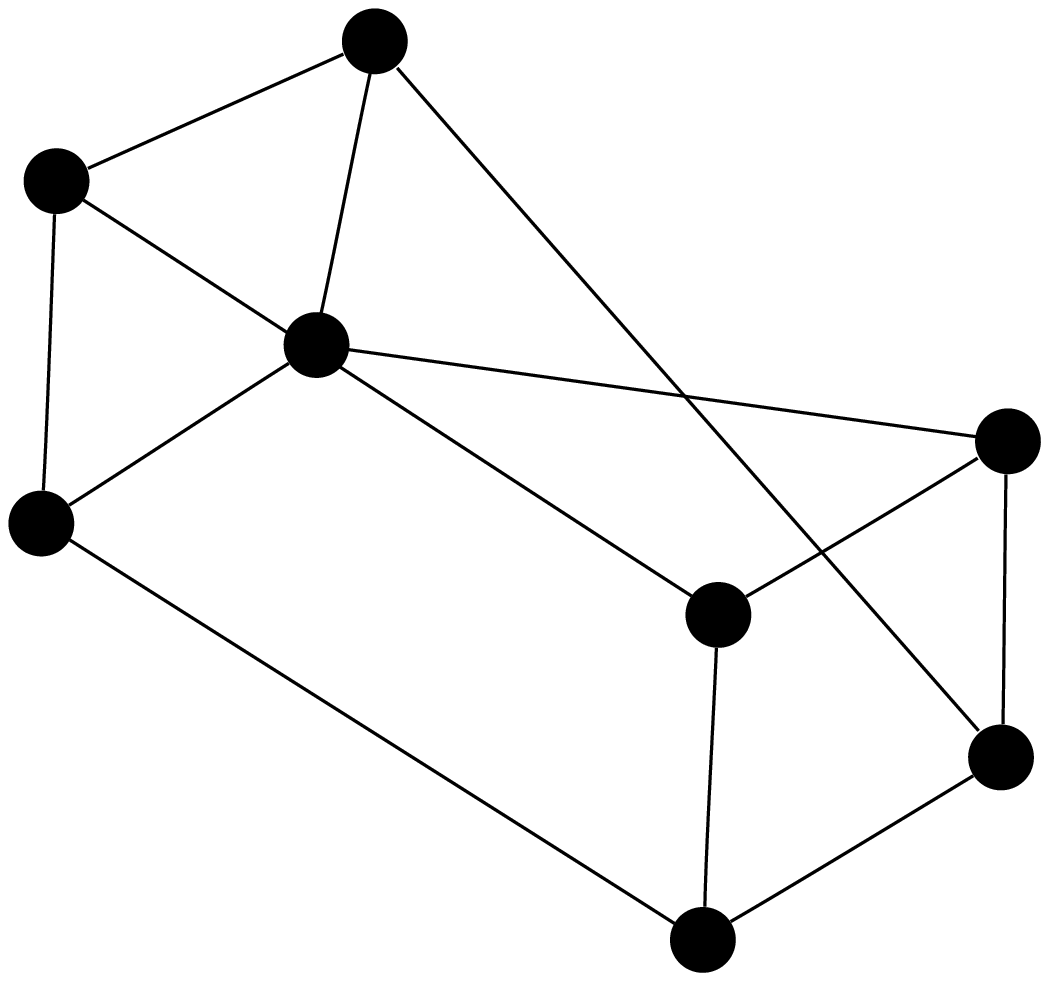}}
  \label{fig:8code}}
\quad
\subfloat[Cubical {$K_2[C_4]$} Graph]
  {\reflectbox{\includegraphics[width=.40\linewidth]{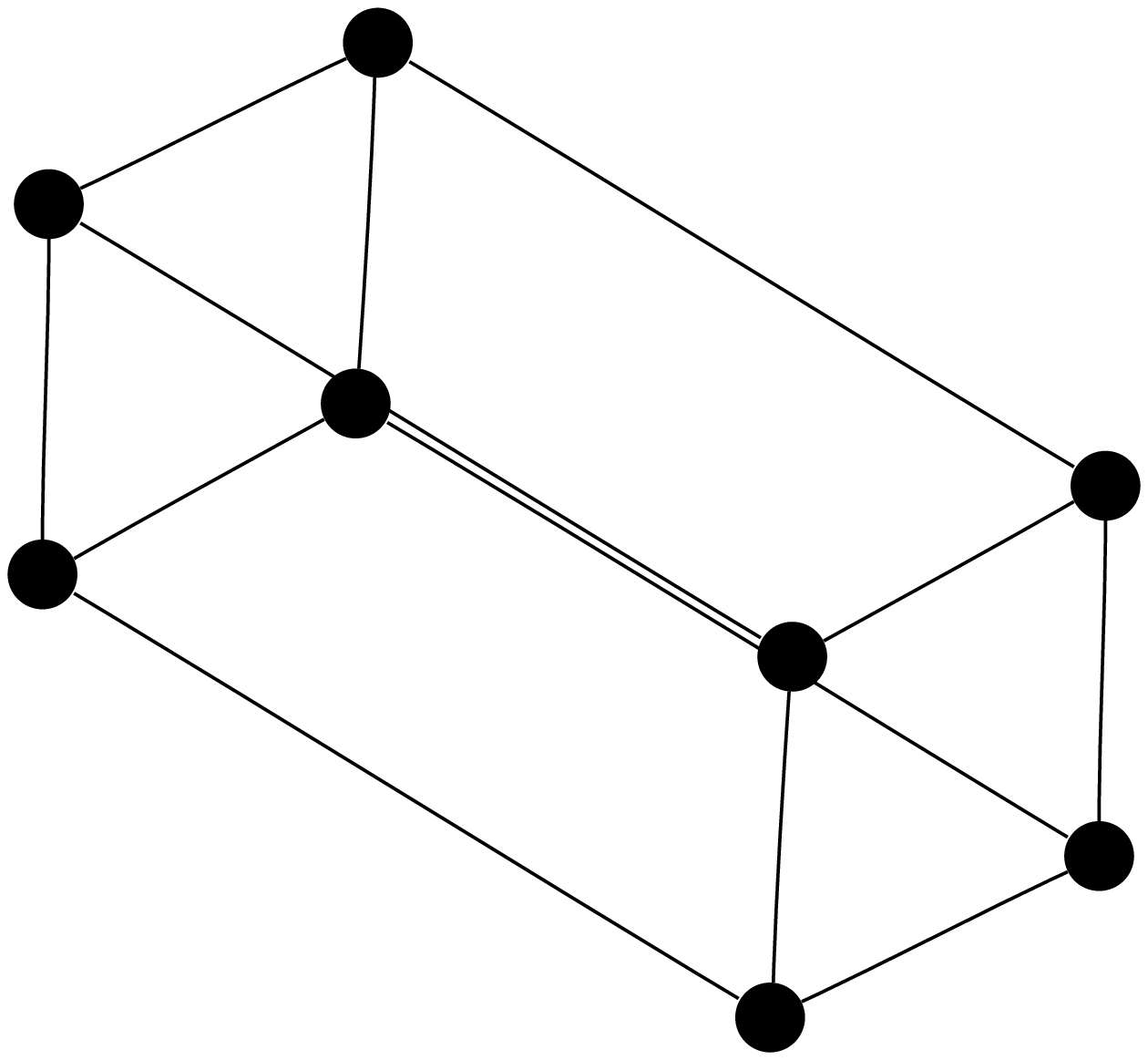}}
  \label{fig:8cube}}
\caption{Two $K_2[C_4]$ Graphs Corresponding to {$[[8,0,4]]$} Codes}
\label{fig:8codegraphs}
\end{figure}

$n=2$, 3, and 4 are not particularly interesting cases, since any
graph with no isolated vertex will correspond to a code with $d=2$.
Graphs with minimum regular vertex degree representing extremal self-dual
quantum codes of length 2, 3, and 4 can be described as 
$K_2$, $K_3$, and $K_2+K_2$ (two unconnected 2-cliques), respectively.
For lengths 5 and 7, the extremal distance is 3. The
minimum regular vertex degree is 2, and the only 2-regular
graph structure is the cycle graph. $C_5$ and $C_7$ correspond
to $[[5,0,3]]$ and $[[7,0,3]]$ codes.
For $n=8$, a graph with minimum regular vertex degree and extremal distance
consists of two 4-cycles which are connected in an ``outer'' 2-clique. 
This graph is depicted in \autoref{fig:8code}. We see
that we need to extend our definition of nested clique graphs to
\emph{nested regular graphs}.
Recall that $R^k_n$ denotes a $k$-regular graph on $n$ vertices,
$R^{n-1}_n = K_n$, and $R^2_n = C_n$.

\begin{definition}
For a general \emph{nested regular graph}, $G=(V,E)$, we use the notation
$R^{k_1}_{n_1}[R^{k_2}_{n_2}[\cdots[R^{k_l}_{n_l}]]]$.
The number of vertices in the graph is $n = n_1 n_2 \cdots n_l$.
It must be possible to partition the vertices into $\frac{n}{n_l}$ disjoint
subsets of size $n_l$, $V=V_1 \cup V_2 \cup \cdots \cup V_{\frac{n}{n_l}}$, 
such that the induced subgraph on each subset is a $k_l$-regular graph, $R^{k_l}_{n_l}$.
Let $E_i$ be the edges of the induced subgraph on $V_i$.
Let $E_{ij}$ be the set of edges $\{u,v\}$ where $u \in V_i$ and $v \in V_j$.
$|E_{ij}|$ must be either 0 or $n_l$. If $|E_{ij}|$ is $n_l$, every vertex
in $V_i$ must be connected to one vertex in $V_j$, and every vertex in $V_j$ 
must be connected to one vertex in $V_i$. If this is the case, we 
say that the sets $V_i$ and $V_j$ are connected.
All edges in $E$ must be part of a subset or a connection between subsets, i.e,
$\bigcup_{i} E_i \bigcup_{i,j} E_{ij} = E$.
We form the graph $G'=(V',E')$ on $\frac{n}{n_l}$ vertices, where each vertex 
$v'_i \in V'$ corresponds to a subset $V_i \subset V$.
Let there be an edge between $v'_i$ and $v'_j$ iff $V_i$ and $V_j$
are connected. $G'$ must be the nested regular graph 
$R^{k_1}_{n_1}[R^{k_2}_{n_2}[\cdots[R^{k_{l-1}}_{n_{l-1}}]]]$
or, if $l=2$, $G'$ must be the regular graph $R^{k_1}_{n_1}$.
\end{definition}

\begin{proposition}
$R^{k_1}_{n_1}[R^{k_2}_{n_2}[\cdots[R^{k_l}_{n_l}]]]$ is a regular graph
with vertex degree $k_1 + k_2 + \cdots + k_l$.
\end{proposition}
\begin{proof}
Every vertex of the graph has $k_l$ neighbours as part of a $k_l$-regular subgraph.
Each vertex must also be connected to one vertex in 
$k_{l-1}$ of the $n_{l-1}-1$ other $k_l$-regular graphs,
which contributes $k_{l-1}$ to its degree.
The same must be true for the other layers of nesting.
\end{proof}

\begin{figure}
\centering
\subfloat[{$K_2[C_5]$} Graph]
  {\reflectbox{\includegraphics[width=.40\linewidth]{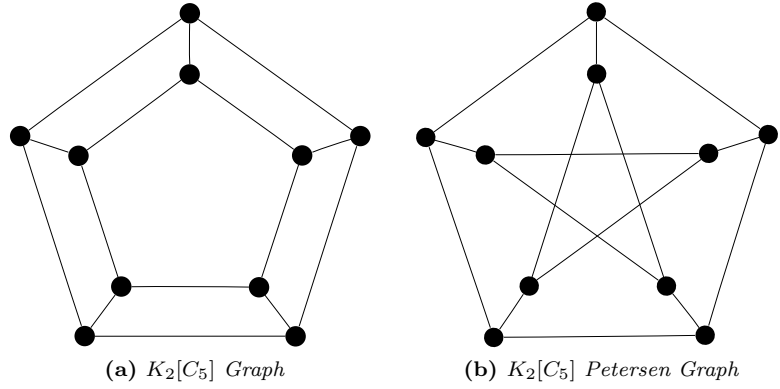}}
  \label{fig:10code}}
\quad
\subfloat[{$K_2[C_5]$} Petersen Graph]
  {\reflectbox{\includegraphics[width=.40\linewidth]{f5.eps}}
  \label{fig:10petersen}}
\caption{Two $K_2[C_5]$ Graphs Corresponding to {$[[10,0,4]]$} Codes}
\label{fig:10codegraphs}
\end{figure}

The cubical graph shown in \autoref{fig:8cube} is another $K_2[C_4]$ graph which 
also corresponds to an $[[8,0,4]]$ code. This graph is not
isomorphic to the one shown in \autoref{fig:8code}, and it does not have
a circulant generator matrix.
For $n=9$, our search reveals a $K_3[K_3]$ graph, as shown in \autoref{l34}.
For $n=10$, we find the $K_2[C_5]$ graph shown in \autoref{fig:10code}.
The famous strongly regular \emph{Petersen graph}, seen in \autoref{fig:10petersen},
can also be described as $K_2[C_5]$. It also corresponds to a $[[10,0,4]]$ code, but
does not have a circulant adjacency matrix, and is not isomorphic to the $K_2[C_5]$ graph
in \autoref{fig:10code}.
A $K_2[K_5]$ graph corresponding to a $[[10,0,4]]$ code also exists, with a
suboptimal regular degree of 4.
For $n=11$, we did not find any circulant code with extremal distance $d=5$.
Since 11 and 13 are primes, there can be no nested regular graphs corresponding
to codes of these lengths.
Were we not able to find a nested clique description of any graph on 14 or 15
vertices corresponding to an extremal code, and 
no graph with minimum regular vertex degree corresponding to a $[[15,0,6]]$ code was found.
For $n=16$, we found the $C_4[K_4]$ graph shown in \autoref{fig:4cl4lc}.
No nested regular graphs exist for the prime lengths 17 and 19.
For $n=18$, the optimal distance is $d=8$, but the best code corresponding to
a circulant adjacency matrix is an $[[18,0,6]]$ code. This code can be described
as a $K_2[K_3[K_3]]$ graph with minimum regular vertex degree. In addition, 
there also exists a $K_3[K_6]$ graph corresponding to an $[[18,0,6]]$ code.
For $n=20$, we discovered a $[[20,0,8]]$ code corresponding to
a $K_5[K_4]$ graph, shown in \autoref{l36}.
There is no circulant extremal code of length $21$, and we did not
find any nested regular graph corresponding to a $[[21,0,7]]$ code.
Neither did we find such a graph description for any circulant $[[22,0,8]]$ code.
For lengths from 23 to 27, the best codes we found all have distance 8.
According to \autoref{l16}, this distance is not extremal, 
but no codes of higher distance are known, except for length 27, where a code 
of distance 9 exists.
We did not find nested regular graphs for $n=23$ or 26, but
for $n=24$ we found a $R^4_6[K_4]$ graph, as seen in \autoref{fig:6x4cl}. 
For $n=25$ we found the $K_5[K_5]$ graph seen in \autoref{l37},
and for $n=27$ we found the $R^6_9[K_3]$ graph seen \autoref{fig:9x3cl}.
We have also found circulant graph codes of extremal distance for lengths
28, 29 and 30, but these graphs do not have minimal regular vertex degree
and can not be described as nested regular graphs.

Gulliver and Kim~\cite{B29} have studied the more general case
of self-dual additive codes over $\GF(4)$ with circulant generator matrices.
As for all self-dual quantum codes, there will be a graph code equivalent to 
any such code, but the graph code will not necessarily be circulant.
Gulliver and Kim~\cite{B29} also classified some other types of
circulant based codes for lengths up to 30, but did not find any codes 
with higher distance than we found in our search of circulant graph codes.

\section{Long Cycles in Nested Regular Graph Codes}

The nested regular description of a graph does not specify the
manner in which the regular subgraphs are connected. It turns out
that these connections are highly structured in nested regular graphs
corresponding to strong self-dual quantum codes.
We have already mentioned that the $K_3[K_4]$ graph representation of the
Dodecacode has a Hamiltonian cycle.
The $K_5[K_5]$ graph, shown in \autoref{l37}, corresponding 
to a $[[25,0,8]]$ code, contains two edge-disjoint Hamiltonian cycles.
We can then account for all the 100 edges of this graph, 50 being part of
inner 5-cliques and 25 of the remaining edges in each Hamiltonian cycle.
But this is still not a description of a unique graph, as the following
example shows.

\begin{example}
Let $G=(V,E)$ be a $K_5[K_5]$ graph.
Let the five inner 5-cliques be on the vertex sets $V_1 = \{v_0$, $v_1$, $v_2$, $v_3$, $v_4\}$,
$V_2 = \{v_5$, $v_6$, $v_7$, $v_8$, $v_9\}$, $V_3 = \{v_{10}$, $v_{11}$, $v_{12}$, $v_{13}$, $v_{14}\}$,
$V_4 = \{v_{15}$, $v_{16}$, $v_{17}$, $v_{18}$, $v_{19}\}$, and
$V_5 = \{v_{20}$, $v_{21}$, $v_{22}$, $v_{23}$, $v_{24}\}$.
Let the remaining 50 edges form two Hamiltonian cycles given by the sequences
of vertices,
$H_1= (v_0$, $v_1$, $v_2$, $v_3$, $v_4$, $v_5$, $v_6$, $v_7$, $v_8$, $v_9$, $v_{10}$, 
      $v_{11}$, $v_{12}$, $v_{13}$, $v_{14}$, $v_{15}$, $v_{16}$, $v_{17}$, $v_{18}$, 
      $v_{19}$, $v_{20}$, $v_{21}$, $v_{22}$, $v_{23}$, $v_{24}$, $v_0)$ and
$H_2= (v_0$, $v_7$, $v_{14}$, $v_{21}$, $v_3$, $v_{10}$, $v_{17}$, $v_{24}$, $v_6$, $v_{13}$, 
      $v_{20}$, $v_2$, $v_9$, $v_{16}$, $v_{23}$, $v_5$, $v_{12}$, $v_{19}$, $v_1$, $v_8$, 
      $v_{15}$, $v_{22}$, $v_4$, $v_{11}$, $v_{18}$, $v_0)$.
This graph corresponds to a $[[25,0,8]]$ self-dual quantum code.
But if we replace the Hamiltonian cycle $H_2$ with
$H'_2= (v_0$, $v_8$, $v_{16}$, $v_{24}$, $v_7$, $v_{15}$, $v_{23}$, $v_6$, $v_{14}$, $v_{22}$, 
       $v_5$, $v_{13}$, $v_{21}$, $v_4$, $v_{12}$, $v_{20}$, $v_3$, $v_{11}$, $v_{19}$, $v_2$, 
       $v_{10}$, $v_{18}$, $v_1$, $v_9$, $v_{17}$, $v_0)$,
we get a $[[25,0,6]]$ code.
\end{example}

The $K_5[K_4]$ graph corresponding to a $[[20,0,8]]$ code, 
shown in \autoref{l36}, contains
one Hamiltonian cycle, in addition to two vertex-disjoint cycles, 
each visiting half the vertices of the graph. We generated the codes corresponding to all 
$K_5[K_4]$ graphs with two Hamiltonian cycles, and the highest distance 
found was 6. Other nested regular graphs listed in \autoref{tab:cycliccodes}
also contain Hamiltonian cycles or long cycles.
Both the nested regular structure and the presence of long cycles seem
to be important characteristics of the graphs corresponding to self-dual
quantum codes of high distance.

Although we have not found a construction technique for nested graph codes
giving a predictable distance, initial results suggest
that the set of nested regular graphs give a small
search space in which strong codes are likely to be found.
Randomly generated nested regular graphs with long disjoint cycles 
typically give codes of higher distance than totally random graphs.
It seems like the long cycles should be arranged
in such a way that no smaller cycles are induced in the graph.
The results shown in \autoref{tab:cycliccodes} suggest that
for codes of length above 25 and distance higher than 8, graph
structures get more complicated. To describe these structures,
a further generalisation of nested regular graphs may be necessary.

\chapter{Orbits of Self-Dual Quantum Codes}\label{chap:orbits}

\section{Local Transformations and Local Complementations}\label{sec:lutlc}

We have seen that an $[[n,0,d]]$ quantum stabilizer code represents a single quantum
state, and that all such codes can be transformed into equivalent
graph codes. The quantum state corresponding to a graph code is
known as a graph state, and in \autoref{sec:qcbool} we will see how a state $\ket{G}$
corresponding to a graph $G$ can be found.
In \autoref{l3} we saw that local unitary transformations are reversible transformations 
that act independently on each qubit in a quantum state.
If there exists a local unitary transformation $U$, such that
$U\ket{G} = \ket{G'}$, the states $\ket{G}$ and $\ket{G'}$ will have the same entanglement 
properties. If $\ket{G}$ and $\ket{G'}$ are graph states, we say that their corresponding graphs, 
$G$ and $G'$, are \emph{LU-equivalent}. $G$ and $G'$ will then represent
equivalent quantum codes, with the same distance, weight distribution, and other properties.

Determining whether two graphs are LU-equivalent seems like a difficult task,
but a sufficient condition for equivalence was given by Hein et~al.~\cite{hein}.
Let the graphs $G=(V,E)$ and $G'=(V,E')$ on $n$ vertices correspond 
to the $n$-qubit graph states $\ket{G}$ and $\ket{G'}$.

\begin{definition}
We define the two $2 \times 2$ unitary matrices,
\[
\tau_x = \sqrt{-i \sigma_x} = \frac{1}{\sqrt{2}} \begin{pmatrix} -1 & i \\ i & -1 \end{pmatrix},\quad
\tau_z = \sqrt{i \sigma_z} = \begin{pmatrix} w & 0 \\ 0 & w^3 \end{pmatrix},
\]
where $w^4 = i^2 = -1$, and $\sigma_x$ and $\sigma_z$ are Pauli matrices.
\end{definition}

\begin{definition}
Given a graph $G=(V=\{0,1,\ldots, n-1\},E)$, corresponding to the graph state $\ket{G}$,
we define a local unitary transformation,
\begin{equation}
U_a = \bigotimes_{i \in N_a} \tau_x^{(i)} \bigotimes_{i \not\in N_a} \tau_z^{(i)},
\end{equation}
where $a \in V$ is any vertex, $N_a \subset V$ is the neighbourhood of $a$, and
$\tau_x^{(i)}$ means that the transform $\tau_x$ should be applied to the qubit 
corresponding to vertex $i$.
\end{definition}

Given a graph $G$, if there exists a finite sequence of vertices 
$(u_0, u_1, \ldots, u_{k-1})$, such that $U_{u_{k-1}} \cdots U_{u_1} U_{u_0}\ket{G} = \ket{G'}$, then
$G$ and $G'$ are LU-equivalent. It was discovered by Hein et~al.~\cite{hein}, and 
by Van~den~Nest et~al.~\cite{nest}, that the sequence of transformations taking $\ket{G}$ to $\ket{G'}$ can 
equivalently be expressed as a sequence of simple graph operations taking $G$ to $G'$.
Exactly the same graph operation, called \emph{vertex neighbourhood complementation} (VNC),
was described by Glynn~et~al.~\cite{B22,B23} as an operation that
maps equivalent self-dual additive codes over $\GF(4)$ to each other.
VNC is another name for \emph{local complementation} (LC), referred
to in the context of \emph{isotropic systems} by Bouchet~\cite{B5,B7}.

\begin{definition}
Given a graph $G=(V,E)$ and a vertex $v \in V$, let $N_v \subset V$ 
be the neighbourhood of $v$. The subgraph induced by $N_v$ is complemented 
to obtain the LC image $G^v$, i.e., $G^v(N_v) = \overline{G(N_v)}$.
It is easy to verify that $(G^v)^v = G$. 
\end{definition}

\begin{figure}
 \centering
 \subfloat[The Graph $G$]
 {\hspace{5pt}\includegraphics[width=85pt]{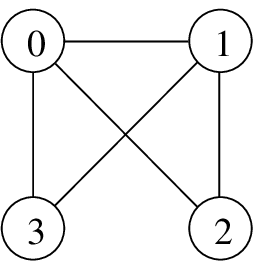}\hspace{5pt}\label{fig:lcexample1}}
 \quad
 \subfloat[The LC Image $G^0$]
 {\hspace{5pt}\includegraphics[width=85pt]{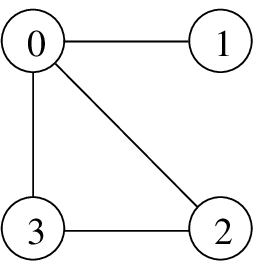}\hspace{5pt}\label{fig:lcexample2}}
 \caption{Example of Local Complementation}\label{fig:lcexample}
\end{figure}

\begin{example}
We will perform local complementation on vertex 0 of the
graph $G$, shown in \autoref{fig:lcexample1}.
We see that the neighbourhood of 0 is $N_0 = \{1,2,3\}$,
and that the induced subgraph on the neighbourhood,
$G(N_0)$, has edges $\{1,2\}$ and $\{1,3\}$. The complement of 
this subgraph, $\overline{G(N_0)}$, contains the single edge $\{2,3\}$.
The resulting LC image, $G^0$, is seen in \autoref{fig:lcexample2}.
\end{example}

\begin{theorem}[Glynn et~al.~\cite{B22,B23}, Hein et~al.~\cite{hein},
                and Van~den~Nest et~al.~\cite{nest}]
The graphs $G$ and $G'$ are \emph{LC-equivalent}, i.e., they correspond
to equivalent self-dual quantum codes, if there is a finite sequence of vertices
$u_0, u_1, \ldots, u_{k-1}$, such that $(((G^{u_0})^{u_1})^{\cdots})^{u_{k-1}} = G'$.
\end{theorem}

LC operations on the graph $G=(V,E)$, represented by the adjacency matrix $\Gamma$,
can also be described in terms of operations on $C = \Gamma + \omega I$,
the generator matrix of a self-dual additive code over $GF(4)$. 
We can then verify that none of the matrix operations will change the properties of the code.
LC on a vertex $a \in V$ corresponds to the following sequence of operations on $C$.
\begin{itemize}
\item For all vertices $i \in N_a$, add row $a$ to row $i$ in $C$.
This operation, which does not change the properties of an additive code, implements
the neighbourhood complementation of the corresponding graph, but also
leaves us with a code that is not a graph code.
The two following steps are needed to restore the code to graph form.
\item Scale column $a$ of $C$ by $\omega$, i.e., multiply coordinate $a$ in all 
rows of $C$ by $\omega$. Scaling the same coordinate in all codewords by some nonzero value 
gives an equivalent code.
\item For all vertices $i \in N_a \cup \{a\}$, conjugate column $i$ of $C$.
Conjugating coordinates does not change the properties of the code.
We now have a generator matrix of the form $C' = \Gamma' + \omega I$.
\end{itemize}

\section{Enumerating LC Orbits\label{sec:enumerate}}

Van~den~Nest et~al.~\cite{B58} report on an efficient algorithm,
first described by Bouchet~\cite{B6}, which determines whether two
graphs are LC-equivalent by solving a set of equations.
This algorithm has complexity $O(n^2)$, where $n$ is the number of vertices in the input graphs.
Note that we can also consider codes corresponding to isomorphic graphs to be equivalent, 
since permuting coordinates of a self-dual additive code over $\GF(4)$ gives an equivalent code.
We therefore want to detect LC-equivalence of graphs \emph{up to isomorphism}, which means that
a permutation of vertex labels is allowed before each LC operation.
Permuting the vertex labels of a graph causes the qubits in the corresponding graph state
to be reordered. Reordering the qubits in a quantum state does not change the
overall entanglement properties, but can not be performed by any local unitary transformation.
The above-mentioned algorithm only considers equivalence via
local unitary transformations, and can therefore not be used to detect
LC-equivalence up to isomorphism.

\begin{definition}
The \emph{LC orbit} $\boldsymbol{L} = [G]$, of a graph $G$, is the set of
all non-isomorphic graphs, including $G$ itself, that can be transformed
into $G$ by any sequence of local complementations and vertex permutations.
\end{definition}

\begin{example}
We consider the ``2-cliques of 3-cliques'' representation of the $[[6,0,4]]$ Hexacode
shown in \autoref{l20} on page~\pageref*{l20}.
An LC operation on any vertex of this graph will produce a graph isomorphic to the ``wheel graph''
shown in \autoref{l21}. An LC operation on the ``centre'' of the ``wheel'' 
will again produce a graph isomorphic to the ``wheel graph'', while an LC operation on any of
the 5 other vertices gives a graph isomorphic to the ``2-clique of 3-cliques''.
These two graphs therefore make up the complete LC orbit of the Hexacode.
\end{example}

Let $\mathcal{G}_n$ be the set of all non-isomorphic simple undirected connected 
graphs on $n$ vertices. (We will later consider a set where unconnected graphs are included.)
Let $\mathcal{L}_n = \{\boldsymbol{L}_1, \boldsymbol{L}_2, \ldots, \boldsymbol{L}_k\}$ be
the set of all distinct LC orbits of graphs in $\mathcal{G}_n$.
All $\boldsymbol{L} \in \mathcal{L}_n$ are disjoint, and $\mathcal{L}_n$ is a partitioning
of $\mathcal{G}_n$, i.e., $\bigcup_{i} \boldsymbol{L}_i = \mathcal{G}_n$.
Two graphs, $G$ and $K$, are equivalent with respect to local complementations and
vertex permutations if one of the graphs is in the LC orbit of the other, for instance, $K \in [G]$.
We will need \autoref{alg:vncorbit}, a recursive algorithm that generates the LC 
orbit of a given graph.
The package \texttt{nauty}, described in \autoref{l19}, is used to implement the procedure
\textsc{NautyCanonise}($G$), which returns a canonical representative 
of the graph $G$. Every isomorphic graph has the same canonical representative.
In our algorithms, we will also require data structures for storage of graphs.
Let $\boldsymbol{T}$ be such a data structure.
The exact implementation of $\boldsymbol{T}$ may vary, but we assume that there is a
procedure \textsc{Add}($\boldsymbol{T}$, $G$) that causes the graph $G$ to be added to $\boldsymbol{T}$.

\begin{algorithm}
\caption{Generating the LC Orbit of a Graph\label{alg:vncorbit}}
\begin{algorithmic}
\State \hskip -6pt
\begin{tabular}{ll}
\textbf{Input} 
& $G$: a graph, $G=(V,E)$\\
\textbf{Output} 
& $\boldsymbol{L}$: data structure containing all graphs in $[G]$
\end{tabular}
\Statex
\Procedure{GenerateOrbit}{$G$}
  \State initialise $\boldsymbol{L}$
  \State \textsc{RecursiveGenerateOrbit}($G$, $\boldsymbol{L}$)
  \State \textbf{return} $\boldsymbol{L}$
\EndProcedure
\Statex
\Procedure{RecursiveGenerateOrbit}{$G$, $\boldsymbol{L}$}
  \If{$G \not\in \boldsymbol{L}$}
    \ForAll{$v \in V$}
      \State $K \gets$ \textsc{NautyCanonise}($G^v$)
      \State \textsc{Add}($\boldsymbol{L}$, $K$)
      \State \textsc{RecursiveGenerateOrbit}($K$, $\boldsymbol{L}$)
    \EndFor
  \EndIf
\EndProcedure
\end{algorithmic} 
\end{algorithm}

\begin{figure}[t]
 \centering
 \vspace{25pt}
 \begin{tabular}{rp{220pt}}
 $\boldsymbol{L}_1$ & \vspace{-25pt}
 \includegraphics[width=45pt]{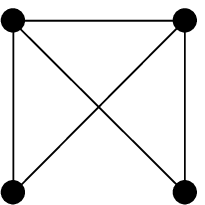}\hspace{10pt}
 \includegraphics[width=45pt]{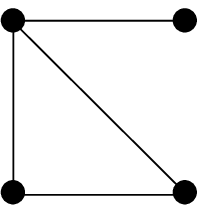}\hspace{10pt}
 \includegraphics[width=45pt]{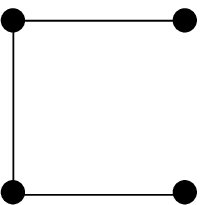}\hspace{10pt}
 {\includegraphics[width=45pt]{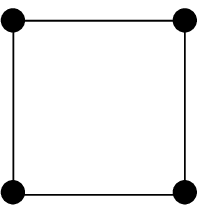} \vspace{15pt}}
 \\
 $\boldsymbol{L}_2$ & \vspace{-25pt}
 \includegraphics[width=45pt]{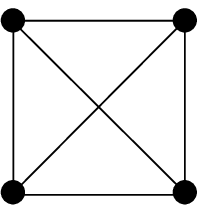}\hspace{10pt}
 \includegraphics[width=45pt]{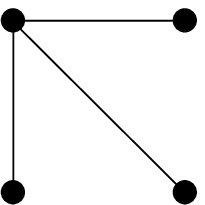}
 \end{tabular}
 \caption{The Two LC Orbits for $n=4$}\label{fig:4orbits}
\end{figure}

\begin{example}
As an example, we will generate $\mathcal{L}_4$, the set of all LC orbits on 4 vertices.
There are $2^{\binom{4}{2}} = 64$ undirected simple graphs on 4 vertices, but
the number of non-isomorphic connected graphs is only $|\mathcal{G}_4| = 6$. We use 
\autoref{alg:vncorbit} on these graphs and find that there
are $|\mathcal{L}_4| = 2$ distinct LC orbits on 4 vertices.
The orbits, $\mathcal{L}_4 = \{\boldsymbol{L}_1, \boldsymbol{L}_2\}$, are
shown in \autoref{fig:4orbits}.
\end{example}

We would like to partition $\mathcal{G}_n$
into a set of LC orbits, $\mathcal{L}_n$, for $n$ as high as possible.
In particular, we want to count the number of LC orbits, $|\mathcal{L}_n|$, 
which is also the number of inequivalent self-dual additive codes over $\GF(4)$ of length~$n$.
If we can also find one representative of each LC orbit, 
we can characterise the properties of all such codes.
Self-dual additive codes over $\GF(4)$ of length~$n$ have previously 
been enumerated by Calderbank~et~al.~\cite{B10} for $n \le 5$,
by Höhn~\cite{hohn} for $n \le 7$, by Hein~et~al.~\cite{hein} for $n \le 7$,
and by Glynn~et~al.~\cite{B23} for $n \le 9$. 
Glynn has also posted his results as sequence 
\href{http://www.research.att.com/projects/OEIS?Anum=A090899}{A090899} in 
\emph{The On-Line Encyclopedia of Integer Sequences}~\cite{B56}.
For higher $n$, only partial classifications of extremal codes have been performed~\cite{B4,B21,B20}.

\autoref{alg:lccanonise}, our first attempt at an algorithm for generating all LC orbits, is
inspired by the concept of canonical representatives, as used by \texttt{nauty}.
We define a procedure \textsc{LCcanonise}, which returns
the same canonical representative for every member of the same LC orbit.
The procedure \textsc{FindOrbits1} with $\mathcal{G}_n$ as input
will canonise every graph in $\mathcal{G}_n$ and remove all duplicates
in the resulting set. We would then have one representative of every LC orbit in $\mathcal{L}_n$.
Generating the set $\mathcal{G}_n$ can be done by the utility \texttt{geng} from the \texttt{nauty} package.
It is not important how the canonical representative of an
LC orbit is chosen, as long as it is done consistently. \autoref{alg:lccanonise} gives an implementation of 
\textsc{LCcanonise}($G$) that first generates the complete orbit $\boldsymbol{L} = [G]$.
The procedure \textsc{First}($\boldsymbol{L}$) then picks the ``first'' graph in 
the set $\boldsymbol{L}$ by some lexicographical ordering. The exact
implementation of this ordering is not important.

\begin{algorithm}[p]
\caption{Finding LC Orbits By Canonisation\label{alg:lccanonise}}
\begin{algorithmic}
\State \hskip -6pt
\begin{tabular}{ll}
\textbf{Input} 
& $\boldsymbol{F}$: a set of graphs\\
\textbf{Output} 
& $\boldsymbol{O}$: a set with one representative of each LC orbit present in $\boldsymbol{F}$
\end{tabular}
\Statex
\Procedure{FindOrbits1}{$\boldsymbol{F}$}
  \State initialise $\boldsymbol{O}$
  \ForAll{$G \in \boldsymbol{F}$}
    \State $K \gets$ \textsc{LCcanonise}($G$)
    \If{$K \not\in$ $\boldsymbol{O}$}
      \State \textsc{add}($\boldsymbol{O}$, $K$)
    \EndIf
  \EndFor
  \State \textbf{return} $\boldsymbol{O}$
\EndProcedure
\Statex
\Procedure{LCcanonise}{$G$}
  \State $\boldsymbol{L} \gets$ \textsc{GenerateOrbit}($G$)
  \State $K \gets$ \textsc{First}($\boldsymbol{L}$)
  \State \textbf{return} $K$
\EndProcedure
\end{algorithmic} 
\end{algorithm}

Finding all LC orbits in $\mathcal{L}_n$ by \autoref{alg:lccanonise}
is clearly not efficient. We are doing much redundant work 
by going through the whole LC orbit of every graph in $\mathcal{G}_n$, since
many orbits will then be generated several times.
A great improvement in running time is achieved by storing
all LC orbits in memory at the same time.
\autoref{alg:lcfast} guarantees that all distinct LC orbits
will only be generated once, at the cost of extra memory requirement.
The procedure will store all members of every LC orbit it generates in
the temporary set $\boldsymbol{T}$. If a graph is already in $\boldsymbol{T}$, its orbit
is not generated again. If we call the procedure \textsc{FindOrbits2} with $\mathcal{G}_n$ 
as input, $\boldsymbol{T}$ will contain all graphs in 
$\mathcal{G}_n$ at the time the procedure terminates.

\begin{algorithm}[p]
\caption{Finding LC Orbits Quickly\label{alg:lcfast}}
\begin{algorithmic}
\State \hskip -6pt
\begin{tabular}{ll}
\textbf{Input} 
& $\boldsymbol{F}$: a set of graphs\\
\textbf{Output} 
& $\boldsymbol{O}$: a set with one representative of each LC orbit present in $\boldsymbol{F}$
\end{tabular}
\Statex
\Procedure{FindOrbits2}{$\boldsymbol{F}$}
  \State initialise $\boldsymbol{O}$ and $\boldsymbol{T}$
  \ForAll{$G \in \boldsymbol{F}$}
    \State $K \gets$ \textsc{NautyCanonise}($G$)
    \If{$K \not\in$ $\boldsymbol{T}$}
      \State \textsc{add}($\boldsymbol{O}$, $K$)
      \State $\boldsymbol{L} \gets$ \textsc{GenerateOrbit}($K$)
      \ForAll{$I \in \boldsymbol{L}$}
        \State \textsc{add}($\boldsymbol{T}$, $I$)
      \EndFor
    \EndIf
  \EndFor
  \State \textbf{return} $\boldsymbol{O}$
\EndProcedure
\end{algorithmic} 
\end{algorithm}

A straightforward implementation of \autoref{alg:lcfast} will work when $n \le 8$.
The total number of undirected graphs on 8 vertices is $2^{\binom{8}{2}} = $ 268,435,456.
Since we have more than 300 MB of memory available, we can simply make $\mathbf{T}$ a binary array
with one bit representing each graph. All bits are initialised to zero, and will 
be set to one once the corresponding graph has been discovered.
This is clearly a waste of memory, since all the graphs we need to store are the
$|\mathcal{G}_8| =$ 11,117 non-isomorphic connected graphs on 8 vertices.
When $n=9$, such an array would require about 10 GB of memory, so a more clever approach is needed.
We therefore use a \emph{binary search tree} as $\mathbf{T}$. Every time we discover a graph
not isomorphic to any graph in $\mathbf{T}$, we add a new node to the tree. 
Since our graphs are undirected, only the lower or upper triangle of the adjacency matrix needs 
to be stored. This means that $\binom{n}{2}$ bits of memory are needed to store each graph. These bits are 
easily interpreted as a numerical value for comparisons in the binary search tree.
With this method, only non-isomorphic graphs will be stored in $\mathbf{T}$, so the memory requirement is
proportional to $|\mathcal{G}_n|$.
Values of $|\mathcal{G}_n|$ are listed in \autoref{tab:setsizes}, and is also sequence 
\href{http://www.research.att.com/projects/OEIS?Anum=A001349}{A001349} in
\emph{The On-Line Encyclopedia of Integer Sequences}~\cite{B56}.

\begin{table}
\centering
\caption{Sizes of Different Sets of Graphs}
\label{tab:setsizes}
\begin{tabular}{crrr}
\toprule
$n$ & $|\mathcal{G}_n|$ & $|\mathcal{E}'_n|$ & $|\mathcal{E}_n|$ \\
\midrule
 1 &                  1 &             - &          - \\ 
 2 &                  1 &             1 &          1 \\ 
 3 &                  2 &             3 &          2 \\ 
 4 &                  6 &             7 &          5 \\
 5 &                 21 &            30 &         14 \\
 6 &                112 &           124 &         48 \\
 7 &                853 &           693 &        228 \\
 8 &             11,117 &         3,302 &      1,338 \\
 9 &            261,080 &        25,755 &     11,309 \\
10 &         11,716,571 &       224,840 &    123,899 \\
11 &      1,006,700,565 &     3,204,036 &  2,138,482 \\
12 &    164,059,830,476 &    82,815,479 & 66,150,188 \\
13 & 50,335,907,869,219 & 5,217,308,460 &          ? \\
\bottomrule
\end{tabular}
\end{table}

While the orbits for $n=9$ are easily computed with the binary search tree implementation of 
\autoref{alg:lcfast}, $n=10$ takes about one hour of running time and
uses more than 100 MB of memory. The memory requirement is the biggest obstacle, and for $n=11$ it
becomes infeasible. To solve this problem, we tried to find an \emph{invariant}, i.e., some property 
that has the same value for all graphs in the same LC orbit, and that can be calculated quickly.
One such property is the weight distribution of the codes corresponding to the graphs, since we can
with certainty say that two codes with different weight distributions are not in the 
same LC orbit. (The converse is however not true, since many LC orbits will have exactly the same
weight distribution.) To reduce the memory requirement of our algorithm, we calculate the 
weight distribution of the codes corresponding to graphs in $\mathcal{G}_n$ and 
store the graphs in $k$ different sets, $\mathbf{F}_1, \mathbf{F}_2, \ldots, \mathbf{F}_k$, 
with the only restriction that codes with the same weight distribution must be in the same set.
We do not necessarily have to compute the complete weight distributions, since the
partial weight distribution $\boldsymbol{w}_p$, i.e., the numbers of codewords of weights up to $p$,
is also an invariant over the LC orbit. $\boldsymbol{w}_p$ can be calculated efficiently
by the method described in \autoref{l26} on page~\pageref*{l26}.
We choose a $p$ that is high enough to give a good separation of the codes, while still 
being computable for all graphs in $\mathcal{G}_n$ in reasonable time. The distribution 
of codes will not be uniform, so the resulting sets will be of various sizes.
When the sets $\mathbf{F}_1, \mathbf{F}_2, \ldots, \mathbf{F}_k$ are generated,
we call the procedure \textsc{FindOrbits2} $k$ times, once with each set as input.
The sets are processed independently, and the data structure $\mathbf{T}$
in procedure \textsc{FindOrbits2} can be reset for each set, reducing the amount of memory needed.
This method also allows us to process the $k$ sets in parallel.
The outputs returned by the $k$ procedure calls are concatenated to form a complete
set of representatives of $\mathcal{L}_n$.
This approach was used to classify all inequivalent $[[11,0,d]]$ codes. Generating all non-isomorphic
graphs and splitting them into 1,000 files according to partial weight distribution 
took about 3 days on an ordinary desktop computer. Many of the resulting files were empty or nearly
empty, whereas the largest were several 100 MB. The total size of all the files after compression
was about 6 GB. The processing of the files was done in parallel on a cluster computer in a 
matter of hours, and 40,457 inequivalent codes were found. Processing the largest file required 
more than 1 GB of memory, so using this method for $n=12$ was not feasible with the available resources.

We have seen that the procedure \textsc{FindOrbits1} is too slow, and that the procedure 
\textsc{FindOrbits2} requires too much memory. In \autoref{alg:lcfinal} we define 
\textsc{FindOrbits3} which, like \textsc{FindOrbits1}, only stores a single LC orbit in memory at any time.
It does not, however, generate the LC orbit of every graph in the input set,
and is therefore faster than \textsc{FindOrbits1}. \textsc{FindOrbits3} also benefits from splitting up
the input set using partial weight distribution. \autoref{alg:lcfinal} uses the procedure 
\textsc{Remove}($\boldsymbol{T}$, $G$) which removes the graph $G$ from $\boldsymbol{T}$, 
and the procedure \textsc{RemoveNext}($\boldsymbol{T}$), which removes some graph from $\boldsymbol{T}$ 
and returns it. The order in which graphs are removed from $\boldsymbol{T}$ is not important here.

\begin{algorithm}
\caption{Finding LC Orbits Quickly Using Less Memory\label{alg:lcfinal}}
\begin{algorithmic}
\State \hskip -6pt
\begin{tabular}{ll}
\textbf{Input} 
& $\boldsymbol{F}$: a set of graphs\\
\textbf{Output} 
& $\boldsymbol{O}$: a set with one representative of each LC orbit present in $\boldsymbol{F}$
\end{tabular}
\Statex
\Procedure{FindOrbits3}{$\boldsymbol{F}$}
  \State initialise $\boldsymbol{O}$
  \While{$\boldsymbol{F}$ is not empty}
    \State $G \gets$ \textsc{RemoveNext}($\boldsymbol{F}$)
    \State $K \gets$ \textsc{NautyCanonise}($G$)
    \State \textsc{add}($\boldsymbol{O}$, $K$)
    \State $\boldsymbol{L} \gets$ \textsc{GenerateOrbit}($K$)
    \ForAll{$I \in \boldsymbol{F}$}
      \If{$I \in \boldsymbol{L}$}
        \State \textsc{remove}($\boldsymbol{F}$, $I$)
      \EndIf
    \EndFor
  \EndWhile
  \State \textbf{return} $\boldsymbol{O}$
\EndProcedure
\end{algorithmic} 
\end{algorithm}

\begin{definition}\label{def:extension}
The $2^n-1$ \emph{extensions} of a graph on $n$ vertices is formed by
adding a new vertex and joining it to all possible combinations of 
at least one of the old vertices.
The set $\mathcal{E}'_{n}$, containing $|\mathcal{L}_{n-1}| \times \left( 2^{n-1}-1 \right)$ graphs,
is formed by making all possible extensions of one representative from each LC orbit in $\mathcal{L}_{n-1}$.
Let $\mathcal{E}_{n}$ contain the same graphs as $\mathcal{E}'_{n}$, except that all isomorphisms are removed.
\end{definition}

\begin{proposition}[Glynn et~al.~\cite{B23}]\label{prop:extendall}
The set $\mathcal{E}'_{n}$ will contain at least one representative from each
LC orbit in $\mathcal{L}_{n}$.
\end{proposition}
\begin{proof}
Let $G=(V,E)$ be any graph on $n$ vertices. Choose any subset $W \subset V$ of $n-1$ vertices.
By doing LC operations on vertices in $W$, we can transform the subgraph $G(W)$ into any member
of the LC orbit $[G(W)]$. One of these members was extended when the set $\mathcal{E}'_{n}$
was constructed. It follows that for all $G \in \mathcal{G}_{n}$, some $G' \in [G]$ must be 
part of $\mathcal{E}'_{n}$.
\end{proof}

\autoref{tab:setsizes} gives the values of $|\mathcal{E}'_{n}|$ and $|\mathcal{E}_{n}|$, 
which are much smaller than the values of $|\mathcal{G}_{n}|$. It will therefore 
be more efficient to use $\mathcal{E}_{n}$ instead of $\mathcal{G}_{n}$ as input to our algorithms,
and, by \autoref{prop:extendall}, we will still find one representative of every LC orbit in $\mathcal{L}_{n}$.
We managed to find all inequivalent self-dual quantum codes of length $n=12$ by
using the set $\mathcal{E}_{12}$ as input to \textsc{FindOrbits3}.
The set of 66,150,188 graphs in $\mathcal{E}_{12}$ was first divided into 1,000 files by the 
partial weight distribution $\boldsymbol{w}_7$, a process that took a few hours.
The files were then processed in parallel on a cluster computer.
A \emph{hash table} was used as the data structure $\mathbf{L}$ in the algorithm,
to allow for fast look-up. The processing took a little more than a week to finish, using a total
of more than 4,000 CPU-hours, and 1,274,068 LC orbits were found.

\autoref{tab:unconnected} shows the value of $|\mathcal{L}_n|$, which is also the
number of inequivalent self-dual additive codes over $\GF(4)$, for $n$ up to 12.
A \href{http://www.ii.uib.no/~larsed/vncorbits/}{database} of orbit representatives with information about
orbit size, distance, and weight distribution is also available~\cite{B14}.
The numbers of inequivalent codes have been added to sequence 
\href{http://www.research.att.com/projects/OEIS?Anum=A090899}{A090899} in 
\emph{The On-Line Encyclopedia of Integer Sequences}~\cite{B56},
which previously only had values for $n$ up to 9.
The numbers of codes of various distances are shown in \autoref{tab:distances}.
\autoref{tab:type2} gives the number of self-dual quantum codes of type II, 
i.e., codes where all codewords have even weight.

\begin{table}
\centering
\caption{Number of Self-Dual Quantum Codes of Length~$n$}
\label{tab:unconnected}
\begin{tabular}{crr}
\toprule
$n$ & $|\mathcal{L}_n|$ & $|\mathcal{L}'_n|$ \\
\midrule
 1 &         1 &         1 \\
 2 &         1 &         2 \\
 3 &         1 &         3 \\
 4 &         2 &         6 \\
 5 &         4 &        11 \\
 6 &        11 &        26 \\
 7 &        26 &        59 \\
 8 &       101 &       182 \\
 9 &       440 &       675 \\ 
10 &     3,132 &     3,990 \\
11 &    40,457 &    45,144 \\
12 & 1,274,068 & 1,323,363 \\
\bottomrule
\end{tabular}
\end{table}

\begin{table}
\centering
\caption{Number of Indecomposable Self-Dual Quantum Codes of Length~$n$ and Distance~$d$}
\label{tab:distances}
\begin{tabular}{crrrrrrrrrrr}
\toprule
 & \multicolumn{11}{c}{$n$} \\
\cmidrule(l){2-12}
$d$ & 2 & 3 & 4 & 5 & 6 & 7 & 8 & 9 & 10 & 11 & 12\\
\midrule
2     & 1 & 1 & 2 & 3 &  9 & 22 &  85 & 363 & 2,436 & 26,750 &   611,036 \\
3     &   &   &   & 1 &  1 &  4 &  11 &  69 &   576 & 11,200 &   467,513 \\
4     &   &   &   &   &  1 &    &   5 &   8 &   120 &  2,506 &   195,455 \\
5     &   &   &   &   &    &    &     &     &       &      1 &        63 \\
6     &   &   &   &   &    &    &     &     &       &        &         1 \\
\midrule
All   & 1 & 1 & 2 & 4 & 11 & 26 & 101 & 440 & 3,132 & 40,457 & 1,274,068 \\
\bottomrule
\end{tabular}
\end{table}

\begin{table}
\centering
\caption{Number of Indecomposable Type II Self-Dual Quantum Codes of Length~$n$ and Distance~$d$}
\label{tab:type2}
\begin{tabular}{crrrrrr}
\toprule
 & \multicolumn{6}{c}{$n$} \\
\cmidrule(l){2-7}
$d$ & 2 & 4 & 6 & 8 & 10 & 12\\
\midrule
2     & 1 & 1 & 3 & 11 &  84 & 2,133 \\
4     &   &   & 1 &  3 &  19 &   792 \\
6     &   &   &   &    &     &     1 \\
\midrule
All   & 1 & 1 & 4 & 14 & 103 & 2,926 \\
\bottomrule
\end{tabular}
\end{table}

\begin{definition}
Let ``$a$'', where $a \in \mathbb{N}$, denote a connected graph on $a$ vertices.
Let ``$a_1^{b_1}a_2^{b_2} \cdots a_k^{b_k}$'', where $a_i, b_i \in \mathbb{N}$, denote
an unconnected graph composed of $b_0$ connected components with $a_0$ vertices,
$b_1$ connected components with $a_1$ vertices, and so on.
\end{definition}

Recall that $\mathcal{L}_n$ only contains LC orbits of connected graphs. Codes
corresponding to connected graphs are called \emph{indecomposable}.
Let the set $\mathcal{L}'_n \subset \mathcal{L}_n$ also include all LC orbits
of unconnected graphs. An unconnected graph on $n$ vertices is composed of 
a set of connected components with less than $n$ vertices. 
Likewise, codes corresponding to unconnected graphs, called \emph{decomposable} codes,
can always be expressed as a combination of indecomposable codes of shorter length.
\autoref{tab:decompose}, which is an extended version of a table
given by Glynn et al.~\cite{B23},
counts the number of non-isomorphic graphs for all possible such combinations.
The values of $|\mathcal{L}'_n|$ for $n$ up to 12 are also shown in 
\autoref{tab:unconnected}, and is a new sequence, 
\href{http://www.research.att.com/projects/OEIS?Anum=A094927}{A094927}, in
\emph{The On-Line Encyclopedia of Integer Sequences}~\cite{B56}.
Note that the number of non-isomorphic graphs of type ``$4^2$'' 
is only 3, and not 4. Likewise, the number of ``$5^2$'' graphs is 10,
the number of ``$4^3$'' graphs is 4, and the number of ``$6^2$'' graphs is 66.
For all other combinations used in \autoref{tab:decompose}, the number
of non-isomorphic unconnected graphs is simply found by multiplying
the numbers of non-isomorphic connected graphs for each connected component,
e.g., the number of ``$641^2$'' graphs is $11 \times 2 \times 1 \times 1 = 22$.

\begin{example}
We will count the number of LC orbits in $\mathcal{L}'_4$.
The unconnected graphs of type ``$2^2$'', i.e., graphs
formed by combining 2 connected graphs on 2 vertices,
must be included. There is only one orbit in $\mathcal{L}_2$, and therefore only one 
inequivalent decomposable code of type ``$2^2$''.
Another way to construct unconnected graphs on 4 vertices is to
combine a connected graph on 3 vertices with an isolated vertex.
This combination is denoted ``$31$''. Since $\mathcal{L}_3 = \mathcal{L}_1 =1$,
there is only one LC orbit of type ``$31$''.
The other combinations that make unconnected graphs on 4 vertices,
``$21^2$'' and ``$1^4$'', also give one orbit each.
$\mathcal{L}'_4$ also contains the 2 orbits of connected graphs in $\mathcal{L}_4$.
(This combination is simply denoted ``4''.)
Thus, when we add all the combinations, $|\mathcal{L}'_4| = 6$.
\end{example}

\begin{table}
\centering\scriptsize
\caption{Numbers of Decomposable Self-Dual Quantum Codes}
\label{tab:decompose}
\setlength{\tabcolsep}{1.4pt}
\resizebox{\linewidth}{!}{
\begin{tabular}{lr|lr|lr|lr|lr|lr|lr|lr|lr|lr|lr|lr}
\toprule
\multicolumn{2}{c}{1} & \multicolumn{2}{c}{2} & \multicolumn{2}{c}{3} &
\multicolumn{2}{c}{4} & \multicolumn{2}{c}{5} & \multicolumn{2}{c}{6} 
&\multicolumn{2}{c}{7} & \multicolumn{2}{c}{8} & \multicolumn{2}{c}{9} &
\multicolumn{2}{c}{10} & \multicolumn{2}{c}{11} & \multicolumn{2}{c}{12} 
\vspace{1pt}\\
\hline
\rule{0pt}{7pt}
$1$&1 & $1^2$&1 & $1^3$&1 & $1^4$&1 & $1^5$&1 & $1^6$&1 & $1^7$&1 & $1^8$&1 & $1^9$&1 & $1^{10}$&1 & $1^{11}$&1 & $1^{12}$&1 \\
& & $2$&1 & $21$&1 & $21^2$&1 & $21^3$&1 & $21^4$&1 & $21^5$&1 & $21^6$&1 & $21^7$&1 & $21^8$&1 & $21^9$&1 & $21^{10}$&1 \\
& & & & $3$&1 & $31$&1 & $31^2$&1 & $31^3$&1 & $31^4$&1 & $31^5$&1 & $31^6$&1 & $31^7$&1 & $31^8$&1 & $31^9$&1 \\
& & & & & & $2^2$&1 & $2^21$&1 & $2^21^2$&1 & $2^21^3$&1 & $2^21^4$&1 & $2^21^5$&1 & $2^21^6$&1 & $2^21^7$&1 & $2^21^8$&1 \\
& & & & & & $4$&2 & $41$&2 & $41^2$&2 & $41^3$&2 & $41^4$&2 & $41^5$&2 & $41^6$&2 & $41^7$&2 & $41^8$&2 \\
& & & & & & & & $32$&1 & $321$&1 & $321^2$&1 & $321^3$&1 & $321^4$&1 & $321^5$&1 & $321^6$&1 & $321^7$&1 \\
& & & & & & & & $5$&4 & $51$&4 & $51^2$&4 & $51^3$&4 & $51^4$&4 & $51^5$&4 & $51^6$&4 & $51^7$&4 \\
& & & & & & & & & & $2^3$&1 & $2^31$&1 & $2^31^2$&1 & $2^31^3$&1 & $2^31^4$&1 & $2^31^5$&1 & $2^31^6$&1 \\
& & & & & & & & & & $42$&2 & $421$&2 & $421^2$&2 & $421^3$&2 & $421^4$&2 & $421^5$&2 & $421^6$&2 \\
& & & & & & & & & & $3^2$&1 & $3^21$&1 & $3^21^2$&1 & $3^21^3$&1 & $3^21^4$&1 & $3^21^5$&1 & $3^21^6$&1 \\
& & & & & & & & & & $6$&11 & $61$&11 & $61^2$&11 & $61^3$&11 & $61^4$&11 & $61^5$&11 & $61^6$&11 \\
& & & & & & & & & & & & $32^2$&1 & $32^21$&1 & $32^21^2$&1 & $32^21^3$&1 & $32^21^4$&1 & $32^21^5$&1 \\
& & & & & & & & & & & & $52$&4 & $521$&4 & $521^2$&4 & $521^3$&4 & $521^4$&4 & $521^5$&4 \\
& & & & & & & & & & & & $43$&2 & $431$&2 & $431^2$&2 & $431^3$&2 & $431^4$&2 & $431^5$&2 \\
& & & & & & & & & & & & $7$&26 & $71$&26 & $71^2$&26 & $71^3$&26 & $71^4$&26 & $71^5$&26 \\
& & & & & & & & & & & & & & $2^4$&1 & $2^41$&1 & $2^41^2$&1 & $2^41^3$&1 & $2^41^4$&1 \\
& & & & & & & & & & & & & & $42^2$&2 & $42^21$&2 & $42^21^2$&2 & $42^21^3$&2 & $42^21^4$&2 \\
& & & & & & & & & & & & & & $3^22$&1 & $3^221$&1 & $3^221^2$&1 & $3^221^3$&1 & $3^221^4$&1 \\
& & & & & & & & & & & & & & $62$&11 & $621$&11 & $621^2$&11 & $621^3$&11 & $621^4$&11 \\
& & & & & & & & & & & & & & $53$&4 & $531$&4 & $531^2$&4 & $531^3$&4 & $531^4$&4 \\
& & & & & & & & & & & & & & $4^2$&3 & $4^21$&3 & $4^21^2$&3 & $4^21^3$&3 & $4^21^4$&3 \\
& & & & & & & & & & & & & & $8$&101 & $81$&101 & $81^2$&101 & $81^3$&101 & $81^4$&101 \\
& & & & & & & & & & & & & & & & $32^3$&1 & $32^31$&1 & $32^31^2$&1 & $32^31^3$&1 \\
& & & & & & & & & & & & & & & & $52^2$&4 & $52^21$&4 & $52^21^2$&4 & $52^21^3$&4 \\
& & & & & & & & & & & & & & & & $432$&2 & $4321$&2 & $4321^2$&2 & $4321^3$&2 \\
& & & & & & & & & & & & & & & & $72$&26 & $721$&26 & $721^2$&26 & $721^3$&26 \\
& & & & & & & & & & & & & & & & $3^3$&1 & $3^31$&1 & $3^31^2$&1 & $3^31^3$&1 \\
& & & & & & & & & & & & & & & & $63$&11 & $631$&11 & $631^2$&11 & $631^3$&11 \\
& & & & & & & & & & & & & & & & $54$&8 & $541$&8 & $541^2$&8 & $541^3$&8 \\
& & & & & & & & & & & & & & & & $9$&440 & $91$&440 & $91^2$&440 & $91^3$&440 \\
& & & & & & & & & & & & & & & & & & $2^5$&1 & $2^51$&1 & $2^51^2$&1 \\
& & & & & & & & & & & & & & & & & & $42^3$&2 & $42^31$&2 & $42^31^2$&2 \\
& & & & & & & & & & & & & & & & & & $3^22^2$&1 & $3^22^21$&1 & $3^22^21^2$&1 \\
& & & & & & & & & & & & & & & & & & $62^2$&11 & $62^21$&11 & $62^21^2$&11 \\
& & & & & & & & & & & & & & & & & & $532$&4 & $5321$&4 & $5321^2$&4 \\
& & & & & & & & & & & & & & & & & & $4^22$&3 & $4^221$&3 & $4^221^2$&3 \\
& & & & & & & & & & & & & & & & & & $82$&101 & $821$&101 & $821^2$&101 \\
& & & & & & & & & & & & & & & & & & $43^2$&2 & $43^21$&2 & $43^21^2$&2 \\
& & & & & & & & & & & & & & & & & & $73$&26 & $731$&26 & $731^2$&26 \\
& & & & & & & & & & & & & & & & & & $64$&22 & $641$&22 & $641^2$&22 \\
& & & & & & & & & & & & & & & & & & $5^2$&10 & $5^21$&10 & $5^21^2$&10 \\
& & & & & & & & & & & & & & & & & & $(10)$&3,132 & $(10)1$&3,132 & $(10)1^2$&3,132 \\
& & & & & & & & & & & & & & & & & & & & $32^4$&1 & $32^41$&1 \\
& & & & & & & & & & & & & & & & & & & & $52^3$&4 & $52^31$&4 \\
& & & & & & & & & & & & & & & & & & & & $432^2$&2 & $432^21$&2 \\
& & & & & & & & & & & & & & & & & & & & $72^2$&26 & $72^21$&26 \\
& & & & & & & & & & & & & & & & & & & & $3^32$&1 & $3^321$&1 \\
& & & & & & & & & & & & & & & & & & & & $632$&11 & $6321$&11 \\
& & & & & & & & & & & & & & & & & & & & $542$&8 & $5421$&8 \\
& & & & & & & & & & & & & & & & & & & & $92$&440 & $921$&440 \\
& & & & & & & & & & & & & & & & & & & & $53^2$&4 & $53^21$&4 \\
& & & & & & & & & & & & & & & & & & & & $4^23$&3 & $4^231$&3 \\
& & & & & & & & & & & & & & & & & & & & $83$&101 & $831$&101 \\
& & & & & & & & & & & & & & & & & & & & $74$&52 & $741$&52 \\
& & & & & & & & & & & & & & & & & & & & $65$&44 & $651$&44 \\
& & & & & & & & & & & & & & & & & & & & $(11)$&40,457 & $(11)1$&40,457 \\
& & & & & & & & & & & & & & & & & & & & & & $2^6$&1 \\
& & & & & & & & & & & & & & & & & & & & & & $42^4$&2 \\
& & & & & & & & & & & & & & & & & & & & & & $3^22^3$&1 \\
& & & & & & & & & & & & & & & & & & & & & & $62^3$&11 \\
& & & & & & & & & & & & & & & & & & & & & & $532^2$&4 \\
& & & & & & & & & & & & & & & & & & & & & & $4^22^2$&3 \\
& & & & & & & & & & & & & & & & & & & & & & $82^2$&101 \\
& & & & & & & & & & & & & & & & & & & & & & $43^22$&2 \\
& & & & & & & & & & & & & & & & & & & & & & $732$&26 \\
& & & & & & & & & & & & & & & & & & & & & & $642$&22 \\
& & & & & & & & & & & & & & & & & & & & & & $5^22$&10 \\
& & & & & & & & & & & & & & & & & & & & & & $(10)2$&3,132 \\
& & & & & & & & & & & & & & & & & & & & & & $3^4$&1 \\
& & & & & & & & & & & & & & & & & & & & & & $63^2$&11 \\
& & & & & & & & & & & & & & & & & & & & & & $543$&8 \\
& & & & & & & & & & & & & & & & & & & & & & $93$&440 \\
& & & & & & & & & & & & & & & & & & & & & & $4^3$&4 \\
& & & & & & & & & & & & & & & & & & & & & & $84$&202 \\
& & & & & & & & & & & & & & & & & & & & & & $75$&104 \\
& & & & & & & & & & & & & & & & & & & & & & $6^2$&66 \\
& & & & & & & & & & & & & & & & & & & & & & $(12)$&1,274,068 \\
\hline
\addlinespace[\belowrulesep]
\multicolumn{2}{r}{1} & \multicolumn{2}{r}{2} & \multicolumn{2}{r}{3} &
\multicolumn{2}{r}{6} & \multicolumn{2}{r}{11} & \multicolumn{2}{r}{26} 
&\multicolumn{2}{r}{59} & \multicolumn{2}{r}{182} & \multicolumn{2}{r}{675} &
\multicolumn{2}{r}{3,990} & \multicolumn{2}{r}{45,144} & \multicolumn{2}{r}{1,323,363}\\
\bottomrule
\end{tabular}
}
\end{table}

\section{The LC Orbits of Some Strong Codes}

In \autoref{l32}, we studied graph codes with circulant generator matrices.
For some lengths we found that the best circulant graph codes had lower distance than the 
best known self-dual quantum codes.
In particular, we did not find any  $[[11,0,5]]$, $[[18,0,8]]$, $[[21,0,8]]$ or $[[27,0,9]]$ codes.
Generator matrices for the best known quantum codes are listed by Grassl~\cite{B26}.
We have transformed some of these codes into graph codes and generated their LC orbits.

\begin{remark}
There are no regular graphs corresponding to $[[11,0,5]]$ or $[[18,0,8]]$ quantum codes.
\end{remark}

The unique $[[11,0,5]]$ code is already known from our complete classification 
of all self-dual quantum codes of length up to 12. This code has 4,742 non-isomorphic
graphs in its LC orbit, and none of these graphs are regular.
The $[[18,0,8]]$ code has been shown (under some constraints) to be unique~\cite{B4}, and it
can be generated as shown in \autoref{l27}. There is no regular graph among
the 3,828 non-isomorphic graphs in the LC orbit of this code, but
the graph representation with the fewest edges contains only
one vertex of degree 9, with the rest of the vertices having degree 7.
This is the closest to minimum regular vertex degree we can get without achieving it, 
since the code is of type II, and all its graph representations must therefore 
have only odd vertex degrees.
We also generate the LC orbit of the $[[21,0,8]]$ code listed by Grassl~\cite{B26}.
It has 77,394 members in its LC orbit, and again no regular graph is found.
Note that neither did Gulliver and Kim~\cite{B29} find any $[[21,0,8]]$ code
with a circulant based generator matrix.
We also tried to generate the LC orbit of a $[[27,0,9]]$ code, but
after finding about 10 million non-isomorphic graphs, the memory resources of the computer
were exhausted. None of the graphs found were regular.
The $[[30,0,12]]$ code we discovered in our search of circulant graph codes
had a regular vertex degree of 17. We could only generate about 10 million
members of its LC orbit, but this sufficed to find a graph with 
regular vertex degree 15. This graph can not be described as a nested
regular graph, however, and the degree is still far from the minimal 11.
The graph representation with the fewest edges found had 171 edges,
an ``average vertex degree'' of $11.4$.

In \autoref{l27} we showed how the $[[30,0,12]]$ code can be constructed as 
a bordered quadratic residue code. The graph corresponding to the
$[[29,0,11]]$ quadratic residue code will have a regular vertex degree of 14.
Bordering this code adds a vertex of degree 29 and increases the degree of
all other vertices to 15. It turns out that, by using
only a single LC operation, this graph can be transformed 
into one with regular vertex degree 15. Furthermore, this LC operation may be performed
on any vertex, except the vertex of degree 29.
The ``wheel graph'' representation of the $[[6,0,4]]$ Hexacode, as shown
in \autoref{l21} on page~\pageref*{l21}, is also a bordered quadratic residue code, and
is also turned into a regular graph by a single LC operation
on any vertex except the ``centre'' of the ``wheel''.

\begin{theorem}\label{thm:bqrlc}
Let $G=(V=\{v_0, v_1, \ldots, v_{m-1}\},E)$, be a Paley graph on $m = 4t+1$ vertices.
Form the graph $K=(V \cup \{v_m\}, E \cup \{ \{u, v_m\} \mid u \in V\})$, i.e., add a
vertex $v_m$ and connect it to all existing vertices. $K$ is a graph where the vertices
$u \in V$ have degree $2t+1$ and vertex $v_m$ has degree $4t+1$.
$K^u$, where local complementation is performed on any vertex $u \in V$, will
be a $2t+1$-regular graph.
\end{theorem}
\begin{proof}
The neighbourhood of $u$ is $N_u = v_m \cup N'_u$, where $N'_u$ consists of
$2t$ vertices. Since the vertex $v_m$ has degree $4t+1$ in $K$ and degree $2t$ 
in $K(N_u)$, its degree in $K^u$ will be $2t+1$.
Since $G$ is a Paley graph, and therefore a strongly regular graph with parameters
$(4t+1, 2t, t-1, t)$, any vertex $w \in N'_u$ will have $t-1$ neighbours in $N'_u$.
Since $w$ is also connected to $v_m$ in $K$, it is connected to
$t$ of the $2t$ vertices in $K(N_u)$. The degree of $w$ in $\overline{K(N_u)}$ will
therefore remain $t$, and thus its degree in $K^u$ will remain $2t+1$.
\end{proof}

\begin{corollary}
The graph corresponding to a bordered quadratic residue code of length $n$ can be 
transformed into an $\frac{n}{2}$-regular graph by a single LC operation on
any vertex, except the one added by the bordering.
\end{corollary}

Note that \autoref{thm:bqrlc} holds for Paley graphs over both prime and
non-prime fields, and could also be extended to other
strongly regular graphs. The result does, however, not hold for the $[[18,0,8]]$ 
code constructed by a technique similar to bordered quadratic residue, since 
it does not contain a strongly regular graph on 17 vertices.
Even though quadratic residue and bordered quadratic residue codes achieve 
high distance and have regular graph representations, their
vertex degree of $n/2$ is far from optimal for high $n$.

\chapter{Quantum Codes and Boolean Functions}\label{chap:boolean}

\section{Introduction to Boolean Functions}

A \emph{Boolean function} of $n$ variables is a function $f: \mathbb{Z}_2^n \to \mathbb{Z}_2$.
There are $2^n$ vectors $\boldsymbol{x} = (x_0, x_1, \ldots, x_{n-1}) \in \mathbb{Z}_2^n$,
and each $\boldsymbol{x}$ can be interpreted as an integer
$2^{n-1} x_{n-1} + \cdots + 2 x_1 + x_0 \in \mathbb{Z}_{2^n}$.
If we evaluate $f(\boldsymbol{x})$ for each $\boldsymbol{x} \in \mathbb{Z}_2^n$ in increasing
order of the corresponding integers, we get a column vector $\boldsymbol{t}$, 
known as the \emph{truth table} of $f$.
A Boolean function can also be represented by the \emph{algebraic normal form} (ANF) which
is a sum of monomials of $n$ variables. Let the $2^n$ monomials of $n$ variables be ordered 
$1, x_0, x_1, x_0x_1, x_2, x_0x_2, \ldots, x_0x_1\cdots{}x_{n-1}$, i.e.,
monomial number $k = 2^{n-1} k_{n-1} + \cdots + 2 k_1 + k_0$ is $x^{(k)} = \prod_{k_i=1} x_i$.
Note that we will sometimes use an abbreviated ANF notation for some many-term Boolean functions, e.g.,
$012,12,0$ is short for $x_0x_1x_2 + x_1x_2 + x_0$.
The ANF of a Boolean function may be represented by the column vector $\boldsymbol{a}$
with $2^n$ binary coefficients, such that $f(\boldsymbol{x}) = \sum_{i \in \mathbb{Z}_{2^n}} a_i x^{(i)}$.
To transform an ANF representation of a Boolean function into a truth table, we could
perform $2^n$ evaluations of the function, but there is a more efficient method.
This method also enables us to do the reverse transformation from truth table to ANF.
The \emph{algebraic normal form transformation} (ANFT) can be expressed as a multiplication
of the vector $\boldsymbol{t}$ or $\boldsymbol{a}$ by a $2^n \times 2^n$ matrix $A_n$, such
that $A_n\boldsymbol{t} = \boldsymbol{a}$ and  $A_n\boldsymbol{a} = \boldsymbol{t}$.

\begin{definition}
The algebraic normal form transformation (ANFT) can be performed using the
matrix $A_n$, which can be decomposed as an $n$-fold tensor product,
\begin{equation}
A_n = \bigotimes_{i=0}^{n-1}
\begin{pmatrix} 1 & 0 \\ 1 & 1 \end{pmatrix}.
\end{equation}
\end{definition}

\begin{figure}
\centering
\includegraphics{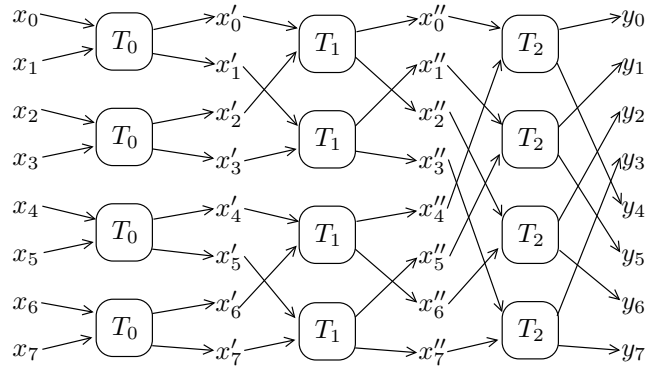}
\caption{Iterations of Algorithm for Tensor-Decomposable Transformations\label{fig:butterfly}}
\end{figure}

The straight-forward way to perform ANFT is to generate the $2^n \times 2^n$ matrix $A_n$ and
then calculate the product $A_n\boldsymbol{t}$ or $A_n\boldsymbol{a}$.
This operation has complexity $O(N^2)$, where $N = 2^n$.
A much more efficient algorithm, with complexity $O(N \log N)$, can
be used for any transformation $T$ that can be decomposed into $2 \times 2$ matrices,
$T = T_0 \otimes T_1 \otimes \cdots \otimes T_{n-1}$, where
\begin{equation}
T_i = \begin{pmatrix}t_i^{(0,0)} & t_i^{(0,1)} \\ t_i^{(1,0)} & t_i^{(1,1)} \end{pmatrix}.
\end{equation}
\autoref{alg:butterfly} is a simple version of this efficient algorithm.
Different optimisation techniques, as described by Fuller et~al.~\cite{B19},
can give faster implementations, but with the same order of complexity.
\autoref{fig:butterfly} illustrates the iterations of the algorithm.

\begin{algorithm}
\caption{Algorithm for Tensor-Decomposable Transformations\label{alg:butterfly}}
\begin{algorithmic}
\State \hskip -6pt
\begin{tabular}{ll}
\textbf{Input} 
& $n$: an integer\\
& $\boldsymbol{x}$: a vector of length $2^n$\\
& $\{T_0, T_1, \ldots, T_{n-1}\}$: $n$ $2 \times 2$ matrices, decomposition of $T$\\
\textbf{Output} 
& $\boldsymbol{y}$: the vector $T\boldsymbol{x}$\\
\end{tabular}
\Statex
\Procedure{Transform}{$n$, $\boldsymbol{x}$, $\{T_0, T_1, \ldots, T_{n-1}\}$}
  \For{$i$}{0}{$n-1$}
    \For{$j$}{0}{$2^n-1$}
      \If{$((j \gg i) \mod 2) = 0$}\Comment{``$\gg i$'' means $i$ right bit shifts}
        \State $y_j \gets t_i^{(0,0)} x_j + t_i^{(0,1)} x_{j + 2^i}$
      \Else
        \State $y_j \gets t_i^{(1,0)} x_{j-2^i} + t_i^{(1,1)} x_j$
      \EndIf
    \EndFor
    \For{$j$}{0}{$2^n-1$}
      \State $x_j \gets y_j$
    \EndFor
  \EndFor
  \State \textbf{return} $\boldsymbol{y}$
\EndProcedure
\end{algorithmic} 
\end{algorithm}

The \emph{degree} of a Boolean function is the degree of the highest order
term in its algebraic normal form. A function of degree 2 is called a \emph{quadratic function}, 
and corresponds to an undirected graph. Functions of higher degree correspond
to undirected \emph{hypergraphs}.
We will only consider simple graphs, i.e., graphs 
with no self-loops, as we will later see that linear and constant terms in Boolean functions can 
be ignored for our applications.
The Boolean function $f(\boldsymbol{x})$ of $n$ variables, $x_0, x_1, \ldots, x_{n-1}$,
corresponds to a hypergraph on $n$ vertices, $v_0, v_1, \ldots, v_{n-1}$,
where the edge $\{v_{i_1}, v_{i_2}, \ldots, v_{i_k}\} \in E$ iff the monomial
$x_{i_1}x_{i_2}\cdots{}x_{i_k}$ occurs in the algebraic normal form of $f$.
In particular, the quadratic function $f$ can be represented by the adjacency matrix $\Gamma$,
where $\Gamma_{i,j} = \Gamma_{j,i} = 1$ if $x_ix_j$ occurs in $f$,
and $\Gamma_{i,j} = \Gamma_{j,i} = 0$ otherwise.
Local complementation on a graph can then be described in terms of the 
corresponding quadratic Boolean function.

\begin{definition}
Let $f$ be a quadratic Boolean function corresponding to the graph $G=(V,E)$.
Let $x_a$ be a variable of $f$ corresponding to the vertex $v_a \in V$.
An LC operation on the variable $x_a$ produces the function
\begin{equation}
f'(\boldsymbol{x}) = f(\boldsymbol{x}) + \sum_{\substack{v_j,v_k \in N_{v_a} \\ j < k}} x_jx_k \pmod{2},
\end{equation}
where $N_{v_a}$ comprises the neighbours of $v_a$ in $G$.
\end{definition}

\begin{definition}
The \emph{Walsh-Hadamard transform} (WHT) of a function $f: \mathbb{Z}_2^n \to \mathbb{R}$ is given by 
the function $\widehat{f}: \mathbb{Z}_2^n \to \mathbb{R}$ defined as
\begin{equation}
\widehat{f}(\boldsymbol{b}) = 2^{-\frac{n}{2}} \sum_{\boldsymbol{x} \in \mathbb{Z}_2^n} 
f(\boldsymbol{x}) \cdot (-1)^{\boldsymbol{b} \cdot \boldsymbol{x}},
\end{equation}
where $\boldsymbol{b} \cdot \boldsymbol{x} = b_0x_0 + b_1x_1 + \cdots + b_{n-1}x_{n-1} \pmod{2}$.
WHT can be performed using the transformation matrix $H_n$, defined by the $n$-fold tensor product
\begin{equation}\label{eq:whtmatrix}
H_n = \bigotimes_{i=0}^{n-1} \frac{1}{\sqrt{2}}
\begin{pmatrix} 1 & 1 \\ 1 & -1 \end{pmatrix}.
\end{equation}
\end{definition}

The WHT of a function can be calculated by \autoref{alg:butterfly}.
Reverse WHT is performed by the same transformation, since $H_n = H_n^{-1}$.

\begin{definition}
Sometimes we prefer to work with the \emph{bipolar representation}, 
$\chi_f: \mathbb{Z}_2^n \to \{-1,1\}$, of a Boolean function $f$. We therefore define 
\begin{equation}
\chi_f(\boldsymbol{x}) = (-1)^{f(\boldsymbol{x})}.
\end{equation}
\end{definition}

\begin{definition}
The WHT of $\chi_f$ is known as the \emph{Walsh spectrum} of $f$, and is given
by the function
\begin{equation}\label{eq:walsh}
\widehat{\chi_f}(\boldsymbol{b}) = 
2^{-\frac{n}{2}} \sum_{\boldsymbol{x} \in \mathbb{Z}_2^n} 
\chi_f(\boldsymbol{x}) \cdot (-1)^{\boldsymbol{b} \cdot \boldsymbol{x}} =
2^{-\frac{n}{2}} \sum_{\boldsymbol{x} \in \mathbb{Z}_2^n} 
(-1)^{f(\boldsymbol{x}) + \boldsymbol{b} \cdot \boldsymbol{x}}.
\end{equation}
\end{definition}

Note that the addition in the exponent of $-1$ in \eqref{eq:walsh} is
modulo 2, and that all additions of binary values will be modulo 2, 
unless otherwise stated.

\begin{theorem}[Parseval's Theorem]\label{thm:parseval}
For any function, $f: \mathbb{Z}_2^n \to \mathbb{R}$, the following relationship holds
\begin{equation}
\sum_{\boldsymbol{b} \in \mathbb{Z}_2^n}  \left. \widehat{f} \right.^2(\boldsymbol{b}) = 
\sum_{\boldsymbol{x} \in \mathbb{Z}_2^n} f^2(\boldsymbol{x}).
\end{equation}
\end{theorem}

\begin{definition}
The \emph{correlation} of two Boolean functions, $f$ and $g$, is
\begin{equation}
\kappa(f,g) = \sum_{\boldsymbol{x} \in \mathbb{Z}_2^n} 
\chi_f(\boldsymbol{x}) \cdot \chi_g(\boldsymbol{x}) = 
\sum_{\boldsymbol{x} \in \mathbb{Z}_2^n} (-1)^{f(\boldsymbol{x}) + g(\boldsymbol{x})}.
\end{equation}
\end{definition}

The Hamming distance between $f$ and $g$, $d(f,g)$, which is the number of positions where 
their truth tables have different values, can be derived from their correlation, since
\begin{equation}
d(f,g) = 2^{n-1} - \frac{\kappa(f,g)}{2}.
\end{equation}
High positive correlation of two functions implies that the distance between them is low,
and therefore that they closely resemble
each other, i.e., that they will often give the same output for the same input.
It is easy to verify that $\kappa(f,g) = - \kappa(f,g+1)$, so
if $f$ has high negative correlation with $g$, it will have equally high positive correlation with 
$g+1$, the complement of $g$.

\begin{definition}\label{def:periodic}
The \emph{periodic autocorrelation} of the Boolean function $f$ is given
by the function $r: \mathbb{Z}_2^n \to \mathbb{R}$ defined as
\begin{equation}
r(\boldsymbol{a}) = \sum_{\boldsymbol{x} \in \mathbb{Z}_2^n}
\chi_f(\boldsymbol{x}) \cdot \chi_f(\boldsymbol{x}+\boldsymbol{a}) = 
\sum_{\boldsymbol{x} \in \mathbb{Z}_2^n} (-1)^{f(\boldsymbol{x}) + f(\boldsymbol{x}+\boldsymbol{a})},
\end{equation}
where $\boldsymbol{x}+\boldsymbol{a} = (x_0+a_0, x_1+a_1, \ldots, x_{n-1}+a_{n-1})$.
\end{definition}

The autocorrelation coefficient $r(\boldsymbol{a})$, 
for some $\boldsymbol{a} \in \mathbb{Z}_2^n$, is the
correlation of the functions $f(\boldsymbol{x})$ and $f(\boldsymbol{x}+\boldsymbol{a})$.
The periodic autocorrelation function therefore gives the correlation of a function and 
all its \emph{periodic shifts}, i.e., all possible combinations of variable inversions.

\begin{theorem}[The Wiener-Khintchine Theorem]\label{thm:wiener}
The periodic autocorrelation and the Walsh spectrum of the Boolean function $f$ are related, since
\begin{equation}
r(\boldsymbol{a}) = 2^{\frac{n}{2}} \widehat{\left. \widehat{\chi_f} \right.^2} (\boldsymbol{a}).
\end{equation}
\end{theorem}

\begin{example}
Consider the graph shown below.
\begin{center}
\includegraphics[width=.2\linewidth]{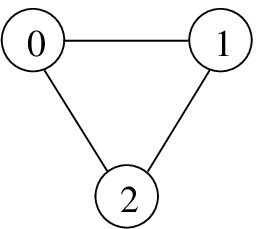}
\end{center}
The graph has edges $E = \{\{0,1\},\{0,2\},\{1,2\}\}$ and
corresponds to the Boolean function $f(\boldsymbol{x}) = x_0x_1 + x_0x_2 + x_1x_2$.
The truth table and ANF of $f$ is given by the following table,
and so are the truth tables of $\widehat{f}$, $\chi_f$ and $\widehat{\chi_f}$.
The table also gives the periodic autocorrelation, $r$, of $f$.
\begin{center}
\begin{tabular}{ccccccccc}
\toprule
$k_2$ & $k_1$ & $k_0$ & ANF & $f(\boldsymbol{k})$ 
& $\widehat{f}(\boldsymbol{k})$ 
& $\chi_f(\boldsymbol{k})$ & $\widehat{\chi_f}(\boldsymbol{k})$ & $r(\boldsymbol{k})$ \\
\cmidrule(r){1-3}\cmidrule(l){4-9}
0 & 0 & 0  & 0 & 0 & $2^{\frac{1}{2}}$   &  1 & 0                  &  8 \\
0 & 0 & 1  & 0 & 0 & $-2^{-\frac{1}{2}}$ &  1 & $2^{\frac{1}{2}}$  &  0 \\
0 & 1 & 0  & 0 & 0 & $-2^{-\frac{1}{2}}$ &  1 & $2^{\frac{1}{2}}$  &  0 \\
0 & 1 & 1  & 1 & 1 & 0                   & -1 & 0                  &  0 \\
1 & 0 & 0  & 0 & 0 & $-2^{-\frac{1}{2}}$ &  1 & $2^{\frac{1}{2}}$  &  0 \\
1 & 0 & 1  & 1 & 1 & 0                   & -1 & 0                  &  0 \\
1 & 1 & 0  & 1 & 1 & 0                   & -1 & 0                  &  0 \\
1 & 1 & 1  & 0 & 1 & $2^{-\frac{1}{2}}$  & -1 & $-2^{\frac{1}{2}}$ &  -8 \\
\bottomrule
\end{tabular}
\end{center}
\end{example}

\section{Propagation Criteria for Boolean Functions}

\begin{definition}
A Boolean function, $f$, is \emph{balanced} if the Hamming weight of its truth table
is $2^{n-1}$, or, equivalently, if
\begin{equation}
\sum_{\boldsymbol{x} \in \mathbb{Z}_2^n} (-1)^{f(\boldsymbol{x})} = 0.
\end{equation}
\end{definition}

\begin{definition}
The bipolar truth tables, $\chi_f$ and $\chi_g$, of the Boolean functions $f$ and $g$ are orthogonal vectors if
\begin{equation}
\sum_{\boldsymbol{x} \in \mathbb{Z}_2^n} \chi_f(\boldsymbol{x}) \cdot \chi_g(\boldsymbol{x}) = 
\sum_{\boldsymbol{x} \in \mathbb{Z}_2^n} (-1)^{f(\boldsymbol{x}) + g(\boldsymbol{x})} = 0,
\end{equation}
which implies that the correlation $\kappa(f,g) = 0$ and that 
$f(\boldsymbol{x}) + g(\boldsymbol{x})$ is a balanced function.
\end{definition}

\begin{definition}
A \emph{linear Boolean function} is a function of degree less than two
that can be written $\boldsymbol{x} \cdot \boldsymbol{b}$ for some $\boldsymbol{b} \in \mathbb{Z}_2^n$.
The set of \emph{affine functions} consists of the linear functions and their
complements, i.e., functions of the form $\boldsymbol{x} \cdot \boldsymbol{b} + c$,
where $\boldsymbol{b} \in \mathbb{Z}_2^n$ and $c \in \mathbb{Z}_2$.
\end{definition}

Boolean functions have important applications in cryptography.
Mappings from $n$ bits to $k$ bits, known as \emph{S-boxes}, are important 
components in \emph{block ciphers}.
Such mappings can be viewed as $k$ mappings from $n$ bits to 1 bit, which
can each be expressed as a Boolean function. These functions, and all their linear 
combinations, must satisfy certain criteria for the cipher to be secure.
Common requirements are high algebraic degree and that the functions are balanced.
We also want functions with a high degree of \emph{nonlinearity}, which
means that no affine functions closely resemble them, since the cipher
could otherwise have been approximated by affine functions in an attack
known as \emph{linear cryptanalysis}.
The nonlinearity of the Boolean function $f$ can be derived from
the Walsh spectrum $\widehat{\chi_f}$, since
\begin{equation}
2^{\frac{n}{2}} \widehat{\chi_f}(\boldsymbol{b}) = 
\kappa(f(\boldsymbol{x}), \boldsymbol{b} \cdot \boldsymbol{x}),
\end{equation}
i.e., the Walsh spectrum gives the correlation of $f$ with
all possible linear functions, $\boldsymbol{x} \cdot \boldsymbol{b}$ 
for all $\boldsymbol{b} \in \mathbb{Z}_2^n$.
The coefficient of the Walsh spectrum with highest magnitude corresponds 
to the affine function closest to $f$.

\begin{definition}
A Boolean function is \emph{correlation immune} of order $m$ if
$\widehat{\chi_f}(\boldsymbol{b}) = 0$ for all $\boldsymbol{b}$ where
$1 \le w_H(\boldsymbol{b}) \le m$. If the function is also balanced, i.e.,
$\widehat{\chi_f}(\boldsymbol{b}) = 0$ for all $\boldsymbol{b}$ where
$0 \le w_H(\boldsymbol{b}) \le m$, then it is called \emph{$m$-resilient}.
The function obtained by fixing the value of $m$ or fewer variables
of an $m$-resilient function will still be a balanced function.
\end{definition}

Boolean functions are also critical components in \emph{stream ciphers},
where they can be used to represent \emph{nonlinear combiners}.
In this context, the correlation immunity or resilience of 
a Boolean function is important, since they give the number
of \emph{linear feedback shift registers} required to realise a
correlation attack on the stream cipher.

\begin{definition}
A vector $\boldsymbol{s}$ of length $2^n$ is \emph{flat} if $|s_i| = |s_j|$ for all 
$i,j \in \mathbb{Z}_{2^n}$, where $s_k$ denotes the $k$th coordinate of $\boldsymbol{s}$.
\end{definition}

It follows from \autoref{thm:parseval} that 
\begin{equation}
\sum_{\boldsymbol{b} \in \mathbb{Z}_2^n}  \left. \widehat{\chi_f} \right.^2(\boldsymbol{b}) = 
\sum_{\boldsymbol{x} \in \mathbb{Z}_2^n} \chi_f^2(\boldsymbol{x}) = 2^n.
\end{equation}
If the Walsh spectrum of a Boolean function is flat, all coefficients of 
the Walsh spectrum must therefore be 1 or $-1$.
Such functions have the highest possible minimum distance to any affine function.

\begin{definition}
A Boolean function with a flat Walsh spectrum is called \emph{perfect nonlinear} or \emph{bent}.
\end{definition}

A vector of length $2^n$ with entries 1 and $-1$ is 
the bipolar truth table of a Boolean function. The Walsh spectrum
of a bent function therefore gives its \emph{dual function}, which will also be bent. 
It follows from \autoref{thm:wiener} that the condition for a function to be bent
can also be expressed in terms of the periodic autocorrelation.
The function $f$ is bent if $r(\boldsymbol{a}) = 0$ for all $\boldsymbol{a}$ 
where $1 \le w_H(\boldsymbol{a}) \le n$, which means that the bipolar truth tables of 
$f(\boldsymbol{x})$ and $f(\boldsymbol{x}+\boldsymbol{a})$ must be orthogonal vectors.
It can be shown that only Boolean functions of an even number of variables can be bent,
and that bent functions can never be balanced.
In general, it is not possible to optimise all desired properties of a Boolean
function, so some compromise must be made. 
As a generalisation of the bent criterion, the \emph{propagation criterion}
of Boolean functions was defined by Preneel et~al.~\cite{B44}
and later studied by Carlet~\cite{B12}.

\begin{definition}
A Boolean function satisfies $\PC(l)$, the \emph{propagation criterion} of degree $l$, if
$r(\boldsymbol{a}) = 0$ for all $\boldsymbol{a}$ where $1 \le w_H(\boldsymbol{a}) \le l$.
\end{definition}

Bent functions satisfy $\PC(n)$, but functions satisfying $\PC(l)$ for $l < n$
can in addition be balanced and are therefore better suited for cryptographic purposes.
Boolean functions used in cryptography should also have good properties
when subsets of their variables are set to fixed values. 

\begin{definition}
The Boolean function $f$ satisfies the \emph{propagation criterion} of degree $l$ and order $m$
if any function obtained by setting $m$ or fewer variables of $f$ to any fixed values satisfies $\PC(l)$.
It is required that the set of fixed bits and the set of modified bits are disjoint, 
and therefore that $l+m \le n$.
\end{definition}

A function satisfying $\PC(l)$ of order $m$ is resistant against attacks where
the attacker knows the value of up to $m$ input bits for a large number of plaintext/ciphertext
pairs. The attacker is further able to modify up to $l$ of the other input bits and 
compare the modified output with the original. This technique is known as
\emph{differential cryptanalysis}.
In a more general scenario, the sets of $m$ known bits and $l$ modified bits need not
be disjoint. Preneel et~al.~\cite{B44} define the \emph{extended propagation
criterion}, where this restriction is removed. Carlet~\cite{B12} reformulated this
criterion as follows.

\begin{proposition}[Carlet~\cite{B12}]
The Boolean function $f$ satisfies $\EPC(l)$ of order $m$, the extended propagation criterion
of degree $l$ and order $m$, if
$f(\boldsymbol{x})+f(\boldsymbol{x}+\boldsymbol{a})$ is $m$-resilient
for any $\boldsymbol{a}$ where $1 \le w_H(\boldsymbol{a}) \le l$.
An equivalent requirement is that
\begin{equation}
\sum_{\boldsymbol{x} \in \mathbb{Z}_2^n} (-1)^{f(\boldsymbol{x}) + 
 f(\boldsymbol{x}+\boldsymbol{a}) + \boldsymbol{b} \cdot \boldsymbol{x}} = 0,
\end{equation}
or that the bipolar truth tables of 
$f(\boldsymbol{x})$ and $f(\boldsymbol{x}+\boldsymbol{a}) + \boldsymbol{b} \cdot \boldsymbol{x}$
must be orthogonal,
for any $\boldsymbol{a}$ and $\boldsymbol{b}$ where 
$w_H(\boldsymbol{a}) \le l$, $w_H(\boldsymbol{b}) \le m$, and
$w_H(\boldsymbol{a})$ and $w_H(\boldsymbol{b})$ are not both zero.
The same criterion can be used for PC, if we add the restriction that 
$\boldsymbol{a}$ and $\boldsymbol{b}$ have disjoint supports, i.e.,
that they never both have the value 1 in the same coordinate.
\end{proposition}

\begin{definition}
Given two vectors $\boldsymbol{u}, \boldsymbol{v} \in \mathbb{Z}_2^n$,
we say that $\boldsymbol{u}$ is \emph{covered by} $\boldsymbol{v}$, denoted
$\boldsymbol{u} \preceq \boldsymbol{v}$, if $u_i \le v_i$ for all $i \in \mathbb{Z}_n$.
We define the negation of $\boldsymbol{u} \in \mathbb{Z}_2^n$ as 
$\overline{\boldsymbol{u}} = (u_0+1, u_1+1, \ldots, u_{n-1}+1)$.
Given a vector $\boldsymbol{u} \in \mathbb{Z}_2^n$, let 
$V_{\boldsymbol{u}} \subseteq \mathbb{Z}_2^n$ be the set 
$V_{\boldsymbol{u}} = \{\boldsymbol{v} \in \mathbb{Z}_2^n \mid \boldsymbol{v} \preceq \boldsymbol{u}\}$.
Given a vector $\boldsymbol{u} \in \mathbb{Z}_2^n$ and a subset
$V \subseteq \mathbb{Z}_2^n$, $\boldsymbol{u} + V = 
\{\boldsymbol{u} + \boldsymbol{v} \mid \boldsymbol{v} \in V\}$ is a \emph{coset} of $V$.
\end{definition}

\begin{definition}\label{def:fixedperiodic}
Let the \emph{fixed-periodic autocorrelation} of a Boolean function $f$ be
defined as
\begin{equation}
r(\boldsymbol{a}, \boldsymbol{\mu}, \boldsymbol{k}) = 
\sum_{\boldsymbol{x} \in \boldsymbol{k} + V_{\overline{\boldsymbol{\mu}}}}
 (-1)^{f(\boldsymbol{x}) + f(\boldsymbol{x}+\boldsymbol{a})},
\end{equation}
where $\boldsymbol{k} \preceq \boldsymbol{\mu}$ and $\boldsymbol{a} \preceq \overline{\boldsymbol{\mu}}$.
The vector $\boldsymbol{\mu}$ indicates a subset of the variables of $f$ which are fixed
to the values given by $\boldsymbol{k}$, while the vector $\boldsymbol{a}$ indicates 
a subset of the remaining variables which are periodically shifted (flipped).
\end{definition}

\begin{proposition}\label{prop:autopc}
It can be shown~\cite{B15} that a Boolean function satisfies $\PC(l)$ of order
$m$ if $r(\boldsymbol{a}, \boldsymbol{\mu}, \boldsymbol{k}) = 0$ for all
$\boldsymbol{a}, \boldsymbol{\mu}, \boldsymbol{k} \in \mathbb{Z}_2^n$ where
$\boldsymbol{k} \preceq \boldsymbol{\mu}$, $\boldsymbol{a} \preceq \overline{\boldsymbol{\mu}}$,
$1 \le w_H(\boldsymbol{a}) \le l$, and $0 \le w_H(\boldsymbol{\mu}) \le m$.
\end{proposition}

\begin{definition}\label{def:aperiodic}
Let the \emph{aperiodic autocorrelation} of a Boolean function $f$ be
defined as
\begin{equation}
s(\boldsymbol{a}, \boldsymbol{k}) = 
\sum_{\boldsymbol{x} \in \boldsymbol{k} + V_{\overline{\boldsymbol{a}}}}
 (-1)^{f(\boldsymbol{x}) + f(\boldsymbol{x}+\boldsymbol{a})},
\end{equation}
where $\boldsymbol{k} \preceq \boldsymbol{a}$. Variables indicated by
$\boldsymbol{a}$ are shifted aperiodically, and are therefore assigned 
the fixed values given by $\boldsymbol{k}$.
\end{definition}

\begin{definition}\label{def:fixedaperiodic}
Let the \emph{fixed-aperiodic autocorrelation} of a Boolean function $f$ be
defined as
\begin{equation}
s(\boldsymbol{a}, \boldsymbol{\mu}, \boldsymbol{k}) = 
\sum_{\boldsymbol{x} \in \boldsymbol{k} + V_{\overline{\boldsymbol{\mu}}}}
 (-1)^{f(\boldsymbol{x}) + f(\boldsymbol{x}+\boldsymbol{a})},
\end{equation}
where $\boldsymbol{a}, \boldsymbol{k} \preceq \boldsymbol{\mu}$.
The vector $\boldsymbol{\mu}$ indicates a subset of the variables of $f$ which are fixed
to the values given by $\boldsymbol{k}$,
including those that are aperiodically shifted as indicated by $\boldsymbol{a}$.
Let $\boldsymbol{\theta} = \boldsymbol{\mu} + \boldsymbol{a}$ indicate the
variables that are fixed but not shifted.
The supports of $\boldsymbol{\theta}$ and $\boldsymbol{a}$ are disjoint by definition.
\end{definition}

Just as PC can be expressed in terms of fixed-periodic autocorrelation,
as described in \autoref{prop:autopc},
the fixed-aperiodic autocorrelation also corresponds to a propagation criterion
which we call the \emph{aperiodic propagation criterion} (APC)~\cite{B15}.

\begin{definition}\label{def:apc}
A Boolean function satisfies $\APC(l)$ of order $m$, the aperiodic propagation criterion
of degree $l$ and order $m$, if $s(\boldsymbol{a}, \boldsymbol{\mu}, \boldsymbol{k}) = 0$
for all
$\boldsymbol{a}, \boldsymbol{\mu}, \boldsymbol{k} \in \mathbb{Z}_2^n$ where
$\boldsymbol{a}, \boldsymbol{k} \preceq \boldsymbol{\mu}$,
$\boldsymbol{\theta} = \boldsymbol{\mu} + \boldsymbol{a}$,
$1 \le w_H(\boldsymbol{a}) \le l$, $0 \le w_H(\boldsymbol{\theta}) \le m$, and
$\boldsymbol{a}$ and $\boldsymbol{\theta}$ have disjoint supports.
\end{definition}

In a cryptographic scenario, a Boolean function satisfying $\APC(l)$ of order $m$ 
is resistant against attacks where the attacker knows up to $m+l$ of the input bits, 
and is further allowed to change up to $l$ of these known bits.

\begin{definition}
A Boolean function $f$ has \emph{APC distance} $d$ if it satisfies $\APC(l)$ of
order $m$ for all positive integers $l$ and $m$ such that $l+m < d$. 
\end{definition}

\begin{definition}\label{def:wgtab}
We define the weight operator
$w_H(\boldsymbol{a}, \boldsymbol{b}) = 
w_H(\boldsymbol{a}) + w_H(\boldsymbol{b}) - w_H(\boldsymbol{a} \wedge \boldsymbol{b})$, 
where $\boldsymbol{a} \wedge \boldsymbol{b} = (a_0b_0, a_1b_1, \ldots, a_{n-1}b_{n-1})$.
\end{definition}

\begin{proposition}\label{prop:apc}
It can be shown~\cite{B15} that the APC distance of the Boolean function $f$ is equal to
the smallest nonzero $w_H(\boldsymbol{a}, \boldsymbol{b})$ where
\begin{equation}\label{eq:apc}
\sum_{\boldsymbol{x} \in \mathbb{Z}_2^n}
(-1)^{f(\boldsymbol{x}) + f(\boldsymbol{x}+\boldsymbol{a}) + \boldsymbol{b} \cdot \boldsymbol{x}} \ne 0.
\end{equation}
\end{proposition}

\begin{proposition}
If the Boolean function $f$ has APC distance $d$, then $f(\boldsymbol{x}) + f(\boldsymbol{x}+\boldsymbol{a})$
must be $(d-w_H(\boldsymbol{a})-1)$-resilient, for all $\boldsymbol{a}$ where $w_H(\boldsymbol{a}) < d$.
\end{proposition}

\section{Quantum Codes as Boolean Functions}\label{sec:qcbool}

\begin{definition}\label{def:funcstate}
The Boolean function $f(\boldsymbol{x})$ corresponds to the vector 
$\boldsymbol{s} = 2^{-\frac{n}{2}}(-1)^{f(\boldsymbol{x})}$, which
can be interpreted as the probability distribution vector of the quantum state 
\begin{equation}
\ket{\psi_f} = 2^{-\frac{n}{2}} \sum_{\boldsymbol{x} \in \mathbb{Z}_2^n} 
(-1)^{f(\boldsymbol{x})} \ket{\boldsymbol{x}}.
\end{equation}
\end{definition}

We can find the single quantum state represented by a self-dual quantum code 
once we know the Boolean function corresponding to an equivalent graph code.
We interpret the Boolean function as a quantum state as described in \autoref{def:funcstate}.
(See \autoref{l3} for the definition of a quantum state.) The normalisation
factor $2^{-\frac{n}{2}}$ ensures that the sum of all probabilities is 1,
but in many cases it can be ignored, and the bipolar truth table of $f$ can then
be interpreted directly as a quantum state.

\begin{example}
The function $f(\boldsymbol{x}) = x_0x_1 + x_0x_2 + x_1x_2$ corresponds to the bipolar vector
$\boldsymbol{s} = 2^{-\frac{3}{2}} (1,1,1,-1,1,-1,-1,-1)^T$. This is the probability distribution
vector of the quantum state
$\ket{\psi_f} = 2^{-\frac{3}{2}}(\ket{000}+\ket{001}+\ket{010}-\ket{011}+\ket{100}-\ket{101}-\ket{110}-\ket{111})$.
\end{example}

We will only consider \emph{bipolar quantum states}, i.e., states where 
all coefficients of the  probability distribution vectors are either $1$ or $-1$.
Parker and Rijmen~\cite{B40} define the more general \emph{algebraic polar form},
$\boldsymbol{s} = m(\boldsymbol{x})(-1)^{p(\boldsymbol{x})}$, where the 
two Boolean functions $m(\boldsymbol{x})$ and $p(\boldsymbol{x})$ describes
magnitude and phase, respectively. All vectors with coefficients from the
set $\{-1,0,1\}$ are covered by this definition. We will only study
the case where $m(\boldsymbol{x}) = 1$.

The self-dual quantum codes studied in the previous chapters have all corresponded 
to \emph{quadratic} Boolean functions.
A bipolar quantum state may correspond to a Boolean function of any degree, but
unlike quadratic functions, functions of degree higher than two do not correspond to 
stabilizer codes or additive codes over $\GF(4)$.
By going back to the original definition of quantum error correcting codes, as
described in \autoref{l12}, it is possible to extend the definition
of zero-dimensional quantum codes to what might be called ``hypergraph codes'', corresponding
to non-quadratic Boolean functions.
To guarantee that an error on a single quantum state can be detected, the
errored state must be orthogonal to the original state. The distance of a zero-dimensional 
quantum code is therefore the weight of the minimum weight quantum error operator that
gives an errored state not orthogonal to the original state.
In quantum error correction we only need to consider the three errors described by 
the Pauli matrices $\sigma_x$, $\sigma_z$ and  $\sigma_y$, corresponding, respectively,
to bit-flip, phase-flip and combined bit-flip and phase-flip.
A Pauli error operating on the hypergraph state $\ket{\psi_f}$ corresponds to an
operation on the corresponding Boolean function $f(\boldsymbol{x})$.
Let the binary vector $\boldsymbol{a}$ indicate which qubits have been bit-flipped, i.e.,
qubit number $k$ has been bit-flipped iff $a_k = 1$. These bit-flips correspond to
a transformation on $\ket{\psi_f}$,
\begin{equation}
\ket{\psi_f} \to \left( \bigotimes_{a_k=1}\sigma_x^{(k)} \bigotimes_{a_k=0}I^{(k)} \right)
\ket{\psi_f}.
\end{equation}
The corresponding bit-flip operation on the Boolean function $f(\boldsymbol{x})$ is
\begin{equation}
f(\boldsymbol{x}) \to f(\boldsymbol{x}+\boldsymbol{a}).
\end{equation}
Similarly, let the binary vector $\boldsymbol{b}$ indicate the qubits which have been phase-flipped.
We get the error operation
\begin{equation}
\ket{\psi_f} \to \left( \bigotimes_{b_k=1}\sigma_z^{(k)} \bigotimes_{b_k=0}I^{(k)} \right)
\ket{\psi_f},
\end{equation}
and we can express the same phase-flips in terms of a Boolean function,
\begin{equation}\label{eq:phaseflip}
f(\boldsymbol{x}) \to f(\boldsymbol{x}) + \boldsymbol{b} \cdot \boldsymbol{x}.
\end{equation}
The third error we must consider is the combined bit-flip and phase-flip.
We have used the definition $\sigma_y = i \sigma_x \sigma_z$, but
the overall phase factor $i$ has no significance and can be ignored.
Neither is the order of the operations important, since 
$\sigma_z \sigma_x = - \sigma_x \sigma_z$, and
this phase factor can also be ignored. 
\begin{equation}
\ket{\psi_f} \to \left( \bigotimes_{\substack{a_k=1 \\ b_k=1}}\sigma_y^{(k)} 
\bigotimes_{a_k=0}I^{(k)} \bigotimes_{b_k=0}I^{(k)} \right) \ket{\psi_f}
\end{equation}
corresponds, up to a global phase factor, to an operation on a Boolean function,
\begin{equation}
f(\boldsymbol{x}) \to f(\boldsymbol{x}+\boldsymbol{a}) + \boldsymbol{b} \cdot \boldsymbol{x} 
+ \boldsymbol{b} \cdot \boldsymbol{a}.
\end{equation}
The linear term $\boldsymbol{b} \cdot \boldsymbol{a}$ can safely be ignored, 
because, as $\boldsymbol{a}$ and $\boldsymbol{b}$ are fixed, it reduces
to a constant and therefore contributes another global phase factor.
We can finally consider a general Pauli error,
\begin{equation}
\ket{\psi_f} \to \left( 
\bigotimes_{\substack{a_k=1 \\ b_k=0}}\sigma_x^{(k)} 
\bigotimes_{\substack{a_k=0 \\ b_k=1}}\sigma_z^{(k)} 
\bigotimes_{\substack{a_k=1 \\ b_k=1}}\sigma_y^{(k)} 
\bigotimes_{\substack{a_k=0 \\ b_k=0}}I^{(k)} \right)
\ket{\psi_f},
\end{equation}
which can also be expressed in terms of a Boolean function,
\begin{equation}
f(\boldsymbol{x}) \to f(\boldsymbol{x}+\boldsymbol{a}) + \boldsymbol{b} \cdot \boldsymbol{x}.
\end{equation}
Note that the error operator $\sigma_y$, although composed of two errors, only counts
as one error. The weight of a quantum error operator in terms of the vectors $\boldsymbol{a}$
and $\boldsymbol{b}$ is therefore given by the weight function $w_H(\boldsymbol{a}, \boldsymbol{b})$,
as defined in \autoref{def:wgtab}.
We can then define the distance of the zero-dimensional quantum code corresponding
to the state $\ket{\psi_f}$ as the smallest nonzero $w_H(\boldsymbol{a}, \boldsymbol{b})$ such
that the bipolar truth tables of $f(\boldsymbol{x})$ and
$f(\boldsymbol{x}+\boldsymbol{a}) + \boldsymbol{b} \cdot \boldsymbol{x}$
are orthogonal vectors.
By comparing this criterion to the definition of APC distance in \autoref{prop:apc},
we see that they are equal.

\begin{theorem}
Let $\ket{\psi_f}$ be a bipolar quantum state with probability distribution vector 
$2^{-\frac{n}{2}}(-1)^{f(\boldsymbol{x})}$, where $f$ is a Boolean function.
Then $\ket{\psi_f}$ corresponds to an $[[n,0,d]]$ quantum code where $d$ is the 
APC distance of $f$. Conversely, any Boolean function with APC distance $d$ corresponds 
to an $[[n,0,d]]$ quantum code.
\end{theorem}

A Boolean function of degree higher than two corresponds to a \emph{non-quadratic quantum code} which
is not equivalent to any stabilizer code or additive code over $\GF(4)$.
While we can quickly find the APC distance of a quadratic Boolean function as the distance
of the corresponding additive code over $\GF(4)$ using \autoref{l25},
no such shortcut is known for higher degree functions. We must therefore check the condition
\eqref{eq:apc} for all errors of increasing weight, 
until an error not satisfying the condition is found, as described in \autoref{alg:distnonquad}.

\begin{algorithm}
\caption{Finding the APC Distance of a Boolean Function}\label{alg:distnonquad} 
\begin{algorithmic}
\State \hskip -6pt
\begin{tabular}{ll}
\textbf{Input} 
& $f$: a Boolean function\\
& $n$: the number of variables of $f$\\
\textbf{Output} 
& $d$: the APC distance of $f$
\end{tabular}
\Statex
\Procedure{FindAPCdistance}{$f$, $n$}
  \State $\boldsymbol{s} \gets \chi_f$, the bipolar truth table of $f(\boldsymbol{x})$
  \ForAll{$\boldsymbol{a} \in \mathbb{Z}_2^n$ and $\boldsymbol{b} \in \mathbb{Z}_2^n$ 
          in increasing order of $w_H(\boldsymbol{a},\boldsymbol{b})$}
    \State $\boldsymbol{s}' \gets$ the bipolar truth table of $f(\boldsymbol{x} + \boldsymbol{a}) + \boldsymbol{b} \cdot \boldsymbol{x}$
    \If{$\boldsymbol{s} \cdot \boldsymbol{s}' \ne 0$}
      \State \textbf{return} $w_H(\boldsymbol{a},\boldsymbol{b})$
    \EndIf
  \EndFor
\EndProcedure
\end{algorithmic} 
\end{algorithm}

\begin{figure}[tb]
\centering
\includegraphics[width=.40\linewidth]{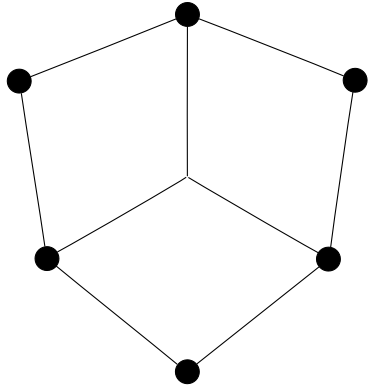}
\caption{A Hypergraph Corresponding to a {$[[6,0,3]]$} Quantum Code\label{fig:hyper6}}
\end{figure}

\begin{example}
The hypergraph shown in \autoref{fig:hyper6} corresponds to the cubic Boolean function
$f(\boldsymbol{x}) = 012,03,04,13,15,24,25$. It can be verified that no error with
$w_H(\boldsymbol{a}, \boldsymbol{b}) \le 2$ satisfies \eqref{eq:apc}.
There are, however, vectors $\boldsymbol{a}$ and $\boldsymbol{b}$ with
$w_H(\boldsymbol{a}, \boldsymbol{b}) =3$ such that \eqref{eq:apc} is satisfied.
One such error is given by $\boldsymbol{a} = (0,0,0,0,0,1)$ and $\boldsymbol{b} = (0,1,1,0,0,0)$,
and it produces the errored state
$f(\boldsymbol{x}+\boldsymbol{a}) + {\boldsymbol{b} \cdot \boldsymbol{x}}
= 012,03,04,12,13,15,24,25$.
We verify that $f(\boldsymbol{x}) + f(\boldsymbol{x}+\boldsymbol{a}) + \boldsymbol{b} \cdot \boldsymbol{x}
= x_1x_2$ is not a balanced function. This means that the APC distance 
of $f$ is 3, and that $f$ corresponds to a $[[6,0,3]]$ non-quadratic quantum code.
\end{example}

Note that \autoref{prop:always2} does not hold for hypergraphs.
Connected hypergraphs where all vertices have high degree may still correspond
to Boolean functions with APC distance 1.

\section[\texorpdfstring{The $\{I,H,N\}^n$ Transform Set}{The \{I,H,N\} \textasciicircum{}n Transform Set}]
{The $\text{\usefont{OT1}{phv}{bc}{it}\{I,\,H,\,N\,\}}^\text{\usefont{OT1}{phv}{bc}{it}n}$ Transform Set}

As described in \autoref{sec:lutlc}, Hein et~al.~\cite{hein} showed that local complementation on a graph
corresponds to a transformation defined by the tensor product of the matrices 
$\tau_x = \sqrt{-i \sigma_x}$ and $\tau_z = \sqrt{i \sigma_z}$.
The LC orbit can then be generated by repeated transformations of this form.

\begin{definition}
Let $\mathcal{T} = \{T_1, T_2, \ldots, T_k\}$ be a set of $k$ $2 \times 2$ unitary
matrices. The transform set $\mathcal{T}^n$ is then
the set of $k^n$ $2^n \times 2^n$ transformation matrices of
the form $U = U_0 \otimes U_1 \otimes \cdots \otimes U_{n-1}$, where
$U_i \in \mathcal{T}$.
Given a vector $\boldsymbol{s}$ of length $2^n$, the $k^n$ transforms
$\boldsymbol{S} = U\boldsymbol{s}$, for all possible choices of $U \in \mathcal{T}^n$,
is a multispectra with $(2k)^n$ spectral points.
We refer to this multispectra as the spectrum with respect to the $\mathcal{T}^n$ transform.
\end{definition}

\begin{definition}
Let $\mathcal{D}$ be the infinite set of all $2 \times 2$ diagonal and anti-diagonal unitary matrices, i.e,
matrices of the form
\[
\begin{pmatrix} a&0\\ 0&b \end{pmatrix}
\text{ and }
\begin{pmatrix} 0&a\\ b&0 \end{pmatrix},
\]
where $a,b \in \mathbb{R}$.
A transform $F \in \mathcal{D}^n$ is then a tensor product of any $n$ matrices from $\mathcal{D}$.
\end{definition}

\begin{proposition}[Riera and Parker~\cite{B51}]
Applying a transformation from $\mathcal{D}^n$ to a vector $\boldsymbol{s}$ will 
not change the magnitudes of the coefficients of $\boldsymbol{s}$.
If two $2 \times 2$ unitary matrices, $A$ and $B$, satisfy $FA = B$ where 
$F \in \mathcal{D}$, they can be considered equivalent, $A \simeq B$.
\end{proposition}

\begin{theorem}[Riera and Parker~\cite{B51}]
To within a subsequent transformation from $\mathcal{D}^n$, all members of the LC orbit
of the Boolean function $f$ can be
found as a subset of the finite set of $3^n$ $\{I,\tau_x,\tau_x\tau_z\}^n$ transforms of $f$.
\end{theorem}

\begin{example}
For a Boolean function $f(\boldsymbol{x})$ of $n=2$ variables, we find the corresponding vector
$\boldsymbol{s} = 2^{-\frac{n}{2}}(-1)^{f(\boldsymbol{x})}$. We then find the 9 vectors
$\boldsymbol{S} = U\boldsymbol{s}$, for all $U \in \{I \otimes I$, $I \otimes \tau_x$,
$I \otimes \tau_x\tau_z$, $\tau_x \otimes I$, $\tau_x \otimes \tau_x$, $\tau_x \otimes \tau_x\tau_z$, 
$\tau_x\tau_z \otimes I$, $\tau_x\tau_z \otimes \tau_x$, $\tau_x\tau_z \otimes \tau_x\tau_z$\}.
By using appropriate transformations from $\mathcal{D}^n$, some of the vectors $\boldsymbol{S}$
may be transformed into bipolar vectors which correspond to truth tables of Boolean functions.
We use ANFT to recover all such functions, and this set will comprise all members 
of the LC orbit of $f$.
\end{example}

\begin{definition}
Let
\[
I =
\begin{pmatrix}
1 & 0 \\
0 & 1
\end{pmatrix},\quad
H = {\frac{1}{\sqrt{2}}}
\begin{pmatrix}
1 & 1 \\
1 & -1
\end{pmatrix},\quad
N = {\frac{1}{\sqrt{2}}}
\begin{pmatrix}
1 & i \\
1 & -i
\end{pmatrix},
\]
where $i^2 = -1$, be the Identity, \emph{Hadamard}, and \emph{Negahadamard} transformations, respectively.
\end{definition}

Let $f$ be a Boolean function on
$n$ variables and $\boldsymbol{s} = 2^{-\frac{n}{2}}(-1)^{f(\boldsymbol{x})}$
be a vector of length $2^n$.
The spectrum with respect to the $\{I,H,N\}^n$ transform is the set of $3^n$
transforms $\boldsymbol{S} = U\boldsymbol{s}$, for all $U \in \{I,H,N\}^n$.

\begin{theorem}[Riera and Parker~\cite{B51}]\label{thm:ihnlc}
It can be shown that $N \simeq \tau_x$ and $H \simeq \tau_x\tau_z$.
It follows that the LC orbit of a Boolean function can be generated 
by the transform set $\{I,H,N\}^n$.
\end{theorem}

There is a good reason for choosing the transform set $\{I,H,N\}^n$ instead if $\{I,\tau_x,\tau_x\tau_z\}^n$.
We have already seen that the Hadamard transformation is used in both the theory of quantum computing and
in the analysis of Boolean functions. 
The $\{H\}^n$ transform, which is simply the Walsh-Hadamard transform as defined in
\eqref{eq:whtmatrix}, is an $n$-dimensional 2-point \emph{discrete Fourier transform}.
We know from \autoref{thm:wiener} that the $\{H\}^n$ spectrum is the ``dual'' of the
periodic autocorrelation, defined in \autoref{def:periodic}.
The $\{I,H\}^n$ spectrum can be shown to be the ``dual'' of the fixed-periodic autocorrelation,
defined in \autoref{def:fixedperiodic}. The Hadamard and Negahadamard 
transformations together define a two times oversampled discrete Fourier transform,
and the $\{H,N\}^n$ spectrum is therefore the ``dual'' of the aperiodic autocorrelation, defined
in \autoref{def:aperiodic}.
Finally, the $\{I,H,N\}^n$ spectrum is the ``dual'' of the fixed-aperiodic autocorrelation, defined in
\autoref{def:fixedaperiodic}, which is related to the aperiodic propagation criterion, as shown 
in \autoref{def:apc}.

\begin{definition}
Let $f$ be a function, $f: \mathbb{Z}_2^n \to \mathbb{Z}_m$, where $m \in 2\mathbb{N}$, i.e.,
$m$ is some even positive integer.
Let $u = e^{\frac{2\pi i}{m}}$, i.e., $u$ is a complex $m$th root of 1.
The vector $\boldsymbol{s} = u^{f(\boldsymbol{x})}$ is \emph{Boolean flat} if it is flat
and there exists some $F \in \mathcal{D}^n$ such that 
$F\boldsymbol{s} = u^{\frac{m}{2} f'(\boldsymbol{x})} = (-1)^{f'(\boldsymbol{x})}$,
where $f'$ is a Boolean function. It follows that $f$ and $f'$ may be interpreted as equivalent functions.
\end{definition}

\begin{definition}
For $k \in \mathbb{N}$, we define the ``$\mathbb{Z}_k$-phase-flip'' matrix
\[
\rho_k = \begin{pmatrix} 1&0\\ 0&e^{\frac{2\pi i}{k}} \end{pmatrix},
\]
\end{definition}

\begin{proposition}\label{prop:affineflat}
Let $\boldsymbol{s} = u^{f(\boldsymbol{x})}$ where 
$u = e^{\frac{2\pi i}{m}}$, $f: \mathbb{Z}_2^n \to \mathbb{Z}_{m}$, and $m \in 2\mathbb{N}$.
The vector $\boldsymbol{s}$ is Boolean flat if we can write 
$\boldsymbol{s} = u^{\frac{m}{2} f'(\boldsymbol{x}) + h(\boldsymbol{x})}$, 
where $f'$ is a Boolean function and $h$ is an affine function from $\mathbb{Z}_2^n$ to $\mathbb{Z}_m$.
\end{proposition}
\begin{proof}
Let $\boldsymbol{s}'$ be the transform
\[
\boldsymbol{s}' = \left( (\rho_m^b)^{(k)} \bigotimes_{l \ne k} I^{(l)} \right) \boldsymbol{s},
\]
where $b \in \mathbb{Z}_m$. It can then be verified that
$\boldsymbol{s}' = u^{f(\boldsymbol{x}) + bx_k}$.
Let $\boldsymbol{s}'' = u^c\boldsymbol{s}$, where $u = e^{\frac{2\pi i}{m}}$ and
$c \in \mathbb{Z}_m$. It can be verified that $\boldsymbol{s}'' = u^{f(\boldsymbol{x}) + c}$.
Thus all linear and constant terms of $h(\boldsymbol{x})$ can be removed by
a transformation $F \in \mathcal{D}^n$ of the form
\begin{equation}
F = u^c \bigotimes_{k=0}^{n-1} \rho_m^{b_k},
\end{equation}
where $u = e^{\frac{2\pi i}{m}}$ and $b_k, c \in \mathbb{Z}_m$.
\end{proof}

\begin{proposition}[Riera and Parker~\cite{B51}]\label{prop:vnvn}
We can perform local complementation on the variable $x_k$ of the Boolean function $f$ with the transformation
\begin{equation}\label{eq:lcui}
U_k = N^{(k)} \bigotimes_{l \ne k} I^{(l)}.
\end{equation}
If $\boldsymbol{s} = 2^{-\frac{n}{2}}(-1)^{f(\boldsymbol{x})}$, then
$\boldsymbol{S} = U_k\boldsymbol{s}$ will be Boolean flat
with coefficients from the set $\{w^a \mid w^4=-1, a \in \mathbb{Z}_8\}$.
Furthermore, $\boldsymbol{S}$ can always be expressed as
$\boldsymbol{S} = w^{4 f'(\boldsymbol{x}) + 2 h(\boldsymbol{x}) + c}$,
where  $f'$ is a Boolean function, $h$ is an affine function from $\mathbb{Z}_2^n$ to $\mathbb{Z}_4$,
and $c \in \mathbb{Z}_8$.
\end{proposition}

It follows from \autoref{prop:affineflat} that the vector $\boldsymbol{S} = U_m\boldsymbol{s}$,
where $U_m$ is of the form given by \eqref{eq:lcui},
can be turned into a bipolar vector by some transformation of the form
\begin{equation}\label{eq:z4form}
F = w^a \bigotimes_{k=0}^{n-1} \rho_4^{b_k},
\end{equation}
where $w^4 = -1$, $a \in \mathbb{Z}_8$, and $b_k \in \mathbb{Z}_4$.
We can then implement a sequence of LC operations by transformations $U_m$ of the form
given by \eqref{eq:lcui}, with appropriate transformations $F$ of the form given by \eqref{eq:z4form}
performed after each $U_m$.

\begin{lemma}\label{lem:IHNBP}
Define the product $U = U_k U_{k-1}\cdots U_1$, where $k \in \mathbb{N}$ and 
$U_i \in \{I,H,N,\sigma_x, \rho_4\}$.
Then $U = U'_3 U'_2 U'_1$, where $U'_1 \in \{I,H,N\}$, $U'_2 \in \{I,\sigma_x\}$,
and $U'_3 = w^a \rho_4^b$, 
where $w^4 = -1$, $a \in \mathbb{Z}_8$, and $b \in \mathbb{Z}_4$.
\end{lemma}
\begin{proof}
The result follows from the identities,
\begin{equation}\label{eq:identities}
\begin{aligned}
HH &= I, \\
HN &= \rho_4, \\
H \rho_4 &= N, \\
H \sigma_x &= \sigma_z H  = \rho_4^2 H, \\
NH &= w \rho_4^3 \sigma_x N, \\
NN &= w \rho_4^3 H, \\
N\rho_4 &= \sigma_x H, \\
N \sigma_x &= -\sigma_y N = i\rho_4^2\sigma_x N, \\
\sigma_x \rho_4 &= i \rho_4^3 \sigma_x, \\
\sigma_x \sigma_x &= I.
\end{aligned}
\end{equation}
\end{proof}

\begin{theorem}\label{prop:ihnbp}
Given any sequence of transformations $U = U_k U_{k-1} \cdots U_1$, where $k \in \mathbb{N}$ and
$U_i \in \{I,H,N,\sigma_x, \rho_4\}^n$. Then 
$U = U'_3 U'_2 U'_1$, where $U'_1 \in \{I,H,N\}^n$, $U'_2 \in \{I,\sigma_x\}^n$,
and $U'_3 = w^a \bigotimes_{k=0}^{n-1} \rho_4^{b_k}$,
where $w^4 = -1$, $a \in \mathbb{Z}_8$, and $b_k \in \mathbb{Z}_4$.
\end{theorem}
\begin{proof}
This result is a simple generalisation of \autoref{lem:IHNBP} to a transformation on $n$ qubits.
\end{proof}

\begin{corollary}
Any sequence of LC operations is equivalent to a single
transformation from $\{I,H,N\}^n$ followed by some transformation from $\mathcal{D}^n$ of the form
\begin{equation}
F = w^a \Biggl( \bigotimes_{k=0}^{n-1} \rho_4^{b_k} \Biggr) \Biggl( \bigotimes_{l=0}^{n-1} \sigma_x^{c_l} \Biggr),
\end{equation}
where $w^4 = -1$, $a \in \mathbb{Z}_8$, $b_k \in \mathbb{Z}_4$, and $c_l \in \mathbb{Z}_2$.
\end{corollary}

\begin{definition}
A function $f: \mathbb{Z}_2^n \to \mathbb{Z}_m$, where $m \in \mathbb{N}$,
can be described by the truth table $\boldsymbol{t}$ or the algebraic normal form
$\boldsymbol{a}$, both being vectors of length $2^n$ with coefficients from $\mathbb{Z}_m$.
The \emph{generalised algebraic normal form transformation}, ANFT$_m$, can
be used to transform $\boldsymbol{t}$ into $\boldsymbol{a}$ and vice versa.
To transform a truth table into ANF, we use
\begin{equation}\label{eq:anft8ta}
\boldsymbol{a} = \left( \bigotimes_{k=0}^{n-1} 
\begin{pmatrix} 1 & 0 \\ m-1 & 1 \end{pmatrix} \right) \boldsymbol{t} \pmod{m}.
\end{equation}
Transformation from ANF to truth table is performed by
\begin{equation}
\boldsymbol{t} = \left( \bigotimes_{k=0}^{n-1} 
\begin{pmatrix} 1 & 0 \\ 1 & 1 \end{pmatrix} \right) \boldsymbol{a} \pmod{m}.
\end{equation}
\end{definition}

If $\boldsymbol{s} = 2^{-\frac{n}{2}}(-1)^{f(\boldsymbol{x})}$ and
$\boldsymbol{S} = U\boldsymbol{s}$ is flat
for some choice of $U \in \{I,H,N\}^n$, then we can 
recover the function $\boldsymbol{S} = w^{g(\boldsymbol{x})}$
by using ANFT$_8$. By separating the monomials of $g(\boldsymbol{x})$
that are divisible by 4 and those that are not, we can write
$\boldsymbol{S} = w^{4f'(\boldsymbol{x}) + h(\boldsymbol{x})}$, where $f'$ is
a Boolean function, $h$ is any function from $\mathbb{Z}_2^n$ to $\mathbb{Z}_8$,
and $w^4 = -1$.
If $h(\boldsymbol{x})$ is an affine function, it can be shown that 
the coefficients of all its linear terms must be divisible by two.
We can then eliminate $h(\boldsymbol{x})$ by post-multiplication
with a transformation $F$ of the form given by \eqref{eq:z4form}, and
get $F\boldsymbol{S} = (-1)^{f'(\boldsymbol{x})}$. 

\begin{definition}
Given a Boolean function $f$, find the set of all Boolean flat
$\{I,H,N\}^n$ transforms of $f$, and recover the corresponding Boolean
functions. The set of all distinct Boolean functions
recovered from the set of all $\{I,H,N\}^n$ transforms of $f$, including
$f$ itself, is called the \emph{$\{I,H,N\}^n$ orbit} of $f$.
\end{definition}

\begin{corollary}
By \autoref{thm:ihnlc}, if $f$ is quadratic then the $\{I,H,N\}^n$ orbit is the LC orbit.
\end{corollary}

\begin{theorem}
It can be shown~\cite{B15} that if two Boolean functions are in the same
$\{I,H,N\}^n$ orbit, then they will have the same APC distance.
\end{theorem}

\begin{example}
Consider the Boolean function $f(\boldsymbol{x}) = x_0x_1 + x_0x_2$.
The bipolar vector describing the corresponding quantum state, 
if we ignore the normalisation factor $2^{-\frac{3}{2}}$, is
\[
\boldsymbol{s} = (-1)^{f(\boldsymbol{x})} = (1,1,1,-1,1,-1,1,1)^T.
\]
We apply the transformation $U = N \otimes I \otimes I$ and get the
result
\[
\boldsymbol{S} = U\boldsymbol{s} = (w,w^7,w^7,w,w^7,w,w,w^7)^T,
\]
where w$^4=-1$. We observe that $|S_k| = 1$ for all $k$, which means that $\boldsymbol{S}$
is flat and can be expressed as $w^{g(x)}$, where the function $g: \mathbb{Z}_2^3 \to \mathbb{Z}_8$
has truth table $(1,7,7,1,7,1,1,7)^T$. Using ANFT$_8$, as described by \eqref{eq:anft8ta},
we find that the ANF of $g$ is $(1,6,6,4,6,4,4,0)^T$, and we can therefore write
\[
\boldsymbol{S} =  w^{4(x_0x_1+x_0x_2+x_1x_2)+(6x_0+6x_1+6x_2+1)}.
\]
We observe that $\boldsymbol{S}$ is Boolean flat, since the terms that are not divisible by 4 
are all linear or constant. We also see that all linear terms are divisible by 2,
which, by \autoref{prop:vnvn}, we should expect. We can eliminate all linear and constant terms
by applying a transformation as described by \eqref{eq:z4form},
in this case we must use the transformation
\[
F = w^7 \begin{pmatrix}1 & 0 \\ 0 & i\end{pmatrix} \otimes 
    \begin{pmatrix}1 & 0 \\ 0 & i\end{pmatrix} \otimes
    \begin{pmatrix}1 & 0 \\ 0 & i\end{pmatrix}.
\]
We get the result
\[
F\boldsymbol{S} = (-1)^{x_0x_1 + x_0x_2 + x_1x_2}.
\]
We can conclude that $f(\boldsymbol{x}) = x_0x_1 + x_0x_2$ and
$f'(\boldsymbol{x}) = x_0x_1 + x_0x_2 + x_1x_2$ are in the same $\{I,H,N\}^n$ orbit, and since
they are quadratic functions, the same LC orbit. This can be verified
by applying the LC operation to the vertex corresponding to the
variable $x_0$ in the graph representation of either function.
\end{example}

\section{Orbits of Boolean Functions}

For quadratic Boolean functions, LC operations on the associated graphs
generate the orbits of equivalent functions. It is unknown whether there exists a similar 
operation on hypergraphs that generates the orbits of equivalent non-quadratic
functions, but the $\{I,H,N\}^n$ orbit can be generated for Boolean functions of any degree.
For non-quadratic functions, there are also other symmetries that must be considered.

We have already seen that permuting the variables of a Boolean function 
gives an equivalent function,
\begin{equation}
f(x_0, x_1, \ldots, x_{n-1}) \simeq f(x_{\pi(0)}, x_{\pi(1)}, \ldots, x_{\pi(n-1)}).
\end{equation}
We can determine whether two functions are equivalent via the 
\emph{permutation symmetry} by checking if their corresponding graphs are
isomorphic. For non-quadratic functions we must check for hypergraph isomorphism.
The package \texttt{nauty} can be used in both cases, as described in \autoref{l19}.

Adding an affine offset to a Boolean function, i.e., adding any linear or constant terms,
will not change the APC distance of the associated quantum state.
We call this the \emph{affine symmetry},
\begin{equation}
f(\boldsymbol{x}) \simeq f(\boldsymbol{x}) + \boldsymbol{b} \cdot \boldsymbol{x} + c,
\quad \boldsymbol{b} \in \mathbb{Z}_2^n, c \in \mathbb{Z}_2.
\end{equation}
Note that the addition of linear terms corresponds to the phase-flip operation 
defined in \eqref{eq:phaseflip}.
Similarly, all bit-flips on a Boolean function will produce an equivalent function.
This is the \emph{bit-flip symmetry},
\begin{equation}
f(\boldsymbol{x}) \simeq f(\boldsymbol{x} + \boldsymbol{a}),
\quad \boldsymbol{a} \in \mathbb{Z}_2^n.
\end{equation}
The affine symmetry may be taken into consideration by simply deleting
any linear or constant terms from a Boolean function.
We have implicitly done this for quadratic functions by only considering
simple graphs.

\begin{remark}
A bit-flip on a Boolean function of degree $m$ can change the coefficients
of terms with degree at most $m-1$.
\end{remark}

Bit-flips on quadratic functions can only produce linear terms, which is also what 
phase-flips do. Therefore the affine symmetry and bit-flip symmetry coincide for quadratic functions.
For non-quadratic functions, bit-flips can produce nonlinear terms, and the bit-flip
symmetry may therefore generate non-trivial orbits of functions.

\begin{definition}
Let $\mathcal{B}_n$ be the set of all non-isomorphic connected Boolean functions of $n$ variables
with no linear or constant terms.
(Two Boolean functions are isomorphic or connected if their corresponding hypergraphs
are isomorphic or connected.)
Given a Boolean function $f \in \mathcal{B}_n$, then the set of all distinct non-isomorphic
Boolean functions recovered from the set of $\{I, \sigma_x\}^n$ transforms of $f$,
with linear and constant terms removed, is called the $\{I, \sigma_x\}^n$ orbit of $f$.
$\mathcal{B}_n$ can be partitioned into a set of disjoint $\{I, \sigma_x\}^n$ orbits,
called $\mathcal{O}_{1,n}$.
$|\mathcal{O}_{1,n}|$ is then the number of inequivalent connected Boolean functions
of $n$ variables with respect to the permutation, affine, and bit-flip symmetries.
\end{definition}

We have generated all orbits in $\mathcal{O}_{1,n}$ for $n \le 5$. 
For 6 variables, we could only generate orbits for all functions of degree up to 3.
For these values of~$n$, \autoref{tab:nonquad} gives the number of orbits, 
while \autoref{tab:bf5} and \autoref{tab:bf6} show the number of 
orbits by APC distance and degree for $n=5$ and $n=6$. A 
\href{http://www.ii.uib.no/~larsed/nonquad/}{database} containing one representative 
of each orbit is also available~\cite{B13}.

\begin{table}
\centering
\caption{Number of Orbits of Boolean Functions of $n$ Variables}
\label{tab:nonquad}
\begin{tabular}{ccrr}
\toprule
$n$ & Degrees & $|\mathcal{O}_{1,n}|$ & $|\mathcal{O}_{2,n}|$ \\
\midrule
3 & All &      3 &      2 \\
4 & All &     33 &     29 \\
5 & All & 22,400 & 22,014 \\
6 & $\le 3$ & 850,705& 746,326 \\
\bottomrule
\end{tabular}
\end{table}

Orbits of non-quadratic functions with respect to $\{I,H,N\}^n$
can not be generated without also considering bit-flips, since
it follows from the identities in~\eqref{eq:identities} that
a sequence of transformations from the set $\{I,H,N\}$ may induce bit-flips.
For instance, applying first $H$ and then $N$ to one variable
is equivalent to performing a bit-flip on the variable.
The set $M$ of Boolean functions found within the set of $\{I,H,N\}^n$ transforms
of the non-quadratic Boolean function $f$ will therefore contain some members of
the $\{I,\sigma_x\}^n$ orbit of $f$. 
If $f'$ is found in $M$, then $\{I,H,N\}^n$ transforms of $f'$ may induce bit-flips 
that produce functions not found in $M$.

\begin{definition}
Given a non-quadratic Boolean function $f$, it follows from \autoref{prop:ihnbp}
that a closed orbit of functions may 
be generated by first finding the set $M$ of all distinct Boolean functions recovered from 
$\{I,H,N\}^n$ transforms of $f$. We then generate $M'$, 
containing all functions in the $\{I,\sigma_x\}^n$ orbits of every function in $M$.
The set of non-isomorphic functions in $M'$, with linear and constant terms removed,
is the $\{I,\sigma_x\}^n\{I,H,N\}^n$ orbit of $f$.
$\mathcal{B}_n$ can be partitioned into a set of disjoint $\{I,\sigma_x\}^n\{I,H,N\}^n$ orbits,
called $\mathcal{O}_{2,n}$.
$|\mathcal{O}_{2,n}|$ is then the number of inequivalent connected Boolean functions
of $n$ variables with respect to the permutation, affine, bit-flip, and $\{I,H,N\}$ symmetries.
\end{definition}

\autoref{alg:nonquadorbit} can be used to generate all members of the 
$\{I,\sigma_x\}^n\{I,H,N\}^n$ orbit of the Boolean function $f$. It first finds
all inequivalent Boolean functions corresponding to $\{I,H,N\}^n$ transforms of $f$, and then
generates the $\{I,\sigma_x\}^n$ orbits of all these functions. When we only want 
to generate the $\{I,\sigma_x\}^n$ orbit of a function, we use the same algorithm but
skip the first loop. When we want to count the orbits in $\mathcal{O}_{1,n}$ or $\mathcal{O}_{2,n}$,
or find a member of each orbit, we use an approach similar to the 
one used in \autoref{alg:lccanonise}. By picking a canonical representative of 
each orbit, we canonise all Boolean functions in $\mathcal{B}_n$ and remove all duplicates.
This can be done much faster by using a generalisation of \autoref{def:extension},
i.e., by generating a set of functions of $n$ variables by extending
one member from each orbit of functions of $n-1$ variables, and then using 
this set of functions as input to the above-mentioned canonisation algorithm.
A function is extended by adding a variable and all combinations
of monomials including this variable.

We use \autoref{alg:butterfly} to find $\{I,H,N\}^n$ transforms or 
$\{I,\sigma_x\}^n$ transforms of a Boolean function. Note that we can skip 
iterations of \autoref{alg:butterfly} where the identity transformation is applied.
It is possible to order any set of transformations, $\mathcal{T}^n$,
such that only one factor of each $n$-fold tensor product changes from
one transformation to the next.
By using some additional temporary transformation matrices, it is
then possible to find all $\mathcal{T}^n$ transforms of a function
by only using $|\mathcal{T}|^n$ single iterations of \autoref{alg:butterfly}.
A \emph{Gray code sequence} gives an ordering of transformations that
makes this possible.

\begin{definition}
Given a sequence $\boldsymbol{s}$ of length $n$, let $\boldsymbol{s}' = \rev(\boldsymbol{s})$ 
be the sequence $\boldsymbol{s}$ in reversed order, i.e., $s'_i = s_{n-1-i}$. 
We also define $\boldsymbol{s}' = \pre(c,\boldsymbol{s})$, where $c \in \mathbb{Z}_n$. If
$c$ is even, then $\boldsymbol{s}'$ consists of all elements from $\boldsymbol{s}$ with
the prefix $c$ added, e.g., $\pre(0,(0,1,2)) = (00,01,02)$. If $c$ is odd, then
$\boldsymbol{s}'$ consists of all elements from $\rev(\boldsymbol{s})$ with
the prefix $c$ added, e.g., $\pre(1,(0,1,2)) = (12,11,10)$.
Let $\conc(\{\boldsymbol{s}_1, \boldsymbol{s}_2, \ldots, \boldsymbol{s}_m\})$ be
the concatenation of all sequences $\boldsymbol{s}_i$ to a single sequence.
\end{definition}

\begin{definition}
The $n$-ary \emph{Gray code sequence} of order $r$, denoted $\boldsymbol{q}_{n,r}$, contains all $n^r$
elements from $\mathbb{Z}_n^r$. The sequence has the property that any
two consecutive elements have the same symbol in $r-1$ positions, so that only one
symbol needs to be changed to jump from one element of $\boldsymbol{q}_{n,r}$ to the next.
A Gray code sequence can be constructed recursively,
\begin{align}
\boldsymbol{q}_{n,1} &= (0,1,\ldots,n-1) \\
\boldsymbol{q}_{n,r} &= \conc\left(\{\pre(c, \boldsymbol{q}_{n,r-1}) \mid c \in \mathbb{Z}_n\}\right).
\end{align}
\end{definition}

\begin{algorithm}[p]
\caption{Generating an $\{I,\sigma_x\}^n\{I,H,N\}^n$ Orbit}\label{alg:nonquadorbit} 
\begin{algorithmic}
\State \hskip -6pt
\begin{tabular}{ll}
\textbf{Input} 
& $f$: a Boolean function\\
& $n$: the number of variables of $f$\\
\textbf{Output} 
& $\boldsymbol{L}$: the set of all Boolean functions in the orbit of $f$
\end{tabular}
\Statex
\Procedure{GenerateOrbit}{$f$, $n$}
  \State $\boldsymbol{s} \gets $ bipolar truth table of $f$
  \ForAll{transformations $U \in \{I,H,N\}^n$} \Comment use Gray code ordering
    \State $\boldsymbol{s}' \gets U\boldsymbol{s}$ \Comment use single iteration of \autoref{alg:butterfly}
    \State $\boldsymbol{s}'' \gets \mathbb{Z}_8$-ANF corresponding to $\boldsymbol{s}'$ \Comment use ANFT$_8$
    \If{$\boldsymbol{s}''$ is Boolean flat}
      \State $f' \gets $ Boolean function corresponding to $\boldsymbol{s}''$
      \State Remove all linear and constant terms from $f'$
      \State $f'' \gets $ \textsc{NautyCanonise}($f'$) \Comment use hypergraph canonisation
      \If{$f'' \not\in \boldsymbol{L}$}
        \State \textsc{Add}($\boldsymbol{L}$, $f''$)
      \EndIf
    \EndIf
  \EndFor
  \ForAll{functions $g \in \boldsymbol{L}$}
    \State $\boldsymbol{s} \gets $ bipolar truth table of $g$
    \ForAll{transformations $U \in \{I,\sigma_x\}^n$} \Comment use Gray code ordering
      \State $\boldsymbol{s}' \gets U\boldsymbol{s}$ \Comment use single iteration of \autoref{alg:butterfly}
      \State $g' \gets $ ANF corresponding to $\boldsymbol{s}'$
      \State Remove all linear and constant terms from $g'$
      \State $g'' \gets $ \textsc{NautyCanonise}($g'$) \Comment use hypergraph canonisation
      \If{$g'' \not\in \boldsymbol{L}$}
        \State \textsc{Add}($\boldsymbol{L}$, $g''$)
      \EndIf
    \EndFor
  \EndFor
  \State \textbf{return} $\boldsymbol{L}$
\EndProcedure
\end{algorithmic} 
\end{algorithm}

\begin{example}
We want to find the 9 $\{I,H,N\}^2$ transforms of some Boolean function $f$ of 2 variables.
We generate the Gray code sequence $\boldsymbol{q}_{3,2} = (00$, $01$, $02$, $12$, $11$, $10$, $20$, $21$, $22)$.
By the mapping $0 \mapsto I$, $1 \mapsto H$, and $2 \mapsto N$, we get the
following sequence of transformations, $(I \otimes I$, $ I \otimes H$, $ I \otimes N$, $ H \otimes N$, $ 
H \otimes H$, $ H \otimes I$, $ N \otimes I$, $ N \otimes H$, $ N \otimes N)$.
To jump from $H$ to $N$, $N$ to $H$, and $N$ to $I$, we need the 
temporary transform matrices $NH^{-1}$, $HN^{-1}$, and $N^{-1}$.
(Note that $H^{-1} = H$.)
The sequence of transformations we will use is then
$(I \otimes I$, $ I \otimes H$, $ I \otimes NH^{-1}$, $ H \otimes I$, $ I \otimes H$, $ I \otimes H$, 
$N \otimes I$, $ I \otimes H$, $ I \otimes NH^{-1})$, where 
only the first transformation is applied to the function $f$, and the
subsequent transformations are applied to the transform from the previous step, so that the
cumulative effect gives the desired result.
\end{example}

Using \autoref{alg:nonquadorbit}, 
we have generated all orbits in $\mathcal{O}_{2,n}$
for $n \le 5$, and also the orbits of functions of 6 variables with degree up to 3.
For these values of~$n$,
\autoref{tab:nonquad} gives the number of orbits, while \autoref{tab:ihnbf5} 
and \autoref{tab:ihnbf6} show the number of orbits by APC distance and degree.
A \href{http://www.ii.uib.no/~larsed/nonquad/}{database} containing 
one representative of each orbit is also available~\cite{B13}.
The numbers of inequivalent functions are not reduced considerably when the 
$\{I,H,N\}$ symmetry is considered in addition to bit-flips, but the difference in orbit size
is larger for functions with low degree and high APC distance.
Recall that for quadratic functions, bit-flips only generate affine offsets.
For functions of high degree, bit-flips account for almost all symmetries.

\begin{remark}
There are no non-quadratic functions of 5 or less variables with
APC distance higher than 2, but there are 11
functions of 6 variables, belonging to different orbits in $\mathcal{O}_{2,6}$,
with degree 3 and APC distance 3.
\end{remark}

A representative of each of the 11 inequivalent non-quadratic $[[6,0,3]]$ quantum codes
is listed in \autoref{tab:nonquadcodes}. The first function in this table
corresponds to the hypergraph shown in \autoref{fig:hyper6}.
The table also gives examples of codes of higher length found
by non-exhaustive searches of functions with a limited number
of non-quadratic terms, or by adding several higher degree terms to
strong quadratic codes.
The functions in \autoref{tab:nonquadcodes} all belong to different 
$\{I,\sigma_x\}^n\{I,H,N\}^n$ orbits,
and the representative from each orbit with fewest monomials is listed.
The table also gives the value of PAR$_{IHN}$ of each function.
PAR$_{IHN}$ will be introduced in the next chapter.

\clearpage

\begin{table}
\centering
\caption{Number of Orbits in $\mathcal{O}_{1,5}$ with APC Distance $d$ and Degree $\delta$}
\label{tab:bf5}
\begin{tabular}{crrrrr}
\toprule
& \multicolumn{5}{c}{$\delta$} \\
\cmidrule(l){2-6}
$d$    & 2 & 3 & 4 & 5 & All \\
\midrule
1 &    & 625 & 10,756 & 10,688 & 22,069 \\
2 & 18 & 109 &    201 &        &    328 \\
3 &  3 &     &        &        &      3 \\
\midrule
All & 21 & 734 & 10,957 & 10,688 & 22,400 \\
\bottomrule
\end{tabular}
\end{table}

\begin{table}
\centering
\caption{Number of Orbits in  $\mathcal{O}_{2,5}$ with APC Distance $d$ and Degree $\delta$}
\label{tab:ihnbf5}
\begin{tabular}{crrrrr}
\toprule
& \multicolumn{5}{c}{$\delta$} \\
\cmidrule(l){2-6}
$d$   & 2 & 3 & 4 & 5 & All \\
\midrule
1 &   & 505 & 10,570 & 10,688 & 21,763 \\
2 & 3 &  61 &    186 &        &    250 \\
3 & 1 &     &        &        &      1 \\
\midrule
All &  4 & 566 & 10,756 & 10,688 & 22,014 \\
\bottomrule
\end{tabular}
\end{table}

\begin{table}
\centering
\caption{Number of Orbits in  $\mathcal{O}_{1,6}$ with APC Distance $d$ and Degree $\delta$}
\label{tab:bf6}
\begin{tabular}{crrrrrr}
\toprule
& \multicolumn{6}{c}{$\delta$} \\
\cmidrule(l){2-7}
$d$    & 2 & 3 & 4 & 5 & 6 & 2 and 3 \\
\midrule
1 &    & 804,326 & ? & ? & ? & 804,326 \\
2 & 94 &  46,243 & ? & ? & ? &  46,337 \\
3 & 16 &      24 & ? & ? & ? &      40 \\
4 &  2 &         & ? & ? & ? &       2 \\
\midrule
All & 112 & 850,593 & ? & ? & ? & 850,705 \\
\bottomrule
\end{tabular}
\end{table}

\begin{table}
\centering
\caption{Number of Orbits in  $\mathcal{O}_{2,6}$ with APC Distance $d$ and Degree $\delta$}
\label{tab:ihnbf6}
\begin{tabular}{crrrrrr}
\toprule
& \multicolumn{6}{c}{$\delta$} \\
\cmidrule(l){2-7}
$d$   & 2 & 3 & 4 & 5 & 6 & 2 and 3 \\
\midrule
1 &   & 717,741 & ? & ? & ? & 717,741 \\
2 & 9 &  28,563 & ? & ? & ? &  28,572 \\
3 & 1 &      11 & ? & ? & ? &      12 \\
4 & 1 &         & ? & ? & ? &       1 \\
\midrule
All & 11 & 746,315 & ? & ? & ? & 746,326 \\
\bottomrule
\end{tabular}
\end{table}

\begin{table}
\centering
\caption{Boolean Functions of $n$~Variables with Degree~$\delta$, 
         APC Distance~$d$, and PAR$_{IHN}$~$p$\label{tab:nonquadcodes}}
\resizebox{\linewidth}{!}{
\begin{tabular}{ccccl}
\toprule
$n$ & $\delta$ & $d$ & $p$ & Function \\
\midrule
6 & 3 & 3 & 8   & $012,03,04,13,15,24,25$\\
6 & 3 & 3 & 4.5 & $012,03,05,14,15,23,24,25,34$\\
6 & 3 & 3 & 4.5 & $023,012,04,05,13,15,23,24,25,34$\\
6 & 3 & 3 & 8   & $123,124,125,01,02,14,25,34,35,45$\\
6 & 3 & 3 & 4.5 & $012,013,03,04,13,15,24,25,34,35,45$\\
6 & 3 & 3 & 4.5 & $012,013,014,03,05,14,15,23,24,25,34$\\
6 & 3 & 3 & 4.5 & $012,014,024,123,134,234,03,13,15,24,25,34,45$\\
6 & 3 & 3 & 8   & $015,012,013,014,03,05,14,15,23,24,25,34,35,45$\\
6 & 3 & 3 & 8   & $025,245,012,124,023,234,04,05,13,15,23,24,35,45$\\
6 & 3 & 3 & 4.5 & $245,235,145,135,024,023,014,013,02,05,14,15,23,34,35,45$\\
6 & 3 & 3 & 8   & $125,145,135,245,235,012,014,013,024,023,05,13,15,24,25,34$\\
\midrule
7 & 3 & 3 &  8    & $012,03,05,14,16,25,26,34$\\
7 & 3 & 3 &  8    & $014,02,05,13,15,26,36,45,46$\\
7 & 3 & 3 &  8    & $014,02,05,13,16,26,35,45,46$\\
7 & 3 & 3 &  8    & $015,02,06,13,16,25,35,45,46$\\
7 & 3 & 3 &  8    & $012,03,06,14,16,25,26,34,35,45$\\
7 & 3 & 3 &  9    & $012,04,05,14,16,25,26,34,35,36$\\
7 & 3 & 3 &  8    & $013,04,06,15,16,23,26,34,35,45$\\
7 & 3 & 3 &  9    & $013,04,06,15,16,24,25,34,35,36$\\
7 & 3 & 3 & 16    & $456,04,05,14,16,25,26,34,35,36$\\
7 & 3 & 3 &  8    & $024,045,01,03,12,26,35,45,46,56$\\
7 & 3 & 3 &  8    & $012,03,06,14,15,24,25,26,35,36,46$\\
7 & 3 & 3 &  9    & $012,05,06,13,15,23,26,34,45,46,56$\\
7 & 3 & 3 &  8    & $012,05,06,13,15,24,26,34,35,46,56$\\
7 & 3 & 3 &  8    & $016,02,03,14,15,24,26,35,36,46,56$\\
7 & 3 & 3 &  8    & $012,03,05,14,15,23,24,26,36,46,56$\\
7 & 3 & 3 &  8    & $035,025,023,06,13,15,16,24,25,26,34,36,45$\\
7 & 3 & 3 &  8    & $013,023,123,012,04,05,14,16,25,26,34,35,36$\\
7 & 4 & 3 & 12.25 & $0123,04,05,14,16,25,26,34,35,36$\\
7 & 4 & 3 &  8    & $0126,03,05,14,15,23,24,26,36,46,56$\\
\midrule
8 & 3 & 4 & 16 & $012,04,05,07,14,16,17,25,26,27,34,35,36$\\
8 & 3 & 4 & 16 & $016,02,03,04,12,13,15,27,36,46,47,56,57,67$\\
8 & 3 & 4 &  8 & $067,01,02,03,14,16,25,27,36,37,46,47,56,57$\\
8 & 3 & 4 &  9 & $234,02,03,05,12,13,14,26,37,46,47,56,57,67$\\
8 & 3 & 4 &  8 & $024,046,01,03,05,12,14,25,27,34,36,46,47,56,57,67$\\
8 & 3 & 4 &  8 & $127,126,067,01,02,03,14,16,25,27,36,37,46,47,56,57$\\
8 & 3 & 4 &  8 & $456,056,246,026,245,025,04,07,15,16,17,23,26,27,35,36,45,46,47,57$\\
8 & 4 & 4 & 16 & $0123,04,05,06,14,15,17,24,26,27,35,36,37$\\
8 & 4 & 4 & 16 & $0167,02,03,04,12,13,15,26,37,46,47,56,57,67$\\
\midrule
9 & 3 & 4 &  8 & $016,02,04,07,13,15,18,23,26,36,45,47,58,67,68,78$\\
9 & 3 & 4 &  9 & $235,234,02,04,06,13,15,16,26,28,36,37,47,48,57,58,78$\\
9 & 3 & 4 &  8 & $037,137,034,134,01,07,08,14,18,23,25,28,36,37,45,46,57,58,67,68$\\
9 & 3 & 4 & 16 & $024,124,047,147,02,05,08,15,17,18,23,26,34,37,38,46,47,48,56,58,67$\\
9 & 3 & 4 & 16 & $234,023,124,012,03,06,08,14,17,18,23,24,25,37,38,46,48,56,57,58,67$\\
\midrule
10 & 3 & 4 & 16 & $016,02,03,08,12,13,17,26,39,45,46,48,57,59,67,68,79,89$\\
\bottomrule
\end{tabular}
}
\end{table}

\chapter{Peak-to-Average Power Ratio}\label{chap:par}

\section{Peaks and Independent Sets}

\begin{definition}
The \emph{peak-to-average power ratio} of the vector $\boldsymbol{s}$
with respect to the $\mathcal{T}^n$ transform, for some transform set
$\mathcal{T} = \{T_1, T_2, \ldots, T_m\}$, is defined as
\begin{equation}
\text{PAR}_{\mathcal{T}}(\boldsymbol{s}) = 
2^n \max_{\substack{ U \in \mathcal{T}^n \\ k \in \mathbb{Z}_{2^n} }} |S_k|^2,
\end{equation}
where $\boldsymbol{S} = U\boldsymbol{s}$. 
In other words, the PAR$_{\mathcal{T}}$ of $\boldsymbol{s}$ is the highest squared magnitude
of the $(2m)^n$ coefficients in the $\mathcal{T}^n$ spectrum of $\boldsymbol{s}$.
\end{definition}

The PAR$_{\mathcal{T}}$ of $\boldsymbol{s}$ can alternatively be
expressed in terms of the \emph{generalised nonlinearity}~\cite{B37}, 
\begin{equation}
\gamma(f) = 2^{\frac{n}{2} - 1} \left( 2^{\frac{n}{2}}
  - {\sqrt{\text{PAR}_{\mathcal{T}} \left( \boldsymbol{s} \right) }} \right),
\end{equation}
but we will use the PAR$_{\mathcal{T}}$ measure.
We will in particular study PAR$_{IHN}$, the peak-to-average power ratio with
respect to the $\{I,H,N\}^n$ transform~\cite{B16}.
If a vector, $\boldsymbol{s}$, has a completely flat $\{I,H,N\}^n$ spectrum,
which is impossible~\cite{B51}, then
PAR$_{IHN}(\boldsymbol{s}) = 1$. If $\boldsymbol{s} = 2^{-\frac{n}{2}}(1,1,\ldots,1,1)^T$ then
PAR$_{IHN}(\boldsymbol{s}) = 2^n$. A typical vector, $\boldsymbol{s}$, will have a PAR$_{IHN}(\boldsymbol{s})$
somewhere between these extremes.
For quadratic functions, PAR$_{IHN}$ will always be a power of~2~\cite{B52}.

Let $\boldsymbol{s} = 2^{-\frac{n}{2}}(-1)^{f(\boldsymbol{x})}$, as before. When we talk about the
PAR$_{IHN}$ of $f$, or its associated graph $G$, we mean PAR$_{IHN}(\boldsymbol{s})$. 
For cryptographic purposes, it is desirable to find
Boolean functions with high generalised nonlinearity and therefore low PAR$_{IHN}$.
PAR$_{IHN}$ is an invariant of the $\{I,\sigma_x\}^n\{I,H,N\}^n$ orbit and, for quadratic functions, the LC orbit.
We observe that Boolean functions from LC orbits associated
with self-dual additive codes over $\GF(4)$ with high
distance typically have low PAR$_{IHN}$. This is not surprising,
since the distance of a self-dual quantum code is equal to 
the APC distance of the associated quadratic Boolean function,
and APC is derived from the fixed-aperiodic autocorrelation which is, in turn,
the autocorrelation ``dual'' of the spectra with respect to $\{I,H,N\}^n$.
\autoref{QuadPARs} shows the value of PAR$_{IHN}$ for every LC orbit of codes
with length $n \le 12$.

\begin{table}
\centering
\caption{Number of LC Orbits with Length $n$ and PAR$_{IHN}$ $p$\label{QuadPARs}}
\begin{tabular}{rrrrrrrrrrrrr}
\toprule
 & \multicolumn{12}{c}{$n$} \\
\cmidrule(l){2-13}
$p$ & 1 & 2 & 3 & 4 & 5 & 6 & 7  & 8  & 9   & 10    & 11     & 12      \\
\midrule
2           & 1 & 1 &   &   &   &   &    &    &     &       &        &         \\     
4           &   &   & 1 & 1 & 1 & 1 &    &    &     &       &        &         \\ 
8           &   &   &   & 1 & 2 & 5 &  6 &  9 &   2 &     1 &        &         \\
16          &   &   &   &   & 1 & 4 & 14 & 52 & 156 &   624 &  3,184 &  12,323 \\
32          &   &   &   &   &   & 1 &  5 & 32 & 212 & 1,753 & 25,018 & 834,256 \\
64          &   &   &   &   &   &   &  1 &  7 &  60 &   639 & 10,500 & 380,722 \\
128         &   &   &   &   &   &   &    &  1 &   9 &   103 &  1,578 &  43,013 \\
256         &   &   &   &   &   &   &    &    &   1 &    11 &    163 &   3,488 \\
512         &   &   &   &   &   &   &    &    &     &     1 &     13 &     249 \\
1024        &   &   &   &   &   &   &    &    &     &       &      1 &      16 \\
2048        &   &   &   &   &   &   &    &    &     &       &        &       1 \\
\bottomrule
\end{tabular}
\end{table}

\begin{definition}
Let $\alpha(G)$ be the independence number of a graph $G$, i.e., the size
of the maximum independent set in $G$. Let $[G]$ be the set of all graphs in
the LC orbit of $G$. We then define 
\begin{equation}
\lambda(G) = \max_{K \in [G]} \alpha(K),
\end{equation}
i.e., $\lambda$ is the size of the maximum independent set over all graphs in the LC orbit of $G$.
\end{definition}
Consider as an example the Hexacode
which has two non-isomorphic graphs in its orbit, as seen in \autoref{l22}
on page~\pageref*{l22}.
It is evident that the independence number of each graph is 2, so $\lambda = 2$.
The values of $\lambda$ for all LC orbits for $n \le 12$ clearly show that
$\lambda$ and $d$, the distance of the associated self-dual additive code over $\GF(4)$, are related. 
LC orbits associated with codes of high distance typically have small values for $\lambda$.
\autoref{MaxIS} summarises this observation by listing the ranges of $\lambda$ observed for 
all LC orbits associated with codes of given lengths and distances. For instance, $[[12,0,2]]$ codes exist with
any value of $\lambda$ between 4 and 11, while $[[12,0,5]]$ and $[[12,0,6]]$ codes only exist
with $\lambda = 4$.

\begin{table}
\centering
\caption{Range of $\lambda$ for Codes of Length $n$ and Distance $d$\label{MaxIS}}
\begin{tabular}{cccccccccccc}
\toprule
 & \multicolumn{11}{c}{$n$} \\
\cmidrule(l){2-12}
$d$ & 2 & 3 & 4 & 5 & 6 & 7 & 8 & 9 & 10 & 11 & 12 \\ 
\midrule
2 & $\boldsymbol{1}$ & $\boldsymbol{2}$ & $\boldsymbol{2}$,3 & 3,4 & 3--5 & 3--6 & 3--7 & 4--8 & 4--9 & 4--10 & 4--11 \\
3 & & & & $\boldsymbol{2}$ & 3 & $\boldsymbol{3}$,4 & 3,4 & 3--5 & 4--6 & 4--7  & 4--8  \\
4 & & & & & $\boldsymbol{2}$ & & $\boldsymbol{3}$,4 & $\boldsymbol{3}$,4 & $\boldsymbol{3}$--5 & 4--6 & 4--7 \\
5 & & & & & & & & & & $\boldsymbol{4}$ & 4 \\
6 & & & & & & & & & & & $\boldsymbol{4}$ \\
\bottomrule
\end{tabular}
\end{table}

\begin{definition}
Let $\Lambda_n$ be the minimum value of $\lambda$ over all LC orbits of graphs on $n$ vertices, i.e,
\begin{equation}
\Lambda_n = \min_{[G] \in \mathcal{L}_n} \lambda(G).
\end{equation}
\end{definition}

From \autoref{MaxIS} we observe that
$\Lambda_n = 2$ for $n$ from 3 to 6, $\Lambda_n = 3$ for $n$ from 7 to 10, and
$\Lambda_n = 4$ when $n$ is 11 or 12. 

\begin{proposition}
$\Lambda_{n+1} \ge \Lambda_n$, i.e.,
$\Lambda_n$ is monotonically nondecreasing when $n$ is increasing.
\end{proposition}
\begin{proof}
Consider a graph $G=(V,E)$ with $n+1$ vertices.
Select a vertex $v$ and let $G'$ be the induced subgraph on
the $n$ vertices $V \backslash \{v\}$.
We generate the LC orbit of $G'$. The LC operations may add or remove
edges between $G'$ and $v$, but the presence of $v$ does not affect the 
LC orbit of $G'$.
The size of the largest independent set in the LC orbit of $G'$ is at least
$\Lambda_n$. This is also an independent set in the LC orbit of $G$, so
$\Lambda_{n+1} \ge \Lambda_n$.
\end{proof}

\begin{definition}
There is a number $r = R(m,n)$, called a \emph{Ramsey number}~\cite{B45},
such that it is guaranteed that all simple undirected graphs on at least $r$ vertices 
will have either an independent set of size $m$ or a clique of size $n$.
\end{definition}

\begin{proposition}\label{prop:ramsey}
If $r$ is the Ramsey number $R(k,k+1)$, then $\Lambda_n \ge k$ for $n \ge r$.
\end{proposition}
\begin{proof}
Consider a graph containing a clique of size $m$. 
An LC operation on any vertex in the clique will produce an independent set of size $m-1$.
Thus the maximum clique in an LC orbit, where the largest independent set has size $\lambda$,
can not be larger than $\lambda + 1$.
Since $r = R(k,k+1)$, it follows that all graphs on at least $r$ vertices must have $\lambda \ge k$,
and therefore that $\Lambda_n \ge k$ for $n \ge r$.
\end{proof}

For instance, the Ramsey number $R(3,4)$ is 9, so by \autoref{prop:ramsey}, 
$\Lambda_n \ge 3$ for $n \ge 9$, which means that no
LC orbit with at least 9 vertices can have $\lambda$ smaller than 3.
Similarly, $R(4,5)=25$, so $\Lambda_n \ge 4$ for $n \ge 25$.
For $n$ from 13 to 21, we have computed the values of $\lambda$ for 
some graphs corresponding to self-dual additive codes over $\GF(4)$ with high distance. 
This gives upper bounds on the value of $\Lambda_n$, 
as shown in \autoref{tab:lambda}. The bounds on $\Lambda_{13}$ and $\Lambda_{14}$ are tight, 
since $\Lambda_{12} = 4$ and $\Lambda_{n+1} \ge \Lambda_n$.
The bounds on $\Lambda_n$ given by \autoref{prop:ramsey} are very loose,
since we can see from \autoref{tab:lambda} that $\Lambda_n \ge 3$ for $n \ge 7$ and that
$\Lambda_n \ge 4$ for $n \ge 11$. The connection to Ramsey theory is
still interesting, and it may be possible to improve the bound.

\begin{table}
\centering
\caption{Values of $\Lambda_n$ for $n \le 14$ and Bounds on $\Lambda_n$ for $n \le 21$\label{tab:lambda}}
\begin{tabular}{cc}
\toprule
$n$ & $\Lambda_n$ \\
\midrule
2  & 1 \\
3  & 2 \\
4  & 2 \\
5  & 2 \\
6  & 2 \\
7  & 3 \\
8  & 3 \\
9  & 3 \\
10 & 3 \\
11 & 4 \\
12 & 4 \\
13 & 4 \\
14 & 4 \\
15 & 4--5 \\
16 & 4--5 \\
17 & 4--5 \\
18 & 4--6 \\
19 & 4--6 \\
20 & 4--6 \\
21 & 4--9 \\
\bottomrule
\end{tabular}
\end{table}

\begin{figure}
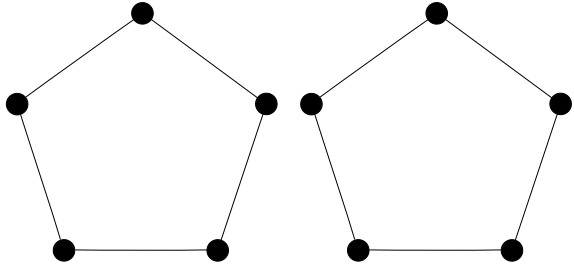

 \centering
 \includegraphics[width=.30\linewidth]{f8.eps}\hspace{5pt}
 \includegraphics[width=.30\linewidth]{f8.eps}
 \caption{The ``Double 5-Cycle'' Graph}
 \label{fivecircle}
\end{figure}

\begin{remark}
For $n=10$, there is a unique LC orbit that satisfies, optimally, $\lambda = 3$,
PAR$_{IHN} = 8$ and $d = 4$.
One of the graphs in this orbit is the \emph{graph complement} of the
``double 5-cycle'' graph, shown in \autoref{fivecircle}.
\end{remark}

\begin{proposition}[Parker and Rijmen~\cite{B40}]\label{th:ihindset}
Given a graph $G=(V,E)$ with a maximum independent set $A \subset V$, 
$|A| = \alpha(G)$. Let $\boldsymbol{s} = (-1)^{f(\boldsymbol{x})}$,
where $f$ is the Boolean function representation of $G$.
Let $U = \bigotimes_{k \in A} H^{(k)} \bigotimes_{k \not\in A} I^{(k)}$, i.e., the
transformation applying $H$ to variables corresponding to vertices $k \in A$
and $I$ to all other variables. Let $\boldsymbol{S} = U\boldsymbol{s}$. 
Then 
\begin{equation}
\max_{m \in \mathbb{Z}_{2^n}} |S_m|^2 = 2^{\alpha(G)}.
\end{equation}
\end{proposition}

Arratia~et~al.~\cite{B2,B3} introduced the \emph{interlace polynomial} $q(G,z)$
of a graph~$G$. Aigner and van der Holst~\cite{B1} later introduced the 
interlace polynomial $Q(G,z)$.
Riera and Parker~\cite{B52} showed that $q(G,z)$ is related
to the $\{I,H\}^n$ spectra of the quadratic Boolean function corresponding to $G$,
and that $Q(G,z)$ is related to the $\{I,H,N\}^n$ spectra.

\begin{proposition}[Aigner and van der Holst~\cite{B1}]\label{prop:recursiveQ}
The interlace polynomial $Q(G,z)$ can be defined recursively as
\begin{equation}
Q(G,z) = Q(G \backslash u, z) + P(G^u \backslash u, z) + P(((G^u)^v)^u \backslash u, z),
\end{equation}
where $G^u$ denotes the LC operation on vertex $u$ of $G$, and
$G \backslash u$ is the graph obtained by removing vertex $u$ and
all edges incident on $u$ from $G$.
\end{proposition}

\begin{proposition}[Riera and Parker~\cite{B52}]\label{th:degree}
Let $f$ be a quadratic Boolean function and $G$ its associated graph.
Then PAR$_{IHN}$ of $f$ is equal to $2^{\deg Q(G,z)}$, where
$\deg Q(G,z)$ is the degree of the interlace polynomial $Q(G,z)$.
\end{proposition}

Aigner and van der Holst~\cite{B1} proved that the degree of $q(G,z)$
is equal to the size of the maximum independent set in the 
\emph{switch-class} of $G$, which is the same as the $\{I,H\}^n$ orbit of $G$.
This proof can be extended to show that 
the degree of $Q(G,z)$ equals $\lambda$, the size of the maximum independent set 
in the $\{I,H,N\}^n$ orbit of $G$.

\begin{theorem}\label{th:indset}
If the maximum independent set over all graphs in the LC orbit $[G]$ has size $\lambda(G)$,
then all functions corresponding to graphs in the orbit will have PAR$_{IHN} = 2^{\lambda(G)}$.
\end{theorem}
\begin{proof}
Let us for brevity define $P(G) = \text{PAR}_{IHN}(\boldsymbol{s})$,
where $\boldsymbol{s} = 2^{-\frac{n}{2}}(-1)^{f(\boldsymbol{x})}$, and
$f(\boldsymbol{x})$ is the Boolean function representation of $G$.
From \autoref{th:ihindset} it follows that $P(G) \ge 2^{\lambda(G)}$.
Choose $K=(V,E) \in [G]$ with $\alpha(K) = \lambda(G)$. If $|V| = 1$ or 2, the
theorem is true. We will prove the theorem for $n > 2$ by induction on $|V|$.
We will show that $P(K) \le 2^{\alpha(K)}$, which is equivalent to saying that
$P(G) \le 2^{\lambda(G)}$.
It follows from \autoref{prop:recursiveQ} and \autoref{th:degree} that
$P(K) = \max \{P(K\backslash u)$, $P(K^u \backslash u)$, $P(((K^u)^v)^u \backslash u)\}$.
Assume, by induction hypothesis, that $P(K \backslash u) = 2^{\lambda(K \backslash u)}$.
Therefore, $P(K\backslash u) = 2^{\alpha(K\backslash u)}$ for some
$K\backslash u \in [K\backslash u]$. Note that
$K\backslash u \in [K\backslash u]$ implies $K \in [K]$.
It must then be true that
$\alpha(K\backslash u) \le \alpha(K) \le \alpha(K)$,
and it follows that $P(K\backslash u) \le 2^{\alpha(K)}$.
Similar arguments hold for $P(K^u \backslash u)$ and $P(((K^u)^v)^u \backslash u)$,
so $P(K) \le 2^{\alpha(K)}$.
\end{proof}

As an example, the Hexacode has $\lambda=2$ and therefore PAR$_{IHN} = 2^2 = 4$.
The vector containing the highest peak can be found by taking either of the two graphs in the LC orbit,
since both have independence number $\lambda$, and then by applying
the $H$ transformation to all variables corresponding to vertices in an independent set of
size $\lambda$ and the $I$ transformation to all other variables.

\begin{corollary}
Any quadratic Boolean function on $n$ or more variables must have PAR$_{IHN} \ge 2^{\Lambda_n}$.
\end{corollary}

\begin{definition}\label{def:parl}
PAR$_{\mathcal{U}}$ is the peak-to-average power ratio with respect to the infinite transform 
set $\mathcal{U}^n$, where $\mathcal{U}$ consists of matrices of the form
\[
U = \begin{pmatrix}
\cos \theta & \sin \theta e^{i\phi}\\
\sin \theta & -\cos \theta e^{i\phi}
\end{pmatrix},
\]
where $i^2 = -1$, and $\theta$ and $\phi$ can take any real values.
$\mathcal{U}$ comprises all $2 \times 2$ unitary transforms to within
a post-multiplication by a matrix from $\mathcal{D}$, 
the set of $2 \times 2$ diagonal and anti-diagonal unitary matrices.
(Note that Parker and Rijmen~\cite{B40} refer to PAR$_{\mathcal{U}}$ as PAR$_{l}$.)
\end{definition}

\begin{theorem}[Parker and Rijmen~\cite{B40}]\label{th:bipartite}
If $\boldsymbol{s}$ corresponds to a bipartite graph, then
PAR$_{\mathcal{U}}(\boldsymbol{s})$ = PAR$_{IH}(\boldsymbol{s})$,
where PAR$_{IH}$ is the peak-to-average power ratio with respect to the
transform set $\{I,H\}^n$.
\end{theorem}

It is obvious that $\{I,H\}^n \subset \{I,H,N\}^n \subset \mathcal{U}^n$,
and it follows that PAR$_{IH} \le \text{PAR}_{IHN} \le \text{PAR}_{\mathcal{U}}$.
We then get the following corollary of \autoref{th:indset} and \autoref{th:bipartite}.

\begin{corollary}
If an LC orbit, $[G]$, contains a bipartite graph, then
all functions corresponding to graphs in the orbit will have PAR$_{\mathcal{U}} = 2^{\lambda(G)}$.
\end{corollary}

All LC orbits with a bipartite member have PAR$_{IHN} = $ PAR$_{\mathcal{U}}$, but
note that these orbits will always have PAR$_{\mathcal{U}} 
\ge 2^{\left\lceil\frac{n}{2}\right\rceil}$~\cite{B40},
and that the fraction of LC orbits which have a bipartite member appears to decrease
exponentially as the number of vertices increases.
In the general case, PAR$_{IHN}$ is only a lower bound on PAR$_{\mathcal{U}}$. For example,
the Hexacode has PAR$_{IHN} = 4$, but a tighter lower bound on PAR$_{\mathcal{U}}$ is $4.486$~\cite{B40}.
(This bound has later been improved to $5.103$~\cite{B39}.)

\section{Constructions for Low PAR}

So far we have only considered the PAR$_{IHN}$ of quadratic Boolean functions.
For cryptographic purposes, we are interested in Boolean functions of degree higher
than $2$. As shown in \autoref{sec:qcbool}, such functions correspond
to hypergraphs and non-quadratic quantum codes.
\autoref{tab:nonquadcodes} on page~\pageref*{tab:nonquadcodes} gives the
value of PAR$_{IHN}$ for some non-quadratic Boolean functions with high APC distance.
Many of these functions have the same PAR$_{IHN}$ as the best quadratic functions,
but no non-quadratic function with lower PAR$_{IHN}$ than the best quadratic 
functions has yet been found.
Exhaustive searching for non-quadratic Boolean functions with low PAR$_{IHN}$
becomes infeasible for more than a few variables. We therefore propose a construction technique
using the best quadratic functions as building blocks~\cite{B16}.
Before we describe our construction we must first state what we mean by ``low PAR$_{IHN}$''. 
For $n = 6$ to $n = 10$ we computed PAR$_{IHN}$ for samples from the space $\mathbb{Z}_2^{2^n}$,
to determine the range of PAR$_{IHN}$ we can expect just by guessing.
\autoref{PARSamples} summarises these results. If we can
construct Boolean functions with PAR$_{IHN}$ lower than the sampled minimum,
we can consider our construction to be somewhat successful.

\begin{table}
\centering
\caption{Sampled Range of PAR$_{IHN}$ for Length ($n$) from 6 to 10\label{PARSamples}}
\begin{tabular}{cr@{,}lr@{--}l}
\toprule
$n$ & \multicolumn{2}{c}{Samples} & \multicolumn{2}{c}{Range of PAR$_{IHN}$} \\
\midrule
 6 & \hphantom{0}50&000 & 6.5&25.0 \\
 7 & 20&000 & 9.0&28.125 \\
 8 &  5&000 & 12.25&28.125 \\
 9 &  2&000 & 14.0625&30.25 \\
10 &  1&000 & 18.0&34.03 \\
\bottomrule
\end{tabular}
\end{table}

Parker and Tellambura~\cite{B42,B41} 
proposed a generalisation of the Maiorana-McFarland construction for
Boolean functions that satisfies a tight upper bound on PAR with respect
to the $\{H,N\}^n$ transform (and other transform sets),
this being a form of Golay Complementary
Set construction and a generalisation of the construction of
Rudin and Shapiro and of Davis and Jedwab~\cite{B17}.

\begin{figure}
 \centering
 \includegraphics{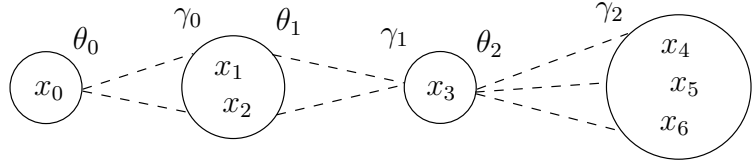}
 \caption{Example of Construction for PAR$_{HN} \le 8$}
 \label{construction1}
\end{figure}

\begin{construction}\label{con:constr1}
Let $p(\boldsymbol{x})$ be a Boolean function on
$n = \sum_{j=0}^{L-1} t_j$
variables, where $T = \{t_0,t_1,\ldots,t_{L-1}\}$ is a set of positive
integers and
$\boldsymbol{x} \in \mathbb{Z}_2^n$. Let $\boldsymbol{y_j} \in \mathbb{Z}_2^{t_j}$, $0 \le j < L$,
such that $\boldsymbol{x} = \boldsymbol{y_0} \times \boldsymbol{y_1} \times \cdots \times \boldsymbol{y_{L-1}}$.
Construct $p(\boldsymbol{x})$ as follows.
\begin{equation}
p(\boldsymbol{x}) = \sum_{j=0}^{L-2} \theta_j(\boldsymbol{y_j})\gamma_j(\boldsymbol{y_{j+1}})
 + \sum_{j=0}^{L-1} g_j(\boldsymbol{y_j}),
\end{equation}
where $\theta_j$ is a
permutation: $\mathbb{Z}_2^{t_j} \rightarrow \mathbb{Z}_2^{t_{j+1}}$,
$\gamma_j$ is a
permutation: $\mathbb{Z}_2^{t_{j+1}} \rightarrow \mathbb{Z}_2^{t_j}$,
and $g_j$ is any Boolean function of $t_j$ variables.
It has been shown~\cite{B42} that the function $p(\boldsymbol{x})$ will
have PAR$_{HN} \le 2^{t_{\text{max}}}$, where
$t_{\text{max}}$ is the largest integer in $T$.
It is helpful to visualise this construction graphically, as in \autoref{construction1}.
In this example, the size of the largest partition is $3$, so PAR$_{HN} \le 8$, regardless
of what choices we make for $\theta_j$, $\gamma_j$, and $g_j$.
\end{construction}

Observe that if we set $L=2$, $t=t_0=t_1$, let $\theta_0$ be the identity permutation,
and $g_0 = 0$,
Construction~\ref{con:constr1} reduces to the Maiorana-McFarland construction over $2t$ variables.
Construction~\ref{con:constr1} can also be viewed as a generalisation of the path graph, 
$f(\boldsymbol{x}) = x_0x_1 + x_1x_2 + \cdots + x_{n-2}x_{n-1}$, 
which has optimal PAR with respect to $\{H,N\}^n$.
Unfortunately, the path graph is not a particularly good construction for low PAR$_{IHN}$. But
as we have seen, graphs corresponding to self-dual additive codes over $\GF(4)$ with
high distance do give us Boolean functions with low PAR$_{IHN}$.

\begin{construction}\label{con:constr2}
We propose the generalised construction,
\begin{equation}
p(\boldsymbol{x}) =
\sum_{i=0}^{L-1}\sum_{j=i+1}^{L-1} \Gamma_{i,j}(\boldsymbol{y_i})\Gamma_{j,i}(\boldsymbol{y_j})
 + \sum_{j=0}^{L-1} g_j(\boldsymbol{y_j}),
\end{equation}
where $\Gamma_{i,j}$ is either a permutation: $\mathbb{Z}_2^{t_i} \rightarrow \mathbb{Z}_2^{t_j}$, or
$\Gamma_{i,j} = 0$, and $g_j$ is any Boolean function on $t_j$ variables.
\end{construction}

It is evident that $\Gamma$ can be thought of as a ``generalised adjacency matrix'', where the entries,
$\Gamma_{i,j}$, are no longer 0 or 1 but, instead, 0 or permutations from
$\mathbb{Z}_2^{t_i}$ to $\mathbb{Z}_2^{t_j}$. Construction~\ref{con:constr1} then becomes a
special case where $\Gamma_{i,j} = 0$ except for when $j = i + 1$, i.e., a ``generalised
adjacency matrix'' of the path graph.
In order to minimise PAR$_{IHN}$ we choose the form of the matrix $\Gamma$
according to the adjacency matrix of a self-dual additive code
over $\GF(4)$ with high distance.
We also choose the ``offset'' functions, $g_j$, to be
Boolean functions corresponding to self-dual additive codes
over $\GF(4)$ with high distance.
Finally for the non-zero $\Gamma_{i,j}$ entries,
we choose selected permutations,
preferably nonlinear to increase the overall degree.
Here are some initial results which demonstrate that, using Construction~\ref{con:constr2},
we can construct
Boolean functions of algebraic degree greater than 2 with low PAR$_{IHN}$.

\begin{figure}
 \centering
 \includegraphics{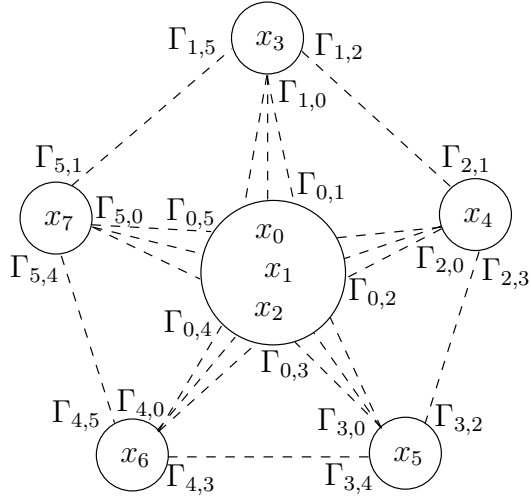}
 \caption{Example of Construction for Low PAR$_{IHN}$}
 \label{construction2}
\end{figure}

\begin{example}[$n=8$]
Use the Hexacode graph
$f = 01,$ $02,$ $03,$ $04,$ $05,$ $12,$ $23,$ $34,$ $45,$ $51$ as a template.
Let $t_0 = 3$, $t_1 = t_2 = t_3 = t_4 = t_5 = 1$. (See \autoref{construction2}.)
We use the following matrix $\Gamma$.
\[
\Gamma = \begin{pmatrix}
0\hspace{3pt} & \hspace{3pt}02,1\hspace{3pt} & \hspace{3pt}02,1\hspace{3pt} & 
  \hspace{3pt}02,1\hspace{3pt} & \hspace{3pt}02,1\hspace{3pt} & \hspace{3pt}02,1\\
3 & 0 & 3 & 0 & 0 & 3\\
4 & 4 & 0 & 4 & 0 & 0\\
5 & 0 & 5 & 0 & 5 & 0\\
6 & 0 & 0 & 6 & 0 & 6\\
7 & 7 & 0 & 0 & 7 & 0
\end{pmatrix}
\]
Let $g_0(\boldsymbol{y_0}) = 01,$ $02,$ $12$
and all other $g_j$ any arbitrary affine functions.
Then, using Construction~\ref{con:constr2} to construct $p(\boldsymbol{x})$ we get
$p(\boldsymbol{x}) = 023,$ $024,$ $025,$ $026,$ $027,$ $01,$ $02,$ $12,$ $13,$ 
 $14,$ $15,$ $16,$ $17,$ $34,$ $37,$ $45,$ $56,$ $67$.
It can be verified that $p(\boldsymbol{x})$ has PAR$_{IHN} = 9.0$.
\end{example}

\begin{example}[$n=8$]
Use the Hexacode graph
$f = 01,$ $02,$ $03,$ $04,$ $05,$ $12,$ $23,$ $34,$ $45,$ $51$ as a template.
Let $t_0 = 3$, $t_1 = t_2 = t_3 = t_4 = t_5 = 1$. (See \autoref{construction2}.)
We use the following matrix $\Gamma$.
\[
\Gamma = \begin{pmatrix}
0\hspace{3pt} & \hspace{3pt}02,1\hspace{3pt} & \hspace{3pt}12,0,1,2\hspace{3pt} & 
  \hspace{3pt}01,02,12,1,2\hspace{3pt} & \hspace{3pt}01,02,12\hspace{3pt} & \hspace{3pt}02,12,1,2\\
3 & 0 & 3 & 0 & 0 & 3\\
4 & 4 & 0 & 4 & 0 & 0\\
5 & 0 & 5 & 0 & 5 & 0\\
6 & 0 & 0 & 6 & 0 & 6\\
7 & 7 & 0 & 0 & 7 & 0
\end{pmatrix}
\]
Let $g_0(\boldsymbol{y_0}) = 01,12$
and all other $g_j$ any arbitrary affine functions.
Then, using Construction~\ref{con:constr2} to construct $p(\boldsymbol{x})$ we get
$p(\boldsymbol{x}) = 015,$ $016,$ $023,$ $025,$ $026,$ $027,$ $124,$ $125,$ $126,$ $127,$ 
$01,$ $04,$ $12,$ $13,$ $14,$ $15,$ $17,$ $24,$ $25,$ $27,$ $34,$ $37,$ $45,$ $56,$ $67$,
where $p(\boldsymbol{x})$ has PAR$_{IHN} = 9.0$.
\end{example}

\begin{figure}
 \centering
 \includegraphics{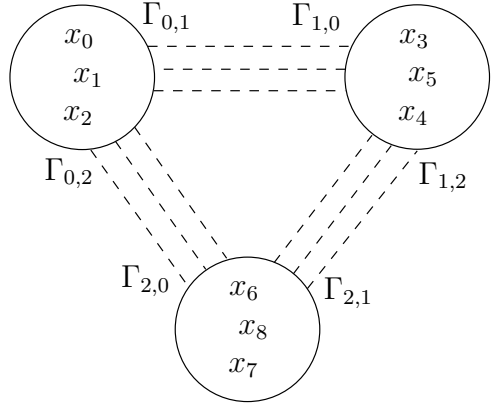}
 \caption{Example of Construction for Low PAR$_{IHN}$}
 \label{construction3}
\end{figure}

\begin{example}[$n=9$]
Use the triangle graph
$f = 01,02,12$ as a template.
Let $t_0 = t_1 = t_2 = 3$. (See \autoref{construction3}.)
Assign the permutations
\begin{align*}
\Gamma_{0,1} = \Gamma_{0,2} &= (12,0,1,2)(01,2)(02,1,2),\\
\Gamma_{1,0} &= (34,5)(35,4,5)(45,3,4,5),\\
\Gamma_{1,2} &= (45,3,4,5)(34,5)(35,4,5),\\
\Gamma_{2,0} &= (68,7,8)(78,6,7,8)(67,8),\\
\Gamma_{2,1} &= (78,6,7,8)(67,8)(68,7,8).
\end{align*}
Let $g_0(\boldsymbol{y_0}) = 01,02,12$, $g_1(\boldsymbol{y_1}) = 34,35,45$, 
and $g_2(\boldsymbol{y_2}) = 67,68,78$.
Then, using Construction~\ref{con:constr2} to construct $p(\boldsymbol{x})$ we get,
$ p(\boldsymbol{x}) = 0135,$ $0178,$ $0245,$ $0267,$ $1234,$ $1268,$ $3467,$ $3568,$ $4578,$ 
$014,$ $015,$ $016,$ $017,$ $018,$ $023,$ $024,$ $025,$ $028,$ $034,$ $068,$ $125,$ 
$127,$ $128,$ $134,$ $145,$ $167,$ $168,$ $234,$ $235,$ $245,$ $267,$ $268,$ $278,$ 
$348,$ $357,$ $358,$ $378,$ $456,$ $457,$ $458,$ $468,$ $478,$ $567,$ $568,$ $578,$ 
$05,$ $07,$ $08,$ $13,$ $14,$ $17,$ $23,$ $25,$ $26,$ $28,$ $36,$ $37,$ $38,$ $46,$ 
$56,$ $58,$ $01,$ $02,$ $12,$ $34,$ $35,$ $45,$ $67,$ $68,$ $78 $,
where $p(\boldsymbol{x})$ has PAR$_{IHN} = 10.25$.
\end{example}

The examples of our construction satisfy a low PAR$_{IHN}$.
Further work should ascertain the proper choice of permutations.
Finally, there is an even more obvious variation of Construction~\ref{con:constr2}, suggested by
the graphs of \autoref{l38}, where the functions $g_j$ are chosen
either to be quadratic cliques or to be further ``nested'' versions of Construction~\ref{con:constr2}.

\section{Quantum Interpretations of Spectral Measures}

A quantum code with good error correcting capability must produce encoded states
that are highly entangled. The single quantum state corresponding to
a zero-dimensional quantum code of high distance must also be a highly entangled
state~\cite{hein}. The distance of self-dual quantum codes, or more generally, the APC distance 
of Boolean functions, may therefore be interpreted as entanglement 
measures of the corresponding quantum states. No entanglement measure is known
that completely quantifies the degree of entanglement in a pure multipartite quantum state
of more than 3 qubits, and APC distance is also only a partial measure.

The PAR$_{\mathcal{U}}$ of a vector $\boldsymbol{s}$ also gives information
about the quantum state with probability distribution vector $\boldsymbol{s}$.
We recall from \autoref{l3} that the values $|s_i|^2$, where $i \in \mathbb{Z}_{2^n}$, 
are the probabilities of observing each of $2^n$ basis states in a particular
measurement basis associated with $\boldsymbol{s}$.
The maximum value of $|s_i|^2$ over all $i \in \mathbb{Z}_{2^n}$ gives the probability of
the most likely outcome of a measurement in this particular measurement basis,
and can therefore be interpreted as a partial measure of the uncertainty of the quantum state. 
If this value is high, the state has low uncertainty in this measurement basis.
A local unitary transformation corresponds to a change of measurement basis.
Local unitary transformations of the vector $\boldsymbol{s}$ are reversible and
do not change the overall entanglement properties of the corresponding quantum state,
but the measurement basis is changed, and the magnitudes of the coefficients of 
$\boldsymbol{s}$, and therefore the uncertainty, may also change.
If we can apply any local unitary transformation to $\boldsymbol{s}$, i.e., use any measurement basis,
what is the lowest uncertainty, i.e., the highest probability of any basis state, we can achieve?
The answer to this question is simply the value of PAR$_{\mathcal{U}}(\boldsymbol{s})$, as 
defined in \autoref{def:parl}. Since PAR$_{IHN}$ is a lower bound on PAR$_{\mathcal{U}}$, it is 
also a lower bound on the uncertainty of a quantum state, where uncertainty means
the highest probability of observing any basis state in any measurement basis.

\begin{definition}
Given an $n$-qubit quantum state $\ket{\psi}$, we can write
\begin{equation}
\ket{\psi} = \sum_{i=1}^R c_i \ket{\phi_i},
\end{equation}
where $c_i \in \mathbb{C}$ and $\ket{\phi_i}$, for $1 \le i \le R$, are $R$ of the $2^n$ basis
states in a particular measurement basis. Let $r$ be the smallest
possible value of $R$ for any measurement basis.
The \emph{Schmidt measure} of $\ket{\psi}$ is then given by
\begin{equation}
E_S(\ket{\psi}) = \log_2(r).
\end{equation}
\end{definition}

The Schmidt measure was defined by 
Eisert and Briegel~\cite{B18} and later studied in the context of 
graph states by Hein, Eisert, and Briegel~\cite{hein}.
We can also define the Schmidt measure in terms of the probability
distribution vector associated with a quantum state.

\begin{definition}
Let $\boldsymbol{s}$ be a complex-valued vector of length $2^n$
corresponding to the quantum state $\ket{\psi}$.
Given a local unitary transform $\boldsymbol{S} = U \boldsymbol{s}$, 
where $U \in \mathcal{U}^n$, we define $R = |\{S_i \ne 0 \mid 0 \le i < n\}|$, i.e., the number
of non-zero coefficients of $\boldsymbol{S}$.
Let $r$ be the smallest possible value of $R$ for any
transformation $U \in \mathcal{U}^n$.
The Schmidt measure of $\ket{\psi}$ is then $E_S(\ket{\psi}) = \log_2(r)$.
\end{definition}

\begin{proposition}
Let $\ket{\psi_f}$ be a bipolar quantum state described by the vector 
$\boldsymbol{s} = 2^{-\frac{n}{2}}(-1)^{f(\boldsymbol{x})}$, where
$f$ is a quadratic Boolean function.
We can then give an upper bound on the Schmidt measure of $\ket{\psi_f}$,
\begin{equation}\label{l39}
E_S(\ket{\psi_f}) \le n - log_2(\text{PAR}_{IHN}(\boldsymbol{s})).
\end{equation}
\end{proposition}
\begin{proof}
It can be shown that the coefficients of any $\{I,H,N\}^n$ transform of a quadratic Boolean
function have only two different magnitudes~\cite{B52}.
It follows that in the $\{I,H,N\}^n$ transform containing the highest
spectral peak, all coefficients must either be zero or have the same magnitude as the peak,
and therefore that there must be $\frac{2^n}{\text{PAR}_{IHN}(\boldsymbol{s})}$ non-zero coefficients.
\end{proof}

By \autoref{th:indset}, \eqref{l39} can also be expressed in terms of $\lambda$,
\begin{equation}
E_S(\ket{\psi_f}) \le n - \lambda(G),
\end{equation}
where $G$ is the graph corresponding to $\boldsymbol{s}$ and $\lambda(G)$ is
the size of the maximum independent set over the LC orbit of $G$.
The size of the \emph{minimum vertex cover} of a graph $G$, denoted $\nu(G)$, is
given by $\nu(G) = n - \alpha(G)$, where $\alpha(G)$ is the size
of the maximum independent set of $G$. It was shown by Hein et~al.~\cite{hein} that 
\begin{equation}
E_S(\ket{\psi_f}) \le \nu(G).
\end{equation}

\begin{proposition}
Let $\ket{\psi_f}$ be a bipolar quantum state described by the vector 
$\boldsymbol{s} = 2^{-\frac{n}{2}}(-1)^{f(\boldsymbol{x})}$, where
$f$ is any Boolean function.
The Schmidt measure of $\ket{\psi_f}$ is then lower bounded by
\begin{equation}
E_S(\ket{\psi_f}) \ge n - log_2(\text{PAR}_{\mathcal{U}}(\boldsymbol{s})).
\end{equation}
\end{proposition}
\begin{proof}
In this case, the transforms may have more than two different magnitudes.
We do therefore not know how many non-zero coefficients the transform with the
highest peak has, but we know that no transform can have any coefficient
with higher magnitude than PAR$_{\mathcal{U}}$, and therefore that
all transforms must have at least 
$\left\lceil\frac{2^n}{\text{PAR}_{\mathcal{U}}(\boldsymbol{s})}\right\rceil$ non-zero coefficients.
\end{proof}

\begin{corollary}
It follows from \autoref{th:bipartite} that if 
$\boldsymbol{s} = 2^{-\frac{n}{2}}(-1)^{f(\boldsymbol{x})}$ corresponds to a bipartite graph $G$, then
\begin{equation}
E_S(\ket{\psi_f}) = n - log_2(\text{PAR}_{\mathcal{U}}(\boldsymbol{s})) =
n - log_2(\text{PAR}_{IH}(\boldsymbol{s})).
\end{equation}
\end{corollary}

When we look at non-quadratic Boolean functions, or when we use other transform
sets than $\{I,H,N\}$, the transforms may have more than two different magnitudes.
The connection between PAR and Schmidt measure in these cases is not clear.

We recall that a Boolean function is called bent if its $\{H\}^n$ transform is flat.
Riera, Petrides, and Parker~\cite{B51,B53} have proposed that this criteria can be generalised
by counting how many of the $6^n$ $\{I,H,N\}^n$ transforms of a Boolean function are flat.
A high number of flat spectra indicates maximal distance to a large subset of generalised affine functions,
and a high number of maximum uncertainty measurement bases within the complete set of
$\{I,H,N\}^n$ measurement bases.
Other entanglement measures can also be derived from the $\{I,H,N\}^n$ spectrum.

\begin{definition}
Given a vector $\boldsymbol{s}$, let $a$ be the sum of the fourth powers of all 
$6^n$ spectral magnitudes in the $\{I,H,N\}^n$ spectrum, i.e.,
\begin{equation}
a = \sum_{\substack{ U \in \{I,H,N\}^n \\ k \in \mathbb{Z}_{2^n}}} |S_k|^4,
\end{equation}
where $\boldsymbol{S} = U\boldsymbol{s}$. 
The \emph{Clifford merit factor} (CMF)~\cite{cmf} of $\boldsymbol{s}$ is then given by
\begin{equation}
\text{CMF}(\boldsymbol{s}) = \frac{6^n}{a - 6^n}.
\end{equation}
\end{definition}

High APC distance, low PAR$_{IHN}$, and high CMF are clearly correlated properties, 
and the best self-dual quantum codes optimise all of them. Non-quadratic Boolean functions with
equally good properties as the best quadratic functions have also been found.
The CMF is an entanglement measure, and it can be shown that CMF remains invariant under any 
local unitary transformation, i.e., 
$\text{CMF}(\boldsymbol{s}) = \text{CMF}(U\boldsymbol{s})$, where $U \in \mathcal{U}^n$.
Since the overall entanglement of a quantum state does not change under local unitary transformations,
all entanglement measures should have this property. The Schmidt measure and 
PAR$_{\mathcal{U}}$ are also invariant under local unitary transformations.
It may be possible to generalise CMF by extending the transform set while increasing
the power of the spectral magnitudes. This could give an infinite sequence of
entanglement measures, all invariant under local unitary transformations, 
which converges towards PAR$_{\mathcal{U}}$.

\chapter{Conclusions and Open Problems}\label{chap:concl}

In this thesis we have studied zero-dimensional quantum codes 
and their interpretations as quantum states, self-dual additive 
codes over $\GF(4)$, graphs, and Boolean functions.
We have looked at different properties and generalisations under
each interpretation. For reasons of computational complexity, 
we have restricted our study to self-dual quantum codes
of length up to 30. But even for codes of such short length, there are many
interesting problems. 

\begin{problem}
As seen in \autoref{l16}, the best achievable distances for
codes of length from 23 to 27 are not known. What are the optimal distances
for these lengths? In particular, does there exist a $[[24,0,10]]$ 
quantum code?
\end{problem}

Many examples have been shown of self-dual quantum codes 
of high minimum distance with highly structured and regular graph representations.
In particular, we have searched all circulant graph codes with
length up to 30 for \emph{nested regular graphs}.
The nested regular description does not completely characterise
the structure of a graph, but we have also shown that nested regular graph
representations of strong codes contain long cycles.
Initial results further suggest that the long cycles should be arranged
in such a way that no smaller cycles are induced.
We have also identified graph representations with \emph{minimum regular vertex degree}
for many self-dual quantum codes. Graphs with minimum regular vertex degree have the lowest 
possible number of edges, and the corresponding generator matrices are therefore 
as sparse as possible. In many applications of classical coding theory, sparsity of
the generator matrix is a desired property.

\begin{problem}
Identify other regular graph structures than those listed in \autoref{tab:cycliccodes},
in particular for codes of length 28 and above.
\end{problem}

\begin{problem}
Is it possible to further generalise and extend the description of 
highly regular graphs corresponding to strong self-dual quantum codes?
\end{problem}

\begin{problem}
Devise a construction technique for nested regular graphs,
giving self-dual quantum codes with a predictable minimum distance.
\end{problem}

\begin{problem}
Find graphs with minimum regular vertex degree corresponding
to self-dual quantum codes of length above 27 and distance higher than 8.
Of particular interest is the existence of a graph with regular vertex 
degree~11 corresponding to a $[[30,0,12]]$ code.
\end{problem}

We investigated all self-dual quantum codes of length up to 30
corresponding to graphs with circulant adjacency matrices, and
found codes of equally high distance to those of
Gulliver and Kim~\cite{B29} in their more general
search of all self-dual additive codes over GF(4) with circulant
generator matrices. 

\begin{problem}
How many $\GF(4)$-circulant codes also have circulant graph
representations, and are codes with circulant graph representations stronger 
than general codes?
\end{problem}

We have seen that the quadratic residue construction produces
codes corresponding to strongly regular Paley graphs.
The largest clique in a Paley graph is known to be very small,
compared to the number of vertices in the graph~\cite{B55}. 
Since it can also be shown that all Paley graphs are isomorphic to their complements, 
their independence numbers must be equally low.
We have seen that graphs corresponding to strong self-dual quantum codes
have small independent sets over their whole \emph{LC orbit}, and
this implies that the largest independent set of a Paley graph remains small
when any sequence of local complementations is applied.

\begin{problem}
Give bounds on the size of the largest independent set
over the LC orbit of a Paley graph. 
\end{problem}

\begin{problem}
Can other families of strongly regular graphs, or families of 
graphs known to have small independence numbers, be used to construct
self-dual quantum codes of high distance?
\end{problem}

We showed that some self-dual quantum codes, in particular
the $[[11,0,5]]$ and $[[18,0,8]]$ codes, have no regular
graph representation. Glynn et al.~\cite{B23} have used finite geometry
to construct and characterise the $[[18,0,8]]$ code.

\begin{problem}
Do strong self-dual quantum codes with no regular graph
representation correspond to graphs that are highly structured in some other way,
and can this structure be generalised?
\end{problem}

We have classified all self-dual additive codes over $\GF(4)$ of length up to 12.
Enumerating all codes of higher length, with a reasonable amount 
of computational resources, is not possible using the algorithms 
in \autoref{sec:enumerate}. 

\begin{problem}
Devise an algorithm for canonising a graph, like \autoref{alg:lccanonise}, 
but without requiring that the whole LC orbit of the graph is generated. 
This could give a much faster method for enumerating LC orbits. 
\end{problem}

\begin{problem}
Devise an efficient algorithm for determining whether 
two graphs are LC-equivalent, like the one described by 
Van~den~Nest et~al.~\cite{B58} and Bouchet~\cite{B6},
but that also considers equivalence via graph isomorphism.
\end{problem}

We have seen that the \emph{APC distance} of a Boolean function is
equal to the distance of the corresponding zero-dimensional quantum code.
For quadratic Boolean functions, which correspond to self-dual
additive codes over $\GF(4)$, we can use the efficient \autoref{l25} 
to find the APC distance. Boolean functions of higher degree correspond to 
non-quadratic quantum codes, and we have used the much more complex 
\autoref{alg:distnonquad} to find their distance. We have found 
several non-quadratic Boolean functions, listed in \autoref{tab:nonquadcodes},
with exactly the same APC distance and PAR$_{IHN}$ as the best quadratic functions.

\begin{problem}
Are there better algorithms for finding the APC distance of a non-quadratic 
Boolean function? For instance, do cubic Boolean functions correspond to some 
generalisation of additive codes over $\GF(4)$ where the distance can be 
found efficiently?
\end{problem}

\begin{problem}
Improve \autoref{alg:nonquadorbit} and classify all orbits
in $\mathcal{O}_{1,n}$ and $\mathcal{O}_{2,n}$ for $n > 5$.
\end{problem}

\begin{problem}
Do there exist non-quadratic Boolean functions with higher APC distance
or lower PAR$_{IHN}$ than the best quadratic functions?
\end{problem}

We have defined $\lambda$, the size of the largest independent set over an LC orbit,
and $\Lambda_n$, the minimum value of $\lambda$ over all LC orbits of graphs on $n$ vertices.
It has also been shown that PAR$_{IHN} = 2^\lambda$ for quadratic Boolean functions,
and we have given bounds on $\Lambda_n$. The bounds on $\Lambda_n$ are also bounds 
on PAR$_{IHN}$ and PAR$_{\mathcal{U}}$.

\begin{problem}
Improve the bounds on $\Lambda_n$, or find exact values of
$\Lambda_n$ for $n \ge 15$.
\end{problem}

\begin{problem}
Can other symmetries of non-quadratic Boolean functions be found
by adding other matrices to the set $\{I,H,N\}$?
\end{problem}

\begin{problem}
Is there an operation, similar to local complementation, that generates
orbits of equivalent hypergraphs?
\end{problem}

\begin{problem}
Is there a relationship between the maximum independent set in the orbit
of a hypergraph and the PAR$_{IHN}$ of the corresponding Boolean function?
Maybe we need to consider PAR with respect to a larger transform set than $\{I,H,N\}$ in order
to find such a relationship.
\end{problem}

We have seen that Construction~\ref{con:constr1} gives Boolean functions of high degree with
predictable PAR$_{HN}$, and we have proposed the more general Construction~\ref{con:constr2}
for Boolean functions of high degree with low PAR$_{IHN}$.

\begin{problem}
What is the best choice of permutations in Construction~\ref{con:constr2}?
\end{problem}

\begin{problem}
Find bounds on the PAR$_{IHN}$ of functions generated by Construction~\ref{con:constr2}
when specific permutations are used.
\end{problem}

Construction~\ref{con:constr1} is a generalisation of the path graph, which have
optimal PAR$_{HN}$ but high PAR$_{IHN}$.
It can be shown that complete graphs (cliques) have optimal PAR$_{IH}$, but that
they also have high PAR$_{IHN}$. Functions that optimise PAR$_{IHN}$ 
should do well for both PAR$_{HN}$ and PAR$_{IH}$, and should
perhaps be some compromise between clique graphs and path graphs.
We have seen that some of the quadratic Boolean functions with 
lowest PAR$_{IHN}$ correspond to nested clique graphs,
which contain both cliques and long disjoint paths.

Zero-dimensional quantum codes of high distance correspond to
highly entangled quantum states, and also to Boolean functions
that satisfy the \emph{aperiodic propagation criterion} of high
order or degree. APC is related to several other criteria which measure
the cryptographic strength of a Boolean function. This
suggests that a highly entangled quantum state may correspond to
a cryptographically strong Boolean function.
We have shown that spectral properties of Boolean functions,
such as APC distance, PAR$_{\mathcal{U}}$, and PAR$_{IHN}$
can be used to measure the degree of entanglement in a quantum state,
as can the Schmidt measure and the Clifford Merit Factor~\cite{cmf}.
In this thesis, we have studied quantum states 
with coefficients from the set $\{-1,1\}$, represented by
the bipolar truth table of a Boolean function. This is, 
of course, only a subset of all possible quantum states.

\begin{problem}
Classify spectral measures of quantum states using other transform sets
than $\{I,H,N\}^n$.
\end{problem}

\begin{problem}
Find efficient techniques for approximating PAR$_{\mathcal{U}}$.
\end{problem}

\begin{problem}
Study the properties of zero-dimensional quantum codes corresponding to 
non-bipolar quantum states, for instance quantum states with coefficients 
from the sets $\{0,1\}$ or $\{-1,0,1\}$.
\end{problem}

\backmatter

\end{document}